\documentclass[11pt]{article}
\usepackage{amssymb}
\usepackage{multirow}
\usepackage{amsmath}
\usepackage{amsfonts}
 \usepackage{pifont}
\usepackage{amsmath,amsfonts,amssymb,theorem,graphicx,listings,txfonts}
\usepackage{graphics}
 \usepackage{caption2}
\usepackage{flafter}
 \usepackage{indentfirst}
\usepackage{epsfig}
  \usepackage{mathrsfs}
  \usepackage{float}
\usepackage{subfigure}
 \usepackage{cite}
\newtheorem{theorem}{Theorem}[section]

\newtheorem{proposition}[theorem]{Proposition}

\newtheorem{definition}[theorem]{Definition}
\newtheorem{algorithm}[theorem]{Algorithm}

\theoremstyle{example}
\newtheorem{example}[theorem]{Example}
\theoremstyle{programme}

\theoremstyle{property}

\theoremstyle{problem}

\renewcommand{\arraystretch}{1}

\title{Incremental approaches to knowledge reduction of covering decision information systems with variations of coverings}

\author
{Guangming Lang$^{1,2,3}$ \hspace{1cm} 
\thanks{Corresponding author.\quad Tel./fax: +86 021 69585800,
\newline\mbox{}\hspace{0.55cm}
E-mail address: langguangming1984@126.com(G.M.Lang), cmjlong@163.com
(M.J.Cai). }\hspace{1cm}
 Mingjie Cai$^{4}$\\
\small {$^{1}$ Department of Computer Science and Technology, Tongji University}\\
\small {Shanghai 201804, P.R. China}\\
\small {$^{2}$ School of Mathematics and Computer Science, Changsha University of Science and Technology}\\
\small {Changsha, Hunan 410114, P.R. China}\\
\small {$^{3}$ The Key Laboratory of Embedded System and Service Computing, Ministry of Education, Tongji University}\\
\small {Shanghai 201804, P.R. China}\\
\small {$^{4}$ College of Mathematics and Econometrics, Hunan University}\\
\small {Changsha, Hunan 410004, P.R. China}}
\date{}

\begin{document}
\maketitle \baselineskip=17pt
\begin{center}
\begin{quote}
{{\bf Abstract.} In practical situations, calculating approximations of concepts is the central step for
knowledge reduction of dynamic covering decision information system, which has received growing interests of researchers in recent years. In this
paper, the second and sixth lower and upper approximations of sets
in dynamic covering information systems with variations of coverings
are computed from the perspective of matrix using incremental approaches. Especially, effective
algorithms are designed for calculating the second and sixth lower and
upper approximations of sets in dynamic covering information systems
with the immigration of coverings. Experimental results demonstrate
that the designed algorithms provide an efficient and effective
method for constructing the second and sixth lower and upper
approximations of sets in dynamic covering information systems. Two
examples are explored to illustrate the process of knowledge
reduction of dynamic covering decision information systems with the covering
immigration.

\\}
\end{quote}
\end{center}
\renewcommand{\thesection}{\arabic{section}}

\section{Introduction}

Covering-based rough set theory\cite{Zakowski} as a generalization of Pawlak's rough sets is a powerful mathematical tool to deal with uncertainty and imprecise information in the field of knowledge discovery and rule acquisition. To handle with uncertainty knowledge, researchers have investigated covering-based rough set theory\cite{Liu7,Chen4,Miao,Qian} and presented three types  approximation operators summarized by Yao\cite{Yao2} as follows: element-based operators, granular-based operators and system-based operators for covering approximation spaces, and  discussed the relationships among them. Additionally, all approximation operators are also classified into dual and non-dual operators and their inner properties are investigated.

Researchers have proposed many lower and upper approximation operators with respect to different backgrounds. Particularly, they\cite{Tan1,Tan2,Wang3,Yang1,Yang2,Yao1,Zhang1,Zhang4,Zhang5,Zhang6,Zhu1,Zhu2,Zhu3} have investigated approximation operators from the view of matrix. For example, Liu\cite{Liu1} provided a new matrix view of rough set theory for Pawlak's lower and upper approximation operators. He also represented a fuzzy equivalence relation using a fuzzy matrix and redefined the pair of lower and upper approximation operators for fuzzy sets using the matrix representation in a fuzzy approximation space.
Wang et al.\cite{Wang3} proposed
the concepts of the type-1 and type-2 characteristic matrices of coverings and transformed
the computation of the second, fifth and sixth lower and upper approximations of a set into products of the type-1 and type-2
characteristic matrices and the characteristic function of the set in covering approximation spaces. Zhang et al.\cite{Zhang1} proposed the matrix characterizations of the lower and upper approximations for set-valued information systems. He\cite{Zhang4,Zhang5,Zhang6} also presented efficient parallel boolean matrix based algorithms for computing rough set approximations in composite information systems and incomplete information systems.
Actually, because of the dynamic characteristic of data collection, there are a lot of dynamic information systems with variations of object sets, attribute sets and attribute values, and researchers\cite{Chen1,Chen2,Chen3,Lang1,Lang2,Lang3,Lang4,Lang5,Li1,Li2,Li3,Li4,Liang1,Liu2,Liu3,Liu4,
Liu5,Liu6,Luo1,Luo2,Luo3,Sang,Shu1,Shu2,Shu3,Wang1,Wang2,Yang3,Zhang1,Zhang2,Zhang3} have focused on knowledge reduction of dynamic information systems.
Especially, researchers\cite{Zhang1,Lang1,Lang2} have computed approximations of sets for knowledge reduction of dynamic information systems from the view of matrix. For instance,
Zhang et al.\cite{Zhang1} provided incremental approaches to updating the relation matrix for computing the lower and upper approximations with dynamic attribute variation in set-valued information systems. They also proposed effective algorithms of computing composite rough set approximations for dynamic data mining.
Lang et al.\cite{Lang1,Lang2} presented incremental algorithms for computing the second and sixth lower and upper approximations of sets from the view of matrix and investigated knowledge reduction of dynamic covering information systems with variations of objects. In practical situations, there are many dynamic covering information systems with the immigration and emigration of coverings, and computing the second and sixth lower and upper approximations of sets is time-consuming using the non-incremental algorithms in these dynamic covering information systems, it also costs more time to conduct knowledge reduction of dynamic covering information systems with variations of coverings. Therefore, it is urgent to propose effective approaches to updating the second and sixth lower and upper approximations of sets for knowledge reduction of dynamic covering decision information systems with the covering variations.

This work is to investigate knowledge reduction of dynamic covering decision information systems.
First, we investigate the basic properties of dynamic covering information systems with variations of coverings. Particularly, we study the properties of the type-1 and type-2 characteristic matrices with the covering variations and the relationship between the original type-1 and type-2 characteristic matrices and the updated type-1 and type-2 characteristic matrices. We also provide incremental algorithms for updating the second and sixth lower and upper approximations of sets using the type-1 and type-2 characteristic matrices, respectively. We employ examples to illustrate how to update the second and sixth lower and upper approximations of sets with variations of coverings. Second, we generate randomly ten dynamic covering information systems with the covering variations randomly and compute the second and sixth lower and upper approximations of sets in these dynamic covering information systems. We also employ experimental results to illustrate the proposed algorithms are effective to update the second and sixth lower and upper approximations of sets in dynamic covering information systems. Third, we employ two examples to demonstrate that the designed algorithms are effective to conduct knowledge reduction of dynamic covering decision information systems with immigrations of coverings, which will enrich covering-based rough set theory from the matrix view.

The rest of this paper is organized as follows: Section 2 briefly
reviews the basic concepts of covering-based rough set theory. In
Section 3, we update the type-1 and type-2 characteristic matrices
in dynamic covering information systems with variations of
coverings. We design the incremental algorithms for computing the
second and sixth lower and upper approximations of sets. We also
provide examples to demonstrate how to calculate the second and sixth
lower and upper approximations of sets. In Section 4, the
experimental results illustrate the incremental algorithms are
effective to construct the second and sixth lower and upper
approximations of sets in dynamic covering information systems with
the covering immigration. In Section 5, we explore two examples
to illustrate how to conduct knowledge reduction of dynamic covering
decision information systems with the covering immigration.
Concluding remarks and further research are given in Section 6.

\section{Preliminaries}

In this section, we  briefly review some concepts related to covering-based rough
sets.

\begin{definition}\cite{Zakowski}
Let $U$ be a finite universe of discourse, and $\mathscr{C}$ a
family of subsets of $U$. Then $\mathscr{C}$ is called a covering of
$U$ if none of elements of $\mathscr{C}$ is empty and
$\bigcup\{C|C\in \mathscr{C}\}=U$. Furthermore, $(U,\mathscr{C})$ is referred to as a covering approximation
space.
\end{definition}

If $U$ is a finite universe of discourse, and $\mathscr{D}=\{\mathscr{C}_{1},\mathscr{C}_{2},...,\mathscr{C}_{m}\}$, where $\mathscr{C}_{i}(1\leq i\leq m)$ is a
covering of $U$, then $(U,\mathscr{D})$ is called a covering information system, which can be viewed as a covering approximation space. Furthermore, if the coverings of $\mathscr{D}$ are classified into two  categories: conditional attribute-based coverings and decision attribute-based coverings, then $(U,\mathscr{D})$ is referred to as a covering decision information system. For convenience, a covering decision information system is denoted as $(U,\mathscr{D}_{C}\cup \mathscr{D}_{D} )$, where $\mathscr{D}_{C}$  and $\mathscr{D}_{D}$ mean conditional attribute-based coverings and decision attribute-based coverings, respectively.

\begin{definition}\cite{Wang3}
Let $(U,\mathscr{C})$ be a covering approximation
space, and $N(x)=\bigcap\{C_{i}|x\in C_{i}\in \mathscr{C}\}$ for $x\in U$. For any
$X\subseteq U$, the second and sixth upper and lower approximations
of $X$ with respect to $\mathscr{C}$ are defined as follows:

$(1)$ $SH_{\mathscr{C}}(X)=\bigcup\{C\in\mathscr{C}|C\cap X\neq
\emptyset\}$, $SL_{\mathscr{C}}(X)=[SH_{\mathscr{C}}(X^{c})]^{c}$;

$(2)$ $XH_{\mathscr{C}}(X)=\{x\in U|N(x)\cap X\neq \emptyset\}$,
$XL_{\mathscr{C}}(X)=\{x\in U|N(x)\subseteq X\}$.
\end{definition}

According to Definition 2.2, the second and sixth lower and upper approximation operators are important standards for knowledge reduction of covering information systems in covering-based rough set theory; they are also typical representatives of approximation operators for covering approximation spaces.

If $U=\{x_{1},x_{2},...,x_{n}\}$ is a finite universe of discourse,
$\mathscr{C}=\{C_{1}, C_{2}, ..., C_{m}\}$ a family of subsets of $U$,
and $M_{\mathscr{C}}=(a_{ij})_{n\times m}$, where $a_{ij}=\left\{
\begin{array}{ccc}
1,&{\rm}& x_{i}\in C_{j};\\
0,&{\rm}& x_{i}\notin C_{j}.
\end{array}
\right.$, then $M_{\mathscr{C}}$ is called a matrix representation
of $\mathscr{C}$. Additionally,
we also have the characteristic function $\mathcal {X}_{X}
=\left[\begin{array}{cccccc}
  a_{1}&a_{2}&.&.&. & a_{n} \\
  \end{array}
\right]^{T}$ for $X\subseteq U$, where $a_{i}=\left\{
\begin{array}{ccc}
1,&{\rm}& x_{i}\in X;\\
0,&{\rm}& x_{i}\notin X.
\end{array}
\right. $.

\begin{definition}\cite{Wang3}
Let $(U,\mathscr{C})$ be a covering approximation
space, $A=(a_{ij})_{n\times m}$ and $B=(b_{ij})_{m\times p}$ Boolean
matrices, and $A\odot B=(c_{ij})_{n\times p}$, where
$c_{ij}=\bigwedge^{m}_{k=1}(b_{kj}-a_{ik}+1).$ Then

$(1)$ $\Gamma(\mathscr{C})=M_{\mathscr{C}}\bullet
M_{\mathscr{C}}^{T}=(d_{ij})_{n\times n}$ is called the type-1 characteristic matrix
of $\mathscr{C}$, where $d_{ij}=\bigvee^{m}_{k=1}(a_{ik}\cdot
a_{jk})$, and $M_{\mathscr{C}}\bullet M_{\mathscr{C}}^{T}$ is the
boolean product of $M_{\mathscr{C}}$ and its transpose $
M_{\mathscr{C}}^{T}$;

$(2)$ $\prod(\mathscr{C})=M_{\mathscr{C}}\odot M_{\mathscr{C}}^{T}=(e_{ij})_{n\times n}$ is called the type-2 characteristic matrix of
$\mathscr{C}$.
\end{definition}

We show the second and sixth lower and upper approximations of sets using the type-1 and type-2 characteristic matrices respectively as follows.

\begin{definition}\cite{Wang3}
Let $(U,\mathscr{C})$ be a covering approximation
space, and
$\mathcal {X}_{X}$ the characteristic function of $X$ in $U$. Then

$(1)$ $\mathcal {X}_{SH_{\mathscr{C}}(X)}=\Gamma(\mathscr{C})\bullet \mathcal
{X}_{X}$, $\mathcal {X}_{SL_{\mathscr{C}}(X)}=\Gamma(\mathscr{C})\odot \mathcal
{X}_{X}$; $(2)$ $\mathcal {X}_{XH_{\mathscr{C}}(X)}=\prod(\mathscr{C})\bullet
\mathcal {X}_{X}$, $\mathcal {X}_{XL_{\mathscr{C}}(X)}=\prod(\mathscr{C})\odot
\mathcal {X}_{X}$.
\end{definition}

We present the concepts of the type-1 and type-2 reducts of covering decision information systems as follows.

\begin{definition}\cite{Lang1}
Let $(U,\mathscr{D}_{C}\cup \mathscr{D}_{D} )$ be a covering decision information
system, where $\mathscr{D}_{C}=\{\mathscr{C}_{i}|i\in I\}$,
$\mathscr{D}_{D}=\{D_{i}|i\in J\}$, I and J are indexed sets. Then
$\mathscr{P}\subseteq \mathscr{D}_{C}$ is called a type-1 reduct of
$(U,\mathscr{D}_{C}\cup \mathscr{D}_{D} )$ if it satisfies $(1)$ and $(2)$ as follows:

$(1)$ $\Gamma(\mathscr{D}_{C})\bullet
M_{\mathscr{D}_{D}}=\Gamma(\mathscr{P})\bullet M_{D_{i}},
\Gamma(\mathscr{D}_{C})\odot M_{\mathscr{D}_{D}}=\Gamma(\mathscr{P})\odot
M_{\mathscr{D}_{D}};$

$(2)$ $\Gamma(\mathscr{D}_{C})\bullet M_{\mathscr{D}_{D}}\neq\Gamma(\mathscr{P^{'}})\bullet M_{\mathscr{D}_{D}},
\Gamma(\mathscr{D}_{C})\odot M_{\mathscr{D}_{D}}\neq\Gamma(\mathscr{P^{'}})\odot M_{\mathscr{D}_{D}},
\forall \mathscr{P^{'}}\subset \mathscr{P}.$
\end{definition}

\begin{definition}\cite{Lang1}
Let $(U,\mathscr{D}_{C}\cup \mathscr{D}_{D} )$ be a covering decision information
system, where $\mathscr{D}_{C}=\{\mathscr{C}_{i}|i\in I\}$,
$\mathscr{D}_{D}=\{D_{i}|i\in J\}$, I and J are indexed sets. Then
$\mathscr{P}\subseteq \mathscr{D}_{C}$ is called a type-2 reduct of
$(U,\mathscr{D}_{C}\cup \mathscr{D}_{D} )$ if it satisfies $(1)$ and $(2)$ as follows:

$(1)$ $\prod(\mathscr{D}_{C})\bullet M_{\mathscr{D}_{D}}=\prod(\mathscr{P})\bullet M_{\mathscr{D}_{D}},
\prod(\mathscr{D}_{C})\odot M_{\mathscr{D}_{D}}=\prod(\mathscr{P})\odot M_{\mathscr{D}_{D}};$

$(2)$ $\prod(\mathscr{D}_{C})\bullet M_{\mathscr{D}_{D}}\neq\prod(\mathscr{P^{'}})\bullet M_{\mathscr{D}_{D}},
\prod(\mathscr{D}_{C})\odot M_{\mathscr{D}_{D}}\neq\prod(\mathscr{P^{'}})\odot M_{\mathscr{D}_{D}},
\forall \mathscr{P^{'}}\subset \mathscr{P}.$
\end{definition}

\section{Update the type-1 and type-2 characteristic matrices
with variations of coverings}

In this section, we present incremental approaches to computing
the type-1 and type-2 characteristic matrices with variations of coverings.

\begin{definition}
Let $(U,\mathscr{D})$ and $(U,\mathscr{D}^{+})$ be covering
information systems, where $U=\{x_{1},x_{2},...,x_{n}\}$,
$\mathscr{D}=\{\mathscr{C}_{1},\mathscr{C}_{2},...,\mathscr{C}_{m}\}$, and
$\mathscr{D}^{+}=\{\mathscr{C}_{1},\mathscr{C}_{2},...,\mathscr{C}_{m},\mathscr{C}_{m+1}\}(m\geq 1)$. Then $(U,\mathscr{D}^{+})$ is called a dynamic covering information system of $(U,\mathscr{D})$.
\end{definition}

In practical situations, the cardinalities of coverings which describes objects in covering information systems are increasing with the development of science and technology. Moreover, $(U,\mathscr{D})$ is referred to as a static covering information system of $(U,\mathscr{D}^{+})$.

\begin{example}
Let $(U,\mathscr{D})$ be a static covering information
system, where $U=\{x_{1},x_{2},x_{3},x_{4},x_{5}\}$,
$\mathscr{D}=\{\mathscr{C}_{1},\mathscr{C}_{2},\mathscr{C}_{3}\}$,
$\mathscr{C}_{1}=\{\{x_{1},x_{2},x_{3},x_{4}\},\{x_{5}\}\}$,
$\mathscr{C}_{2}=\{\{x_{1},x_{2}\},\{x_{3},x_{4},x_{5}\}\}$, and
$\mathscr{C}_{3}=\{\{x_{1},x_{2},x_{5}\},\{x_{3},x_{4}\}\}$.
By adding $\mathscr{C}_{4}=\{\{x_{1},x_{2}\},\{x_{3},x_{4}\},\{x_{5}\}\}$ into $\mathscr{D}$, we obtain a dynamic covering information system $(U,\mathscr{D}^{+})$ of $(U,\mathscr{D})$, where
$\mathscr{D}^{+}=\{\mathscr{C}_{1},\mathscr{C}_{2},\mathscr{C}_{3},\mathscr{C}_{4}\}$.
\end{example}

In what follows, we show how to construct $\Gamma(\mathscr{D}^{+})$
based on $\Gamma(\mathscr{D})$. For convenience, we denote
$M_{\mathscr{D}}=\left[\begin{array}{cccccc}
  M_{\mathscr{C}_{1}}&M_{\mathscr{C}_{2}}&.&.&. & M_{\mathscr{C}_{m}}
  \end{array}
\right]$,
$M_{\mathscr{D}^{+}}=\left[\begin{array}{ccccccc}
  M_{\mathscr{C}_{1}}&M_{\mathscr{C}_{2}}&.&.&. & M_{\mathscr{C}_{m}}&M_{\mathscr{C}_{m+1}}
  \end{array}
\right]$, $M_{\mathscr{C}_{k}}=(a^{k}_{ij})_{n\times |\mathscr{C}_{k}|}$,
$\Gamma(\mathscr{D})=(b_{ij})_{n\times n}$, and
$\Gamma(\mathscr{D}^{+})=(c_{ij})_{n\times n}$, where $|\ast|$ denotes the cardinality of $\ast$.

\begin{theorem}
Let $(U,\mathscr{D}^{+})$ be a dynamic covering information system
of $(U,\mathscr{D})$, $\Gamma(\mathscr{D})$ and
$\Gamma(\mathscr{D}^{+})$ the type-1 characteristic matrices of
$\mathscr{D}$ and $\mathscr{D}^{+}$, respectively. Then
 \begin{eqnarray*}
 \Gamma(\mathscr{D}^{+})=
    \Gamma(\mathscr{D})\bigvee
    \Gamma(\mathscr{C}_{m+1}),
\end{eqnarray*} where
  $\Gamma(\mathscr{C}_{m+1})=M_{\mathscr{C}_{m+1}}\bullet M^{T}_{\mathscr{C}_{m+1}}.$
\end{theorem}

\noindent\textbf{Proof.} By Definitions 2.3 and 3.1, we get
$\Gamma(\mathscr{C})$ and $\Gamma(\mathscr{C}^{+})$ as follows:
\begin{eqnarray*}
\Gamma(\mathscr{D})&=&M_{\mathscr{D}}\bullet
M_{\mathscr{D}}^{T}\\&=&\left[
  \begin{array}{ccccccccccccccc}
    a^{1}_{11} & a^{1}_{12} & . & a^{1}_{1|\mathscr{C}_{1}|} & a^{2}_{11} & a^{2}_{12} & . & a^{2}_{1|\mathscr{C}_{2}|} & . & . & . & a^{m}_{11} & a^{m}_{12} & . & a^{m}_{1|\mathscr{C}_{m}|}\\
    a^{1}_{21} & a^{1}_{22} & . & a^{1}_{2|\mathscr{C}_{1}|} & a^{2}_{21} & a^{2}_{22} & . & a^{2}_{2|\mathscr{C}_{2}|} & . & . & . & a^{m}_{21} & a^{m}_{22} & . & a^{m}_{2|\mathscr{C}_{m}|}\\
    . & . & . & . & . & . & . & . & . & . & . & . & . & . & . \\
    . & . & . & . & . & . & . & . & . & . & . & . & . & . & . \\
    . & . & . & . & . & . & . & . & . & . & . & . & . & . & . \\
    a^{1}_{n1} & a^{1}_{n2} & . & a^{1}_{n|\mathscr{C}_{1}|} & a^{2}_{n1} & a^{2}_{n2}& . & a^{2}_{n|\mathscr{C}_{2}|}& . & . & . & a^{m}_{n1} & a^{m}_{n2} & . & a^{m}_{n|\mathscr{C}_{m}|}
  \end{array}
\right] \bullet \\&&\left[
  \begin{array}{ccccccccccccccc}
    a^{1}_{11} & a^{1}_{12} & . & a^{1}_{1|\mathscr{C}_{1}|} & a^{2}_{11} & a^{2}_{12} & . & a^{2}_{1|\mathscr{C}_{2}|} & . & . & . & a^{m}_{11} & a^{m}_{12} & . & a^{m}_{1|\mathscr{C}_{m}|}\\
    a^{1}_{21} & a^{1}_{22} & . & a^{1}_{2|\mathscr{C}_{1}|} & a^{2}_{21} & a^{2}_{22} & . & a^{2}_{2|\mathscr{C}_{2}|} & . & . & . & a^{m}_{21} & a^{m}_{22} & . & a^{m}_{2|\mathscr{C}_{m}|}\\
    . & . & . & . & . & . & . & . & . & . & . & . & . & . & . \\
    . & . & . & . & . & . & . & . & . & . & . & . & . & . & . \\
    . & . & . & . & . & . & . & . & . & . & . & . & . & . & . \\
    a^{1}_{n1} & a^{1}_{n2} & . & a^{1}_{n|\mathscr{C}_{1}|} & a^{2}_{n1} & a^{2}_{n2}& . & a^{2}_{n|\mathscr{C}_{2}|}& . & . & . & a^{m}_{n1} & a^{m}_{n2} & . & a^{m}_{n|\mathscr{C}_{m}|}
  \end{array}
\right]^{T}
\\&=&\left[
  \begin{array}{cccccc}
    b_{11} & b_{12} & . & . & . & b_{1n} \\
    b_{21} & b_{22} & . & . & . & b_{2n} \\
    . & . & . & . & . & . \\
    . & . & . & . & . & . \\
    . & . & . & . & . & . \\
    b_{n1} & b_{n2} & . & . & . & b_{nn}
  \end{array}
\right],\\
\Gamma(\mathscr{D}^{+})&=&M_{\mathscr{D}^{+}}\bullet
M_{\mathscr{D}^{+}}^{T}\\&=&\left[
  \begin{array}{ccccccccccccccccccc}
    a^{1}_{11} & a^{1}_{12} & . & a^{1}_{1|\mathscr{C}_{1}|} & a^{2}_{11} & a^{2}_{12} & . & a^{2}_{1|\mathscr{C}_{2}|} & . & . & . & a^{m}_{11} & a^{m}_{12} & . & a^{m}_{1|\mathscr{C}_{m}|}& a^{m+1}_{11} & a^{m+1}_{12} & . & a^{m+1}_{1|\mathscr{C}_{m+1}|} \\
    a^{1}_{21} & a^{1}_{22} & . & a^{1}_{2|\mathscr{C}_{1}|} & a^{2}_{21} & a^{2}_{22} & . & a^{2}_{2|\mathscr{C}_{2}|} & . & . & . & a^{m}_{21} & a^{m}_{22} & . & a^{m}_{2|\mathscr{C}_{m}|}& a^{m+1}_{21} & a^{m+1}_{22} & . & a^{m+1}_{2|\mathscr{C}_{m+1}|} \\
    . & . & . & . & . & . & . & . & . & . & . & . & . & . & . & . & . & . & .\\
    . & . & . & . & . & . & . & . & . & . & . & . & . & . & . & . & . & . & .\\
    . & . & . & . & . & . & . & . & . & . & . & . & . & . & . & . & . & . & .\\
    a^{1}_{n1} & a^{1}_{n2} & . & a^{1}_{n|\mathscr{C}_{1}|} & a^{2}_{n1} & a^{2}_{n2}& . & a^{2}_{n|\mathscr{C}_{2}|}& . & . & . & a^{m}_{n1} & a^{m}_{n2} & . & a^{m}_{n|\mathscr{C}_{m}|}& a^{m+1}_{n1} & a^{m+1}_{n2} & . & a^{m+1}_{n|\mathscr{C}_{m+1}|}
  \end{array}
\right] \bullet \\ && \left[
  \begin{array}{ccccccccccccccccccc}
    a^{1}_{11} & a^{1}_{12} & . & a^{1}_{1|\mathscr{C}_{1}|} & a^{2}_{11} & a^{2}_{12} & . & a^{2}_{1|\mathscr{C}_{2}|} & . & . & . & a^{m}_{11} & a^{m}_{12} & . & a^{m}_{1|\mathscr{C}_{m}|}& a^{m+1}_{11} & a^{m+1}_{12} & . & a^{m+1}_{1|\mathscr{C}_{m+1}|} \\
    a^{1}_{21} & a^{1}_{22} & . & a^{1}_{2|\mathscr{C}_{1}|} & a^{2}_{21} & a^{2}_{22} & . & a^{2}_{2|\mathscr{C}_{2}|} & . & . & . & a^{m}_{21} & a^{m}_{22} & . & a^{m}_{2|\mathscr{C}_{m}|}& a^{m+1}_{21} & a^{m+1}_{22} & . & a^{m+1}_{2|\mathscr{C}_{m+1}|} \\
    . & . & . & . & . & . & . & . & . & . & . & . & . & . & . & . & . & . & .\\
    . & . & . & . & . & . & . & . & . & . & . & . & . & . & . & . & . & . & .\\
    . & . & . & . & . & . & . & . & . & . & . & . & . & . & . & . & . & . & .\\
    a^{1}_{n1} & a^{1}_{n2} & . & a^{1}_{n|\mathscr{C}_{1}|} & a^{2}_{n1} & a^{2}_{n2}& . & a^{2}_{n|\mathscr{C}_{2}|}& . & . & . & a^{m}_{n1} & a^{m}_{n2} & . & a^{m}_{n|\mathscr{C}_{m}|}& a^{m+1}_{n1} & a^{m+1}_{n2} & . & a^{m+1}_{n|\mathscr{C}_{m+1}|}
  \end{array}
\right]^{T} \\&=&\left[
  \begin{array}{cccccc}
    c_{11} & c_{12} & . & . & . & c_{1n} \\
    c_{21} & c_{22} & . & . & . & c_{2n} \\
    . & . & . & . & . & . \\
    . & . & . & . & . & . \\
    . & . & . & . & . & . \\
    c_{n1} & c_{n2} & . & . & . & c_{nn}
  \end{array}
\right].
\end{eqnarray*}

According to Definition 2.3, we have
\begin{eqnarray*}
b_{ij}&=&\left[
  \begin{array}{ccccccccccccccc}
    a^{1}_{i1} & a^{1}_{i2} & . & a^{1}_{i|\mathscr{C}_{1}|} & a^{2}_{i1} & a^{2}_{i2} & . & a^{2}_{i|\mathscr{C}_{2}|} & . & . & . & a^{m}_{i1} & a^{m}_{i2} & . & a^{m}_{i|\mathscr{C}_{m}|}\\
  \end{array}
\right]\bullet \\&&\left[
  \begin{array}{ccccccccccccccc}
   a^{1}_{j1} & a^{1}_{j2} & . & a^{1}_{j|\mathscr{C}_{1}|} & a^{2}_{j1} & a^{2}_{j2} & . & a^{2}_{j|\mathscr{C}_{2}|} & . & . & . & a^{m}_{j1} & a^{m}_{j2} & . & a^{m}_{j|\mathscr{C}_{m}|}\\
  \end{array}
\right]^{T}\\&=&[(a^{1}_{i1}\cdot a^{1}_{j1})\vee (a^{1}_{i2}\cdot a^{1}_{j2})\vee...\vee (a^{1}_{i|\mathscr{C}_{1}|}\cdot a^{1}_{j|\mathscr{C}_{1}|})]\vee [(a^{2}_{i1}\cdot a^{2}_{j1})\vee (a^{2}_{i2}\cdot a^{2}_{j2})\vee...\vee (a^{2}_{i|\mathscr{C}_{2}|}\cdot a^{2}_{j|\mathscr{C}_{2}|})]\vee ...\vee \\&& [(a^{m}_{i1}\cdot a^{m}_{j1})\vee (a^{m}_{i2}\cdot a^{m}_{j2})\vee...\vee (a^{m}_{i|\mathscr{C}_{m}|}\cdot a^{m}_{j|\mathscr{C}_{m}|})],\\
c_{ij}&=&\left[
  \begin{array}{ccccccccccccccccccc}
    a^{1}_{i1} & a^{1}_{i2} & . & a^{1}_{i|\mathscr{C}_{1}|} & a^{2}_{i1} & a^{2}_{i2} & . & a^{2}_{i|\mathscr{C}_{2}|} & . & . & . & a^{m}_{i1} & a^{m}_{i2} & . & a^{m}_{i|\mathscr{C}_{m}|}& a^{m+1}_{i1} & a^{m+1}_{i2} & . & a^{m+1}_{i|\mathscr{C}_{m+1}|} \\
  \end{array}
\right]\bullet \\
&&\left[
  \begin{array}{ccccccccccccccccccc}
    a^{1}_{j1} & a^{1}_{j2} & . & a^{1}_{j|\mathscr{C}_{1}|} & a^{2}_{j1} & a^{2}_{j2} & . & a^{2}_{j|\mathscr{C}_{2}|} & . & . & . & a^{m}_{j1} & a^{m}_{j2} & . & a^{m}_{j|\mathscr{C}_{m}|}& a^{m+1}_{j1} & a^{m+1}_{j2} & . & a^{m+1}_{j|\mathscr{C}_{m+1}|} \\
  \end{array}
\right]^{T}\\&=&[(a^{1}_{i1}\cdot a^{1}_{j1})\vee (a^{1}_{i2}\cdot a^{1}_{j2})\vee...\vee (a^{1}_{i|\mathscr{C}_{1}|}\cdot a^{1}_{j|\mathscr{C}_{1}|})]\vee [(a^{2}_{i1}\cdot a^{2}_{j1})\vee (a^{2}_{i2}\cdot a^{2}_{j2})\vee...\vee (a^{2}_{i|\mathscr{C}_{2}|}\cdot a^{2}_{j|\mathscr{C}_{2}|})]\vee ...\vee \\&&[(a^{m}_{i1}\cdot a^{m}_{j1})\vee (a^{m}_{i2}\cdot a^{m}_{j2})\vee...\vee (a^{m}_{i|\mathscr{C}_{m}|}\cdot a^{m}_{j|\mathscr{C}_{m}|})]\vee [(a^{m+1}_{i1}\cdot a^{m+1}_{j1})\vee (a^{m+1}_{i2}\cdot a^{m+1}_{j2})\vee...\vee (a^{m+1}_{i|\mathscr{C}_{m+1}|}\cdot a^{m}_{j|\mathscr{C}_{m+1}|})]\\&=&b_{ij}\vee \left[
  \begin{array}{ccccccccccc}
    a^{m+1}_{i1} & a^{m+1}_{i2} & . & . & . & a^{m+1}_{i|\mathscr{C}_{m+1}|}
  \end{array}
\right]\bullet\left[
  \begin{array}{ccccccccccc}
     a^{m+1}_{j1} & a^{m+1}_{j2} & . & . & . & a^{m+1}_{j|\mathscr{C}_{m+1}|}
  \end{array}
\right]^{T}.
\end{eqnarray*}

To obtain
$\Gamma(\mathscr{D}^{+})$, we only need to compute
$\Gamma(\mathscr{C}_{m+1})$ as follows:
  \begin{eqnarray*}
  \Gamma(\mathscr{C}_{m+1})&=&\left[\begin{array}{cccccc}
  a^{m+1}_{11}&a^{m+1}_{12}&.&.&. & a^{m+1}_{1|\mathscr{C}_{m+1}|}\\
  a^{m+1}_{21}&a^{m+1}_{22}&.&.&. & a^{m+1}_{2|\mathscr{C}_{m+1}|}\\
  .&.&.&.&. & .\\
  .&.&.&.&. & .\\
  .&.&.&.&. & .\\
  a^{m+1}_{n1}&a^{m+1}_{n2}&.&.&. & a^{m+1}_{n|\mathscr{C}_{m+1}|}
  \end{array}
\right]\bullet \left[\begin{array}{cccccc}
  a^{m+1}_{11}&a^{m+1}_{12}&.&.&. & a^{m+1}_{1|\mathscr{C}_{m+1}|}\\
  a^{m+1}_{21}&a^{m+1}_{22}&.&.&. & a^{m+1}_{2|\mathscr{C}_{m+1}|}\\
  .&.&.&.&. & .\\
  .&.&.&.&. & .\\
  .&.&.&.&. & .\\
  a^{m+1}_{n1}&a^{m+1}_{n2}&.&.&. & a^{m+1}_{n|\mathscr{C}_{m+1}|}
  \end{array}
\right]^{T}.
\end{eqnarray*}

Therefore, we have
 \begin{eqnarray*}
 \Gamma(\mathscr{D}^{+})=
    \Gamma(\mathscr{D})\bigvee
    \Gamma(\mathscr{C}_{m+1}),
\end{eqnarray*} where
  $\Gamma(\mathscr{C}_{m+1})=M_{\mathscr{C}_{m+1}}\bullet M^{T}_{\mathscr{C}_{m+1}}.$ $\Box$

We present the non-incremental and incremental algorithms for computing $SH_{\mathscr{D}^{+}}(X)$ and $SL_{\mathscr{D}^{+}}(X)$ in dynamic covering information systems.

\begin{algorithm}(Non-incremental algorithm of computing $SH_{\mathscr{D}^{+}}(X)$ and $SL_{\mathscr{D}^{+}}(X)\bf{(NIS)}$)

Step 1: Input $(U,\mathscr{D}^{+})$;

Step 2: Construct
$\Gamma(\mathscr{D}^{+})=M_{\mathscr{D}^{+}}\bullet M^{T}_{\mathscr{D}^{+}};$

Step 3: Compute $\mathcal {X}_{SH_{\mathscr{D}^{+}}(X)}=\Gamma(\mathscr{D}^{+})\bullet \mathcal
{X}_{X}$ and $\mathcal {X}_{SL_{\mathscr{D}^{+}}(X)}=\Gamma(\mathscr{D}^{+})\odot \mathcal
{X}_{X}$;

Step 4: Output $SH_{\mathscr{D}^{+}}(X)$ and $SL_{\mathscr{D}^{+}}(X)$.
\end{algorithm}

\begin{algorithm}(Incremental algorithm of computing $SH_{\mathscr{D}^{+}}(X)$ and $SL_{\mathscr{D}^{+}}(X)\bf{(IS)}$)

Step 1: Input $(U,\mathscr{D})$ and $(U,\mathscr{D}^{+})$;

Step 2: Calculate $\Gamma(\mathscr{D})=M_{\mathscr{D}}\bullet
M^{T}_{\mathscr{D}}$;

Step 3: Construct $\Gamma(\mathscr{D}^{+})=
    \Gamma(\mathscr{D})\bigvee
    \Gamma(\mathscr{C}_{m+1})$, where
$
 \Gamma(\mathscr{C}_{m+1})=M_{\mathscr{C}_{m+1}}\bullet M^{T}_{\mathscr{C}_{m+1}};
$

Step 4:  Obtain $
\mathcal {X}_{SH_{\mathscr{D}^{+}}(X)}=\Gamma(\mathscr{D}^{+})\bullet \mathcal
{X}_{X}\text{ and } \mathcal {X}_{SL_{\mathscr{D}^{+}}(X)}=\Gamma(\mathscr{D}^{+})\odot \mathcal
{X}_{X};
$

Step 5: Output $SH_{\mathscr{D}^{+}}(X)$ and $SL_{\mathscr{D}^{+}}(X)$.
\end{algorithm}

The time complexity of computing the second lower and upper
approximations of sets is $O(2n^{2}\ast\sum^{m+1}_{i=1}|\mathscr{C}_{i}|+2n^{2})$ using
Algorithm 3.4. Furthermore, $O(2n^{2}\ast|\mathscr{C}_{m+1}|+3n^{2})$ is
the time complexity of Algorithm 3.5. Therefore, the
time complexity of the incremental algorithm is lower than that of
the non-incremental algorithm.

\begin{example}(Continued from Example 3.2) Taking $X=\{x_{2},x_{3},x_{4}\}$. According to Definition 2.3, we first have
\begin{eqnarray*}
\Gamma(\mathscr{D})=M_{\mathscr{D}}\bullet
M_{\mathscr{D}}^{T}=\left[
\begin{array}{ccccc}
1 & 1 & 1 & 1 &1 \\
1 & 1 & 1 & 1 &1 \\
1 & 1 & 1 & 1 &1 \\
1 & 1 & 1 & 1 &1 \\
1 & 1 & 1 & 1 &1 \\
\end{array}
\right],
\Gamma(\mathscr{C}_{4})=M_{\mathscr{C}_{4}}\bullet
M_{\mathscr{C}_{4}}^{T}=\left[
\begin{array}{ccccc}
1 & 1 & 0 & 0 &0 \\
1 & 1 & 0 & 0 &0 \\
0 & 0 & 1 & 1 &0 \\
0 & 0 & 1 & 1 &0 \\
0 & 0 & 0 & 0 &1 \\
\end{array}
\right].
\end{eqnarray*}

Second, by Theorem 3.3, we obtain
\begin{eqnarray*}
 \Gamma(\mathscr{D}^{+})=
    \Gamma(\mathscr{D}) \bigvee \Gamma(\mathscr{C}_{4})&=&\left[
\begin{array}{ccccc}
1 & 1 & 1 & 1 &1 \\
1 & 1 & 1 & 1 &1 \\
1 & 1 & 1 & 1 &1 \\
1 & 1 & 1 & 1 &1 \\
1 & 1 & 1 & 1 &1 \\
\end{array}
\right]\bigvee \left[
\begin{array}{ccccc}
1 & 1 & 0 & 0 &0 \\
1 & 1 & 0 & 0 &0 \\
0 & 0 & 1 & 1 &0 \\
0 & 0 & 1 & 1 &0 \\
0 & 0 & 0 & 0 &1 \\
\end{array}
\right]\\&=& \left[
\begin{array}{ccccc}
1 & 1 & 1 & 1 &1 \\
1 & 1 & 1 & 1 &1 \\
1 & 1 & 1 & 1 &1 \\
1 & 1 & 1 & 1 &1 \\
1 & 1 & 1 & 1 &1 \\
\end{array}
\right].
\end{eqnarray*}

Third, by Definition 2.4, we get
\begin{eqnarray*}
\mathcal {X}_{SH_{\mathscr{D}^{+}}(X)}&=&\Gamma(\mathscr{D}^{+})\bullet \mathcal {X}_{X}
=\left[
\begin{array}{ccccc}
1 & 1 & 1 & 1 &1 \\
1 & 1 & 1 & 1 &1 \\
1 & 1 & 1 & 1 &1 \\
1 & 1 & 1 & 1 &1 \\
1 & 1 & 1 & 1 &1 \\
\end{array}
\right]\bullet\left[
\begin{array}{c}
0 \\
1 \\
1 \\
1 \\
0 \\
\end{array}
\right] =\left[
\begin{array}{ccccc}
1 & 1 & 1 & 1 & 1
\end{array}
\right]^{T},
\\
 \mathcal {X}_{SL_{\mathscr{D}^{+}}(X)}&=&\Gamma(\mathscr{D}^{+})\odot
\mathcal {X}_{X}=\left[
\begin{array}{ccccc}
1 & 1 & 1 & 1 &1 \\
1 & 1 & 1 & 1 &1 \\
1 & 1 & 1 & 1 &1 \\
1 & 1 & 1 & 1 &1 \\
1 & 1 & 1 & 1 &1 \\
\end{array}
\right]\odot \left[
\begin{array}{c}
0 \\
1 \\
1 \\
1 \\
0 \\
\end{array}
\right]=\left[
\begin{array}{cccccc}
0 & 0 & 0 & 0 & 0
\end{array}
\right]^{T}.
\end{eqnarray*}

Therefore, $SH_{\mathscr{D}^{+}}(X)=\{x_{1},x_{2},x_{3},x_{4},x_{5}\}$ and
$SL_{\mathscr{D}^{+}}(X)=\emptyset$.
\end{example}

In Example 3.6, we only need to calculate elements of
$\Gamma(\mathscr{C}_{4})$ for computing $SH_{\mathscr{D}^{+}}(X)$ and
$SL_{\mathscr{D}^{+}}(X)$ using Algorithm 3.5. But we must construct $\Gamma(\mathscr{D}^{+})$ for computing $SH_{\mathscr{D}^{+}}(X)$ and
$SL_{\mathscr{D}^{+}}(X)$ using Algorithm 3.4. Thereby, the
incremental approach is more effective to compute the second lower
and upper approximations of sets.

\begin{theorem}
Let $(U,\mathscr{D}^{+})$ be a dynamic covering information system
of $(U,\mathscr{D})$, $\Gamma(\mathscr{D})=(b_{ij})_{n\times n}$ and
$\Gamma(\mathscr{D}^{+})=(c_{ij})_{n\times n}$ the type-1 characteristic matrices of
$\mathscr{D}$ and $\mathscr{D}^{+}$, respectively. Then
 \begin{eqnarray*}
 c_{ij}=\left\{
\begin{array}{ccc}
1,&{\rm}& b_{ij}=1;\\
\left[
  \begin{array}{ccccccccccc}
    a^{m+1}_{i1} & a^{m+1}_{i2} & . & . & . & a^{m+1}_{i|\mathscr{C}_{m+1}|}
  \end{array}
\right]\bullet\left[
  \begin{array}{ccccccccccc}
     a^{m+1}_{j1} & a^{m+1}_{j2} & . & . & . & a^{m+1}_{j|\mathscr{C}_{m+1}|}
  \end{array}
\right]^{T},&{\rm}& b_{ij}=0.
\end{array}
\right.
\end{eqnarray*}
\end{theorem}

\noindent\textbf{Proof.} It is straightforward by Theorem 3.3.$\Box$

\begin{example}(Continued from Example 3.6) According to Definition 2.3, we have
\begin{eqnarray*}
\Gamma(\mathscr{D})=M_{\mathscr{D}}\bullet
M_{\mathscr{D}}^{T}=\left[
\begin{array}{cccccc}
1 & 1 & 1 & 1 &1 \\
1 & 1 & 1 & 1 &1 \\
1 & 1 & 1 & 1 &1 \\
1 & 1 & 1 & 1 &1 \\
1 & 1 & 1 & 1 &1 \\
\end{array}
\right].
\end{eqnarray*}

Therefore, by Theorem 3.7, we get
\begin{eqnarray*}
\Gamma(\mathscr{D}^{+})=\Gamma(\mathscr{D})=\left[
\begin{array}{cccccc}
1 & 1 & 1 & 1 &1 \\
1 & 1 & 1 & 1 &1 \\
1 & 1 & 1 & 1 &1 \\
1 & 1 & 1 & 1 &1 \\
1 & 1 & 1 & 1 &1 \\
\end{array}
\right].
\end{eqnarray*}
\end{example}

\begin{proposition}
Let $(U,\mathscr{D}^{+})$ be a dynamic covering information system
of $(U,\mathscr{D})$, $\Gamma(\mathscr{D})$ and
$\Gamma(\mathscr{D}^{+})$ the type-1 characteristic matrices of
$\mathscr{D}$ and $\mathscr{D}^{+}$, respectively.

$(1)$ If $\Gamma(\mathscr{D})=(1)_{n\times n}$, then $\Gamma(\mathscr{D}^{+})=(1)_{n\times n}$;

$(2)$ If $\Gamma(\mathscr{D})=(0)_{n\times n}$, then $\Gamma(\mathscr{D}^{+})=\Gamma(\mathscr{C}_{m+1})$.
\end{proposition}

\noindent\textbf{Proof.} It is straightforward by Theorem 3.7.$\Box$

Subsequently, we construct $\prod(\mathscr{C}^{+})$ based on
$\prod(\mathscr{C})$. For convenience, we denote
$\prod(\mathscr{C})=(d_{ij})_{n\times n}$ and
$\prod(\mathscr{C}^{+})=(e_{ij})_{n\times n}$.

\begin{theorem}
Let $(U,\mathscr{D}^{+})$ be a dynamic covering information system
of $(U,\mathscr{D})$, $\prod(\mathscr{D})$ and
$\prod(\mathscr{D}^{+})$ the type-2 characteristic matrices of
$\mathscr{D}$ and $\mathscr{D}^{+}$, respectively. Then
 \begin{eqnarray*}
 \prod(\mathscr{D}^{+})=
    \prod(\mathscr{D})\bigwedge
    \prod(\mathscr{C}_{m+1}),
\end{eqnarray*} where
  $\prod(\mathscr{C}_{m+1})=M_{\mathscr{C}_{m+1}}\odot M^{T}_{\mathscr{C}_{m+1}}.$
\end{theorem}

\noindent\textbf{Proof.} By Definitions 2.3 and 3.1, we get
$\prod(\mathscr{D})$ and $\prod(\mathscr{D}^{+})$ as follows:
\begin{eqnarray*}
\prod(\mathscr{D})&=&M_{\mathscr{D}}\odot
M_{\mathscr{D}}^{T}\\&=&\left[
  \begin{array}{ccccccccccccccc}
     a^{1}_{11} & a^{1}_{12} & . & a^{1}_{1|\mathscr{C}_{1}|} & a^{2}_{11} & a^{2}_{12} & . & a^{2}_{1|\mathscr{C}_{2}|} & . & . & . & a^{m}_{11} & a^{m}_{12} & . & a^{m}_{1|\mathscr{C}_{m}|}\\
    a^{1}_{21} & a^{1}_{22} & . & a^{1}_{2|\mathscr{C}_{1}|} & a^{2}_{21} & a^{2}_{22} & . & a^{2}_{2|\mathscr{C}_{2}|} & . & . & . & a^{m}_{21} & a^{m}_{22} & . & a^{m}_{2|\mathscr{C}_{m}|}\\
    . & . & . & . & . & . & . & . & . & . & . & . & . & . & . \\
    . & . & . & . & . & . & . & . & . & . & . & . & . & . & . \\
    . & . & . & . & . & . & . & . & . & . & . & . & . & . & . \\
    a^{1}_{n1} & a^{1}_{n2} & . & a^{1}_{n|\mathscr{C}_{1}|} & a^{2}_{n1} & a^{2}_{n2}& . & a^{2}_{n|\mathscr{C}_{2}|}& . & . & . & a^{m}_{n1} & a^{m}_{n2} & . & a^{m}_{n|\mathscr{C}_{m}|}
  \end{array}
\right] \odot \\&&\left[
  \begin{array}{ccccccccccccccc}
     a^{1}_{11} & a^{1}_{12} & . & a^{1}_{1|\mathscr{C}_{1}|} & a^{2}_{11} & a^{2}_{12} & . & a^{2}_{1|\mathscr{C}_{2}|} & . & . & . & a^{m}_{11} & a^{m}_{12} & . & a^{m}_{1|\mathscr{C}_{m}|}\\
    a^{1}_{21} & a^{1}_{22} & . & a^{1}_{2|\mathscr{C}_{1}|} & a^{2}_{21} & a^{2}_{22} & . & a^{2}_{2|\mathscr{C}_{2}|} & . & . & . & a^{m}_{21} & a^{m}_{22} & . & a^{m}_{2|\mathscr{C}_{m}|}\\
    . & . & . & . & . & . & . & . & . & . & . & . & . & . & . \\
    . & . & . & . & . & . & . & . & . & . & . & . & . & . & . \\
    . & . & . & . & . & . & . & . & . & . & . & . & . & . & . \\
    a^{1}_{n1} & a^{1}_{n2} & . & a^{1}_{n|\mathscr{C}_{1}|} & a^{2}_{n1} & a^{2}_{n2}& . & a^{2}_{n|\mathscr{C}_{2}|}& . & . & . & a^{m}_{n1} & a^{m}_{n2} & . & a^{m}_{n|\mathscr{C}_{m}|}
  \end{array}
\right]^{T}
\\&=&\left[
  \begin{array}{cccccc}
    d_{11} & d_{12} & . & . & . & d_{1n} \\
    d_{21} & d_{22} & . & . & . & d_{2n} \\
    . & . & . & . & . & . \\
    . & . & . & . & . & . \\
    . & . & . & . & . & . \\
    d_{n1} & d_{n2} & . & . & . & d_{nn}
  \end{array}
\right],\\
\prod(\mathscr{D}^{+})&=&M_{\mathscr{D}^{+}}\odot
M_{\mathscr{D}^{+}}^{T}\\&=&\left[
  \begin{array}{ccccccccccccccccccc}
    a^{1}_{11} & a^{1}_{12} & . & a^{1}_{1|\mathscr{C}_{1}|} & a^{2}_{11} & a^{2}_{12} & . & a^{2}_{1|\mathscr{C}_{2}|} & . & . & . & a^{m}_{11} & a^{m}_{12} & . & a^{m}_{1|\mathscr{C}_{m}|}& a^{m+1}_{11} & a^{m+1}_{12} & . & a^{m+1}_{1|\mathscr{C}_{m+1}|} \\
    a^{1}_{21} & a^{1}_{22} & . & a^{1}_{2|\mathscr{C}_{1}|} & a^{2}_{21} & a^{2}_{22} & . & a^{2}_{2|\mathscr{C}_{2}|} & . & . & . & a^{m}_{21} & a^{m}_{22} & . & a^{m}_{2|\mathscr{C}_{m}|}& a^{m+1}_{21} & a^{m+1}_{22} & . & a^{m+1}_{2|\mathscr{C}_{m+1}|} \\
    . & . & . & . & . & . & . & . & . & . & . & . & . & . & . & . & . & . & .\\
    . & . & . & . & . & . & . & . & . & . & . & . & . & . & . & . & . & . & .\\
    . & . & . & . & . & . & . & . & . & . & . & . & . & . & . & . & . & . & .\\
    a^{1}_{n1} & a^{1}_{n2} & . & a^{1}_{n|\mathscr{C}_{1}|} & a^{2}_{n1} & a^{2}_{n2}& . & a^{2}_{n|\mathscr{C}_{2}|}& . & . & . & a^{m}_{n1} & a^{m}_{n2} & . & a^{m}_{n|\mathscr{C}_{m}|}& a^{m+1}_{n1} & a^{m+1}_{n2} & . & a^{m+1}_{n|\mathscr{C}_{m+1}|}
  \end{array}
\right] \odot\\ && \left[
  \begin{array}{ccccccccccccccccccc}
    a^{1}_{11} & a^{1}_{12} & . & a^{1}_{1|\mathscr{C}_{1}|} & a^{2}_{11} & a^{2}_{12} & . & a^{2}_{1|\mathscr{C}_{2}|} & . & . & . & a^{m}_{11} & a^{m}_{12} & . & a^{m}_{1|\mathscr{C}_{m}|}& a^{m+1}_{11} & a^{m+1}_{12} & . & a^{m+1}_{1|\mathscr{C}_{m+1}|} \\
    a^{1}_{21} & a^{1}_{22} & . & a^{1}_{2|\mathscr{C}_{1}|} & a^{2}_{21} & a^{2}_{22} & . & a^{2}_{2|\mathscr{C}_{2}|} & . & . & . & a^{m}_{21} & a^{m}_{22} & . & a^{m}_{2|\mathscr{C}_{m}|}& a^{m+1}_{21} & a^{m+1}_{22} & . & a^{m+1}_{2|\mathscr{C}_{m+1}|} \\
    . & . & . & . & . & . & . & . & . & . & . & . & . & . & . & . & . & . & .\\
    . & . & . & . & . & . & . & . & . & . & . & . & . & . & . & . & . & . & .\\
    . & . & . & . & . & . & . & . & . & . & . & . & . & . & . & . & . & . & .\\
    a^{1}_{n1} & a^{1}_{n2} & . & a^{1}_{n|\mathscr{C}_{1}|} & a^{2}_{n1} & a^{2}_{n2}& . & a^{2}_{n|\mathscr{C}_{2}|}& . & . & . & a^{m}_{n1} & a^{m}_{n2} & . & a^{m}_{n|\mathscr{C}_{m}|}& a^{m+1}_{n1} & a^{m+1}_{n2} & . & a^{m+1}_{n|\mathscr{C}_{m+1}|}
  \end{array}
\right]^{T} \\&=&\left[
  \begin{array}{cccccc}
    e_{11} & e_{12} & . & . & . & e_{1n} \\
    e_{21} & e_{22} & . & . & . & e_{2n} \\
    . & . & . & . & . & . \\
    . & . & . & . & . & . \\
    . & . & . & . & . & . \\
    e_{n1} & e_{n2} & . & . & . & e_{nn}
  \end{array}
\right].
\end{eqnarray*}

According to Definition 2.3, we have
\begin{eqnarray*}
d_{ij}&=&\left[
  \begin{array}{ccccccccccccccc}
    a^{1}_{i1} & a^{1}_{i2} & . & a^{1}_{i|\mathscr{C}_{1}|} & a^{2}_{i1} & a^{2}_{i2} & . & a^{2}_{i|\mathscr{C}_{2}|} & . & . & . & a^{m}_{i1} & a^{m}_{i2} & . & a^{m}_{i|\mathscr{C}_{m}|}\\
  \end{array}
\right]\odot \\&&\left[
  \begin{array}{ccccccccccccccc}
   a^{1}_{j1} & a^{1}_{j2} & . & a^{1}_{j|\mathscr{C}_{1}|} & a^{2}_{j1} & a^{2}_{j2} & . & a^{2}_{j|\mathscr{C}_{2}|} & . & . & . & a^{m}_{j1} & a^{m}_{j2} & . & a^{m}_{j|\mathscr{C}_{m}|}\\
  \end{array}
\right]^{T}\\&=&[(a^{1}_{j1}- a^{1}_{i1}+1)\wedge (a^{1}_{j2}- a^{1}_{i2}+1)\wedge...\wedge (a^{1}_{j|\mathscr{C}_{1}|}- a^{1}_{i|\mathscr{C}_{1}|}+1)]\wedge \\&& [(a^{2}_{j1}- a^{2}_{i1}+1)\wedge (a^{2}_{j2}- a^{2}_{i2}+1)\wedge...\wedge (a^{2}_{j|\mathscr{C}_{2}|}- a^{2}_{i|\mathscr{C}_{2}|}+1)]\wedge ...\wedge \\&& [(a^{m}_{j1}- a^{m}_{i1}+1)\wedge (a^{m}_{j2}- a^{m}_{i2}+1)\wedge...\wedge (a^{m}_{j|\mathscr{C}_{m}|}- a^{m}_{i|\mathscr{C}_{m}|}+1)],\\
e_{ij}&=&\left[
  \begin{array}{ccccccccccccccccccc}
    a^{1}_{i1} & a^{1}_{i2} & . & a^{1}_{i|\mathscr{C}_{1}|} & a^{2}_{i1} & a^{2}_{i2} & . & a^{2}_{i|\mathscr{C}_{2}|} & . & . & . & a^{m}_{i1} & a^{m}_{i2} & . & a^{m}_{i|\mathscr{C}_{m}|}& a^{m+1}_{i1} & a^{m+1}_{i2} & . & a^{m+1}_{i|\mathscr{C}_{m+1}|} \\
  \end{array}
\right]\odot \\
&&\left[
  \begin{array}{ccccccccccccccccccc}
    a^{1}_{j1} & a^{1}_{j2} & . & a^{1}_{j|\mathscr{C}_{1}|} & a^{2}_{j1} & a^{2}_{j2} & . & a^{2}_{j|\mathscr{C}_{2}|} & . & . & . & a^{m}_{j1} & a^{m}_{j2} & . & a^{m}_{j|\mathscr{C}_{m}|}& a^{m+1}_{j1} & a^{m+1}_{j2} & . & a^{m+1}_{j|\mathscr{C}_{m+1}|} \\
  \end{array}
\right]^{T}\\&=&[(a^{1}_{j1}- a^{1}_{i1}+1)\wedge (a^{1}_{j2}- a^{1}_{i2}+1)\wedge...\wedge (a^{1}_{j|\mathscr{C}_{1}|}- a^{1}_{i|\mathscr{C}_{1}|}+1)]\wedge \\&& [(a^{2}_{j1}- a^{2}_{i1}+1)\wedge (a^{2}_{j2}- a^{2}_{i2}+1)\wedge...\wedge (a^{2}_{j|\mathscr{C}_{2}|}- a^{2}_{i|\mathscr{C}_{2}|}+1)]\wedge . . . \wedge \\&& [(a^{m}_{j1}- a^{m}_{i1}+1)\wedge (a^{m}_{j2}- a^{m}_{i2}+1)\wedge...\wedge (a^{m}_{j|\mathscr{C}_{m}|}- a^{m}_{i|\mathscr{C}_{m}|}+1)]\wedge\\&&[(a^{m+1}_{j1}- a^{m+1}_{i1}+1)\wedge (a^{m+1}_{j2}- a^{m+1}_{i2}+1)\wedge...\wedge (a^{m+1}_{j|\mathscr{C}_{m+1}|}- a^{m+1}_{i|\mathscr{C}_{m+1}|}+1)]
\\&=&d_{ij}\wedge \left[
  \begin{array}{ccccccccccc}
    a^{m+1}_{i1} & a^{m+1}_{i2} & . &. & . & a^{m+1}_{i|\mathscr{C}_{m+1}|}
  \end{array}
\right]\odot\left[
  \begin{array}{ccccccccccc}
     a^{m+1}_{j1} & a^{m+1}_{j2} & . & . & . & a^{m+1}_{j|\mathscr{C}_{m+1}|}
  \end{array}
\right]^{T}.
\end{eqnarray*}

To obtain
$\prod(\mathscr{D}^{+})$, we only need to compute
$\prod(\mathscr{C}_{m+1})$ as follows:
  \begin{eqnarray*}
  \prod(\mathscr{C}_{m+1})=\left[\begin{array}{cccccc}
  a^{m+1}_{11}&a^{m+1}_{12}&.&.&. & a^{m+1}_{1|\mathscr{C}_{m+1}|}\\
  a^{m+1}_{21}&a^{m+1}_{22}&.&.&. & a^{m+1}_{2|\mathscr{C}_{m+1}|}\\
  .&.&.&.&. & .\\
  .&.&.&.&. & .\\
  .&.&.&.&. & .\\
  a^{m+1}_{n1}&a^{m+1}_{n2}&.&.&. & a^{m+1}_{n|\mathscr{C}_{m+1}|}
  \end{array}
\right]\odot \left[\begin{array}{cccccc}
  a^{m+1}_{11}&a^{m+1}_{12}&.&.&. & a^{m+1}_{1|\mathscr{C}_{m+1}|}\\
  a^{m+1}_{21}&a^{m+1}_{22}&.&.&. & a^{m+1}_{2|\mathscr{C}_{m+1}|}\\
  .&.&.&.&. & .\\
  .&.&.&.&. & .\\
  .&.&.&.&. & .\\
  a^{m+1}_{n1}&a^{m+1}_{n2}&.&.&. & a^{m+1}_{n|\mathscr{C}_{m+1}|}
  \end{array}
\right]^{T}.
\end{eqnarray*}

Therefore, we have
 \begin{eqnarray*}
 \prod(\mathscr{D}^{+})=
    \prod(\mathscr{D})\bigwedge
    \prod(\mathscr{C}_{m+1}),
\end{eqnarray*} where
  $\prod(\mathscr{C}_{m+1})=M_{\mathscr{C}_{m+1}}\odot M^{T}_{\mathscr{C}_{m+1}}.$ $\Box$

We also provide the non-incremental and incremental algorithms for computing $XH_{\mathscr{D}^{+}}(X)$ and $XL_{\mathscr{D}^{+}}(X)$ in dynamic covering information systems.

\begin{algorithm}(Non-incremental algorithm of computing $XH_{\mathscr{D}^{+}}(X)$ and $XL_{\mathscr{D}^{+}}(X)\bf{(NIX)}$)

Step 1: Input $(U,\mathscr{D}^{+})$;

Step 2: Construct
$\prod(\mathscr{D}^{+})=M_{\mathscr{D}^{+}}\odot M^{T}_{\mathscr{D}^{+}};$

Step 3: Compute $\mathcal {X}_{XH_{\mathscr{D}^{+}}(X)}=\prod(\mathscr{D}^{+})\bullet \mathcal
{X}_{X}$ and $\mathcal {X}_{XL_{\mathscr{D}^{+}}(X)}=\prod(\mathscr{D}^{+})\odot \mathcal
{X}_{X}$;

Step 4: Output $XH_{\mathscr{D}^{+}}(X)$ and $XL_{\mathscr{D}^{+}}(X)$.
\end{algorithm}

\begin{algorithm}(Incremental algorithm of computing $XH_{\mathscr{D}^{+}}(X)$ and $XL_{\mathscr{D}^{+}}(X)\bf{(IX)}$)

Step 1: Input $(U,\mathscr{D})$ and $(U,\mathscr{D}^{+})$;

Step 2: Construct
$\prod(\mathscr{D})=M_{\mathscr{D}}\odot M^{T}_{\mathscr{D}};$

Step 3: Calculate $
 \prod(\mathscr{D}^{+})=
    \prod(\mathscr{D})\bigwedge
    \prod(\mathscr{C}_{m+1}),
$ where
  $\prod(\mathscr{C}_{m+1})=M_{\mathscr{C}_{m+1}}\odot M^{T}_{\mathscr{C}_{m+1}};$

Step 4:  Get $
XH_{\mathscr{D}^{+}}(X)=\prod(\mathscr{D}^{+})\cdot \mathcal
{X}_{X}\text{ and }
XL_{\mathscr{D}^{+}}(X)=\prod(\mathscr{D}^{+})\odot \mathcal
{X}_{X};
$

Step 5: Output $XH_{\mathscr{D}^{+}}(X)$ and $XL_{\mathscr{D}^{+}}(X)$.
\end{algorithm}

The time complexity of computing the sixth lower and upper
approximations of sets is $O(2n^{2}\ast\sum^{m+1}_{i=1}|\mathscr{C}_{i}|+2n^{2})$ by
Algorithm 3.11. Furthermore, $O(2n^{2}\ast|\mathscr{C}_{m+1}|+3n^{2})$ is the time
complexity of Algorithm 3.12. Therefore, the time complexity of the
incremental algorithm is lower than that of the non-incremental
algorithm.

\begin{example}(Continued from Example 3.2) Taking $X=\{x_{2},x_{3},x_{4}\}$. By Definition 2.3, we first have
\begin{eqnarray*}
\prod(\mathscr{D})=M_{\mathscr{D}}\odot
M_{\mathscr{D}}^{T}=\left[
\begin{array}{ccccc}
1 & 1 & 0 & 0 &0\\
1 & 1 & 0 & 0 &0 \\
0 & 0 & 1 & 1 &0 \\
0 & 0 & 1 & 1 &0 \\
0 & 0 & 0 & 0 &1 \\
\end{array}
\right].
\end{eqnarray*}

Second, by Theorem 3.10, we get
\begin{eqnarray*}
 \prod(\mathscr{D}^{+})=
    \prod(\mathscr{D}) \bigwedge \prod(\mathscr{C}_{4})&=&\left[
\begin{array}{ccccc}
1 & 1 & 0 & 0 &0\\
1 & 1 & 0 & 0 &0 \\
0 & 0 & 1 & 1 &0 \\
0 & 0 & 1 & 1 &0 \\
0 & 0 & 0 & 0 &1 \\
\end{array}
\right]\bigwedge \left[
\begin{array}{ccccc}
1 & 1 & 0 & 0 &0\\
1 & 1 & 0 & 0 &0 \\
0 & 0 & 1 & 1 &0 \\
0 & 0 & 1 & 1 &0 \\
0 & 0 & 0 & 0 &1 \\
\end{array}
\right]\\&=& \left[
\begin{array}{ccccc}
1 & 1 & 0 & 0 &0\\
1 & 1 & 0 & 0 &0 \\
0 & 0 & 1 & 1 &0 \\
0 & 0 & 1 & 1 &0 \\
0 & 0 & 0 & 0 &1 \\
\end{array}
\right].
\end{eqnarray*}

Third, according to Definition 2.4, we obtain
\begin{eqnarray*}
\mathcal {X}_{XH_{\mathscr{D}^{+}}(X)}&=&\prod(\mathscr{D}^{+})\bullet \mathcal {X}_{X}
=\left[
\begin{array}{ccccc}
1 & 1 & 0 & 0 &0\\
1 & 1 & 0 & 0 &0 \\
0 & 0 & 1 & 1 &0 \\
0 & 0 & 1 & 1 &0 \\
0 & 0 & 0 & 0 &1 \\
\end{array}
\right]\bullet\left[
\begin{array}{c}
0 \\
1 \\
1 \\
1 \\
0 \\
\end{array}
\right] =\left[
\begin{array}{ccccc}
1 & 1 & 1 & 1 & 0
\end{array}
\right]^{T},
\\
 \mathcal {X}_{XL_{\mathscr{D}^{+}}(X)}&=&\prod(\mathscr{D}^{+})\odot
\mathcal {X}_{X}=\left[
\begin{array}{ccccc}
1 & 1 & 0 & 0 &0\\
1 & 1 & 0 & 0 &0 \\
0 & 0 & 1 & 1 &0 \\
0 & 0 & 1 & 1 &0 \\
0 & 0 & 0 & 0 &1 \\
\end{array}
\right]\odot \left[
\begin{array}{c}
0 \\
1 \\
1 \\
1 \\
0 \\
\end{array}
\right]=\left[
\begin{array}{ccccc}
0 & 0 & 1 & 1 & 0
\end{array}
\right]^{T}.
\end{eqnarray*}

Therefore, $XH_{\mathscr{D}^{+}}(X)=\{x_{1},x_{2},x_{3},x_{4}\}$ and
$XL_{\mathscr{D}^{+}}(X)=\{x_{3},x_{4}\}$.
\end{example}

In Example 3.11, we must compute
$\prod(\mathscr{C}^{+})$ for constructing $XH_{\mathscr{D}^{+}}(X)$ and
$XL_{\mathscr{D}^{+}}(X)$
using algorithm 3.11. But we only need to calculate
$\prod(\mathscr{C}_{4})$ for computing $XH_{\mathscr{D}^{+}}(X)$ and
$XL_{\mathscr{D}^{+}}(X)$ using Algorithm 3.12. Thereby, the incremental
approach is more effective to compute the sixth lower and upper approximations of sets.

\begin{theorem}
Let $(U,\mathscr{D}^{+})$ be a dynamic covering information system
of $(U,\mathscr{D})$, $\prod(\mathscr{C})=(d_{ij})_{n\times n}$ and
$\prod(\mathscr{C}^{+})=(e_{ij})_{n\times n}$ the type-2 characteristic matrices of
$\mathscr{D}$ and $\mathscr{D}^{+}$, respectively. Then
 \begin{eqnarray*}
 e_{ij}=\left\{
\begin{array}{ccc}
0,&{\rm}& d_{ij}=0;\\
\left[
  \begin{array}{ccccccccccc}
    a^{m+1}_{i1} & a^{m+1}_{i2} & . & . & . & a^{m+1}_{i|\mathscr{C}_{m+1}|}
  \end{array}
\right]\odot\left[
  \begin{array}{ccccccccccc}
     a^{m+1}_{j1} & a^{m+1}_{j2} & . & . & . & a^{m+1}_{j|\mathscr{C}_{m+1}|}
  \end{array}
\right]^{T},&{\rm}& d_{ij}=1.
\end{array}
\right.
\end{eqnarray*}
\end{theorem}

\noindent\textbf{Proof.} It is straightforward by Theorem 3.10.$\Box$

\begin{example}(Continued from Example 3.2) According to Definition 2.3, we have
\begin{eqnarray*}
\prod(\mathscr{D})=M_{\mathscr{D}}\odot
M_{\mathscr{D}}^{T}=\left[
\begin{array}{cccccc}
1 & 1 & 0 & 0 &0\\
1 & 1 & 0 & 0 &0 \\
0 & 0 & 1 & 1 &0 \\
0 & 0 & 1 & 1 &0 \\
0 & 0 & 0 & 0 &1 \\
\end{array}
\right].
\end{eqnarray*}

Therefore, by Theorem 3.14, we
obtain
\begin{eqnarray*}
 \prod(\mathscr{D}^{+})= \left[
\begin{array}{cccccc}
1 & 1 & 0 & 0 &0\\
1 & 1 & 0 & 0 &0 \\
0 & 0 & 1 & 1 &0 \\
0 & 0 & 1 & 1 &0 \\
0 & 0 & 0 & 0 &1 \\
\end{array}
\right].
\end{eqnarray*}
\end{example}

\begin{proposition}
Let $(U,\mathscr{D}^{+})$ be a dynamic covering information system
of $(U,\mathscr{D})$, $\prod(\mathscr{D})$ and
$\prod(\mathscr{D}^{+})$ the type-2 characteristic matrices of
$\mathscr{D}$ and $\mathscr{D}^{+}$, respectively.

$(1)$ If $\prod(\mathscr{D})=(0)_{n\times n}$, then $\prod(\mathscr{D}^{+})=(0)_{n\times n}$;

$(2)$ If $\prod(\mathscr{D})=(1)_{n\times n}$, then $\prod(\mathscr{D}^{+})=\prod(\mathscr{C}_{m+1})$.
\end{proposition}

\noindent\textbf{Proof.} It is straightforward by Theorem 3.14.$\Box$

In practical situations, there are some dynamic covering information systems because of the emigration of coverings, which are presented as follows.

\begin{definition}
Let $(U,\mathscr{D})$ and $(U,\mathscr{D}^{-})$ be covering
information systems, where $U=\{x_{1},x_{2},...,x_{n}\}$,
$\mathscr{D}=\{\mathscr{C}_{1},\mathscr{C}_{2},...,\mathscr{C}_{m}\}$, and
$\mathscr{D}^{-}=\{\mathscr{C}_{1},\mathscr{C}_{2},...,\mathscr{C}_{m-1}\}(m\geq 2 )$. Then $(U,\mathscr{D}^{-})$ is called a dynamic covering information system of $(U,\mathscr{D})$.
\end{definition}

In other words, $(U,\mathscr{D})$ is also referred to as a static covering information system of $(U,\mathscr{D}^{-})$. Furthermore, we employ an example to illustrate dynamic covering information systems given by Definition 3.17 as follows.

\begin{example}
Let $(U,\mathscr{D})$ be a static covering information
system, where $U=\{x_{1},x_{2},x_{3},x_{4},x_{5}\}$,
$\mathscr{D}=\{\mathscr{C}_{1},\mathscr{C}_{2},\mathscr{C}_{3},\mathscr{C}_{4}\}$,
$\mathscr{C}_{1}=\{\{x_{1},x_{2},x_{3},x_{4}\},\{x_{5}\}\}$,
$\mathscr{C}_{2}=\{\{x_{1},x_{2}\},\{x_{3},x_{4},x_{5}\}\}$,
$\mathscr{C}_{3}=\{\{x_{1},x_{2},x_{5}\},\{x_{3},x_{4}\}\}$, and
$\mathscr{C}_{4}=\{\{x_{1},x_{2}\},\{x_{3},x_{4}\},\{x_{5}\}\}$.
If we delete $\mathscr{C}_{4}$ from $\mathscr{D}$, then we obtain a dynamic covering information system $(U,\mathscr{D}^{-})$ of $(U,\mathscr{D})$, where
$\mathscr{D}^{-}=\{\mathscr{C}_{1},\mathscr{C}_{2},\mathscr{C}_{3}\}$.
\end{example}

We also show how to construct $\Gamma(\mathscr{C}^{-})$
based on $\Gamma(\mathscr{C})$. For convenience, we denote
$\Gamma(\mathscr{D})=(b_{ij})_{n\times n}$ and
$\Gamma(\mathscr{D}^{-})=(c^{-}_{ij})_{n\times n}$.

\begin{theorem}
Let $(U,\mathscr{D}^{-})$ be a dynamic covering information system
of $(U,\mathscr{D})$, $\Gamma(\mathscr{D})=(b_{ij})_{n\times n}$ and
$\Gamma(\mathscr{D}^{-})=(c^{-}_{ij})_{n\times n}$ the type-1 characteristic matrices of
$\mathscr{D}$ and $\mathscr{D}^{+}$, respectively. Then
 \begin{eqnarray*}
 c^{-}_{ij}=\left\{
\begin{array}{ccc}
0,&{\rm}& b_{ij}=0;\\
1,&{\rm}& b_{ij}=1\wedge \triangle c_{ij}=0;\\
c^{\ast}_{ij},&{\rm}& b_{ij}=1\wedge \triangle c_{ij}=1.
\end{array}
\right.
\end{eqnarray*}
where
\begin{eqnarray*}
\triangle c_{ij}&=&\left[
  \begin{array}{ccccccccccc}
    a^{m+1}_{i1} & a^{m+1}_{i2} & . &  . & . & a^{m+1}_{i|\mathscr{C}_{m+1}|}
  \end{array}
\right]\bullet\left[
  \begin{array}{ccccccccccc}
     a^{m+1}_{j1} & a^{m+1}_{j2} & . & . & . & a^{m+1}_{j|\mathscr{C}_{m+1}|}
  \end{array}
\right]^{T};\\
c^{\ast}_{ij}&=&\left[
  \begin{array}{ccccccccccccccc}
    a^{1}_{i1} & a^{1}_{i2} & . & a^{1}_{i|\mathscr{C}_{1}|} & a^{2}_{i1} & a^{2}_{i2} & . & a^{2}_{i|\mathscr{C}_{2}|} & . & . & . & a^{m-1}_{i1} & a^{m-1}_{i2} & . & a^{m-1}_{i|\mathscr{C}_{m-1}|}\\
  \end{array}
\right]\bullet\\&&\left[
  \begin{array}{ccccccccccccccc}
   a^{1}_{j1} & a^{1}_{j2} & . & a^{1}_{j|\mathscr{C}_{1}|} & a^{2}_{j1} & a^{2}_{j2} & . & a^{2}_{j|\mathscr{C}_{2}|} & . & . & . & a^{m-1}_{j1} & a^{m-1}_{j2} & . & a^{m-1}_{j|\mathscr{C}_{m-1}|}\\
  \end{array}
\right]^{T}.
\end{eqnarray*}
\end{theorem}

\noindent\textbf{Proof.} It is straightforward by Theorem 3.3.$\Box$

 \begin{example}  (Continued from Example 3.14) Taking $X=\{x_{2},x_{3},x_{4}\}$.
According to Definition 2.3, we first obtain
\begin{eqnarray*}
\Gamma(\mathscr{D})=M_{\mathscr{D}}\bullet
M_{\mathscr{D}}^{T}=\left[
\begin{array}{ccccc}
1 & 1 & 1 & 1 &1 \\
1 & 1 & 1 & 1 &1 \\
1 & 1 & 1 & 1 &1 \\
1 & 1 & 1 & 1 &1 \\
1 & 1 & 1 & 1 &1 \\
\end{array}
\right].
\end{eqnarray*}

Second, by Theorem 3.19, we get
\begin{eqnarray*}
\Gamma(\mathscr{D}^{-})=\left[
\begin{array}{ccccc}
1 & 1 & 1 & 1 &1 \\
1 & 1 & 1 & 1 &1 \\
1 & 1 & 1 & 1 &1 \\
1 & 1 & 1 & 1 &1 \\
1 & 1 & 1 & 1 &1 \\
\end{array}
\right].
\end{eqnarray*}

Third, by Definition 2.4, we have
\begin{eqnarray*}
\mathcal {X}_{SH_{\mathscr{D}^{-}}(X)}&=&\Gamma(\mathscr{D}^{-})\bullet \mathcal {X}_{X}
=\left[
\begin{array}{ccccc}
1 & 1 & 1 & 1 &1 \\
1 & 1 & 1 & 1 &1 \\
1 & 1 & 1 & 1 &1 \\
1 & 1 & 1 & 1 &1 \\
1 & 1 & 1 & 1 &1 \\
\end{array}
\right]\bullet\left[
\begin{array}{c}
0 \\
1 \\
1 \\
1 \\
0 \\
\end{array}
\right] =\left[
\begin{array}{ccccc}
1 & 1 & 1 & 1 & 1
\end{array}
\right]^{T},
\\
 \mathcal {X}_{SL_{\mathscr{D}^{-}}(X)}&=&\Gamma(\mathscr{D}^{-})\odot
\mathcal {X}_{X}=\left[
\begin{array}{ccccc}
1 & 1 & 1 & 1 &1 \\
1 & 1 & 1 & 1 &1 \\
1 & 1 & 1 & 1 &1 \\
1 & 1 & 1 & 1 &1 \\
1 & 1 & 1 & 1 &1 \\
\end{array}
\right]\odot \left[
\begin{array}{c}
0 \\
1 \\
1 \\
1 \\
0 \\
\end{array}
\right]=\left[
\begin{array}{ccccc}
0 & 0 & 0 & 0 & 0
\end{array}
\right]^{T}.
\end{eqnarray*}

Therefore, $SH_{\mathscr{D}^{-}}(X)=\{x_{1},x_{2},x_{3},x_{4},x_{5}\}$ and
$SL_{\mathscr{D}^{-}}(X)=\emptyset$.
\end{example}

We also show how to construct $\prod(\mathscr{C}^{-})$
based on $\prod(\mathscr{C})$. For convenience, we denote
$\prod(\mathscr{D})=(d_{ij})_{n\times n}$ and
$\prod(\mathscr{D}^{-})=(e^{-}_{ij})_{n\times n}$.

\begin{theorem}
Let $(U,\mathscr{D}^{-})$ be a dynamic covering information system
of $(U,\mathscr{D})$, $\prod(\mathscr{D})=(d_{ij})_{n\times n}$ and
$\prod(\mathscr{D}^{-})=(e^{-}_{ij})_{n\times n}$ the type-2 characteristic matrices of
$\mathscr{D}$ and $\mathscr{D}^{-}$, respectively. Then
 \begin{eqnarray*}
 e^{-}_{ij}=\left\{
\begin{array}{ccc}
1,&{\rm}& d_{ij}=1\wedge \triangle e_{ij}=1;\\
0,&{\rm}& d_{ij}=0\wedge \triangle e_{ij}=1;\\
e^{\ast}_{ij},&{\rm}& d_{ij}=0\wedge \triangle e_{ij}=0.
\end{array}
\right.
\end{eqnarray*}
where
\begin{eqnarray*}
\triangle e_{ij}&=&\left[
  \begin{array}{ccccccccccc}
    a^{m+1}_{i1} & a^{m+1}_{i2} & . & . & . & a^{m+1}_{i|\mathscr{C}_{m+1}|}
  \end{array}
\right]\odot\left[
  \begin{array}{ccccccccccc}
     a^{m+1}_{j1} & a^{m+1}_{j2} & . & . & . & a^{m+1}_{j|\mathscr{C}_{m+1}|}
  \end{array}
\right]^{T},\\
e^{\ast}_{ij}&=&\left[
  \begin{array}{ccccccccccccccc}
    a^{1}_{i1} & a^{1}_{i2} & . & a^{1}_{i|\mathscr{C}_{1}|} & a^{2}_{i1} & a^{2}_{i2} & . & a^{2}_{i|\mathscr{C}_{2}|} & . & . & . & a^{m-1}_{i1} & a^{m-1}_{i2} & . & a^{m-1}_{i|\mathscr{C}_{m-1}|}\\
  \end{array}
\right]\odot\\&&\left[
  \begin{array}{ccccccccccccccc}
   a^{1}_{j1} & a^{1}_{j2} & . & a^{1}_{j|\mathscr{C}_{1}|} & a^{2}_{j1} & a^{2}_{j2} & . & a^{2}_{j|\mathscr{C}_{2}|} & . & . & . & a^{m-1}_{j1} & a^{m-1}_{j2} & . & a^{m-1}_{j|\mathscr{C}_{m-1}|}\\
  \end{array}
\right]^{T}.
\end{eqnarray*}
\end{theorem}

\noindent\textbf{Proof.} It is straightforward by Theorem 3.10.$\Box$

\begin{example} (Continued from Example 3.18)
According to Definition 2.3, we first have
\begin{eqnarray*}
\prod(\mathscr{D})=M_{\mathscr{D}}\odot
M_{\mathscr{D}}^{T}=\left[
\begin{array}{ccccc}
1 & 1 & 0 & 0 &0\\
1 & 1 & 0 & 0 &0 \\
0 & 0 & 1 & 1 &0 \\
0 & 0 & 1 & 1 &0 \\
0 & 0 & 0 & 0 &1 \\
\end{array}
\right].
\end{eqnarray*}

Second, by Theorem 3.21, we get
\begin{eqnarray*}
 \prod(\mathscr{D}^{-})=\left[
\begin{array}{ccccc}
1 & 1 & 0 & 0 &0\\
1 & 1 & 0 & 0 &0 \\
0 & 0 & 1 & 1 &0 \\
0 & 0 & 1 & 1 &0 \\
0 & 0 & 0 & 0 &1 \\
\end{array}
\right].
\end{eqnarray*}

Third, by Definition 2.4, we obtain
\begin{eqnarray*}
\mathcal {X}_{XH_{\mathscr{D}^{-}}(X)}&=&\prod(\mathscr{D}^{-})\bullet \mathcal {X}_{X}
=\left[
\begin{array}{ccccc}
1 & 1 & 0 & 0 &0\\
1 & 1 & 0 & 0 &0 \\
0 & 0 & 1 & 1 &0 \\
0 & 0 & 1 & 1 &0 \\
0 & 0 & 0 & 0 &1 \\
\end{array}
\right] \bullet \left[
\begin{array}{c}
0 \\
1 \\
1 \\
1 \\
0 \\
\end{array}
\right] =\left[
\begin{array}{ccccc}
1 & 1 & 1 & 1 & 0
\end{array}
\right]^{T},\\
 \mathcal {X}_{XL_{\mathscr{D}^{-}}(X)}&=&\prod(\mathscr{D}^{-})\odot
\mathcal {X}_{X}=\left[
\begin{array}{ccccc}
1 & 1 & 0 & 0 &0\\
1 & 1 & 0 & 0 &0 \\
0 & 0 & 1 & 1 &0 \\
0 & 0 & 1 & 1 &0 \\
0 & 0 & 0 & 0 &1 \\
\end{array}
\right] \odot \left[
\begin{array}{c}
0 \\
1 \\
1 \\
1 \\
0 \\
\end{array}
\right]=\left[
\begin{array}{ccccc}
0 & 0 & 1 & 1 & 0
\end{array}
\right]^{T}.
\end{eqnarray*}

Therefore, $XH_{\mathscr{D}^{-}}(X)=\{x_{1},x_{2},x_{3},x_{4}\}$ and
$XL_{\mathscr{D}^{-}}(X)=\{x_{3},x_{4}\}$.
\end{example}

In practical situations, we compute the type-1 and type-2 characteristic matrices for dynamic covering information systems with the
immigrations and emigrations of covering simultaneously using
two steps as follows:
(1) compute the  type-1 and type-2 characteristic
matrices by Theorems 3.3 and 3.10, respectively; (2)
construct the type-1 and type-2 characteristic matrices by Theorems 3.19 and 3.21,
respectively. Actually, there are more dynamic covering information
systems given by Definition 3.1 than those defined by Definition
3.17. Therefore, the following discussion focuses on the dynamic covering information
systems given by Definition 3.1.

\section{Experimental analysis}

In this section, we perform experiments to illustrate the
effectiveness of Algorithms 3.5 and 3.12 for computing the second and
sixth lower and upper approximations of concepts, respectively, in dynamic covering
information systems with the immigration of coverings.

To test Algorithms 3.5 and 3.12, we generated randomly ten artificial
covering information systems
$\{(U_{i},\mathscr{D}_{i})|i=1,2,3,...,10\}$, which are outlined in
Table 1, where  $|U_{i}|$ means the number of objects in $U_{i}$ and
$|\mathscr{D}_{i}|$ denotes the cardinality of the covering set
$\mathscr{D}_{i}$. For convenience, each covering here contains five
elements in each covering information system
$(U_{i},\mathscr{D}_{i})$. Moreover, we conducted all computations
on a PC with a Intel(R) Dual-Core(TM) i5-4590 CPU $@$ 3.30 GHZ and 8
GB memory, running 64-bit Windows 7; the software used was 64-bit
Matlab R2009b.

\begin{table}[H]\renewcommand{\arraystretch}{1.1}
\caption{Covering information systems for experiments.
} \tabcolsep0.6in
\begin{tabular}{cccc}
\hline
No.&Name&$|U_{i}|$ &$|\mathscr{D}_{i}|$\\\hline
1  & $(U_{1},\mathscr{D}_{1})$ &2000& 1000\\
2  & $(U_{2},\mathscr{D}_{2})$ &4000& 1000\\
3  & $(U_{3},\mathscr{D}_{3})$ &6000& 1000\\
4  & $(U_{4},\mathscr{D}_{4})$ &8000& 1000\\
5  & $(U_{5},\mathscr{D}_{5})$ &10000& 1000\\
6  & $(U_{6},\mathscr{D}_{6})$ &12000& 1000\\
7  & $(U_{7},\mathscr{D}_{7})$ &14000& 1000\\
8  & $(U_{8},\mathscr{D}_{8})$ &16000& 1000\\
9  & $(U_{9},\mathscr{D}_{9})$ &18000& 1000\\
10  & $(U_{10},\mathscr{D}_{10})$ &20000& 1000\\
\hline
\end{tabular}
\end{table}

\subsection{The stability of Algorithms 3.4, 3.5, 3.11 and 3.12}

In this section, we illustrate the stability of Algorithms 3.4, 3.5, 3.11 and 3.12 with the experimental results. First, we present the concept of sub-covering information system as follows.

\begin{definition}
Let $(U,\mathscr{D})$ be a covering information
system, and $\mathscr{D}^{j}\subseteq \mathscr{D}$. Then $(U,\mathscr{D}^{j})$ is called a sub-covering information system of $(U,\mathscr{D})$.
\end{definition}

According to Definition 4.1, we see that a sub-covering information system is a covering information system. Furthermore, the number of sub-covering covering information systems is $2^{|\mathscr{D}|}-1$ for the covering information system $(U,\mathscr{D})$.

\begin{example}
Let $(U,\mathscr{D})$ be a covering information
system, where $U=\{x_{1},x_{2},x_{3},x_{4},x_{5}\}$,
$\mathscr{D}=\{\mathscr{C}_{1},\mathscr{C}_{2},\mathscr{C}_{3}\}$,
$\mathscr{C}_{1}=\{\{x_{1},x_{2},x_{3},x_{4}\},\{x_{5}\}\}$,
$\mathscr{C}_{2}=\{\{x_{1},x_{2}\},\{x_{3},x_{4},x_{5}\}\}$, and
$\mathscr{C}_{3}=\{\{x_{1},x_{2},x_{5}\},\{x_{3},x_{4}\}\}$. Then we obtain a sub-covering information system $(U,\mathscr{D}^{1})$ by taking
$\mathscr{D}^{1}=\{\mathscr{C}_{1},\mathscr{C}_{2}\}$. Furthermore, $(U,\mathscr{D}^{2})$ is also a sub-covering information system of $(U,\mathscr{D})$, where
$\mathscr{D}^{2}=\{\mathscr{C}_{1},\mathscr{C}_{3}\}$.
\end{example}

Second, according to Definition 4.1, we obtain ten sub-covering information systems
$\{(U_{i},\mathscr{D}^{j}_{i})| j=1,2,3,...,10\}$ for covering information system $(U_{i},\mathscr{D}_{i})$ outlined in Table 1, and show these sub-covering information systems
in Table 2, where $\mathscr{D}^{j}_{i}\subseteq\mathscr{D}_{i}$.

\begin{table}[H]\renewcommand{\arraystretch}{1.1}
\caption{Sub-covering information systems for experiments. }
\tabcolsep0.143in
\begin{tabular}{ccccccccccccc}
\hline
$(U,\mathscr{D})$
&$|\mathscr{D}^{1}_{i}|$&$|\mathscr{D}^{2}_{i}|$&$|\mathscr{D}^{3}_{i}|$
&$|\mathscr{D}^{4}_{i}|$&$|\mathscr{D}^{5}_{i}|$&$|\mathscr{D}^{6}_{i}|$&$|\mathscr{D}^{7}_{i}|$
&$|\mathscr{D}^{8}_{i}|$&$|\mathscr{D}^{9}_{i}|$&$|\mathscr{D}^{10}_{i}|$\\\hline
$(U_{1},\mathscr{D}^{j}_{1})$& 100& 200& 300& 400& 500& 600& 700& 800& 900& 1000\\
$(U_{2},\mathscr{D}^{j}_{2})$& 100& 200& 300& 400& 500& 600& 700& 800& 900& 1000\\
$(U_{3},\mathscr{D}^{j}_{3})$& 100& 200& 300& 400& 500& 600& 700& 800& 900& 1000\\
$(U_{4},\mathscr{D}^{j}_{4})$& 100& 200& 300& 400& 500& 600& 700& 800& 900& 1000\\
$(U_{5},\mathscr{D}^{j}_{5})$& 100& 200& 300& 400& 500& 600& 700& 800& 900& 1000\\
$(U_{6},\mathscr{D}^{j}_{6})$& 100& 200& 300& 400& 500& 600& 700& 800& 900& 1000\\
$(U_{7},\mathscr{D}^{j}_{7})$& 100& 200& 300& 400& 500& 600& 700& 800& 900& 1000\\
$(U_{8},\mathscr{D}^{j}_{8})$& 100& 200& 300& 400& 500& 600& 700& 800& 900& 1000\\
$(U_{9},\mathscr{D}^{j}_{9})$& 100& 200& 300& 400& 500& 600& 700& 800& 900& 1000\\
$(U_{10},\mathscr{D}^{j}_{10})$& 100& 200& 300& 400& 500& 600& 700& 800& 900& 1000\\
\hline
\end{tabular}
\end{table}

Third, to demonstrate the stability of Algorithms 3.4, 3.5,
3.11, and 3.12, we compute the second and sixth lower and upper
approximations of sets in sub-covering information systems
$\{(U_{i},\mathscr{D}^{j}_{i})|i,j=1,2,3,...,10\}$. For example, we
show the process of computing the second and sixth lower and upper
approximations of sets in covering information system
$(U_{1},\mathscr{D}^{1}_{1})$, where $|U_{1}|=2000$ and
$|\mathscr{D}^{1}_{1}|=100$ as follows:

(1) By adding a covering into
$\mathscr{D}^{1}_{1}$, we obtain the dynamic covering information
system $(U_{1},\mathscr{D}^{1+}_{1})$, where $|U_{1}|=2000$ and
$|\mathscr{D}^{1+}_{1}|=101$.

(2) Taking any $X\subseteq
U_{1}$, we compute the second lower and upper approximations of $X$
in dynamic covering information system
$(U_{1},\mathscr{D}^{1+}_{1})$ using Algorithms 3.4 and 3.5. Furthermore,
we also compute the sixth lower and upper approximations of
$X$ in dynamic covering information system
$(U_{1},\mathscr{D}^{1+}_{1})$ using Algorithms 3.11 and 3.12. To
confirm the accuracy of the experiment results, we conduct each
experiment ten times and show the average time of ten experimental
results in Table 3, where $t(s)$ denotes that the measure of time is
in seconds.

\begin{table}[H]\renewcommand{\arraystretch}{1.1}
\caption{Computational times using NIS, NIX, IS, and IX in $(U_{i},\mathscr{D}^{j}_{i})$} \tabcolsep0.046in
\label{bigtable}
\begin{center}
\begin{tabular}{cccccccccccc}
\hline
$(U,\mathscr{D})$ & Algo. & $\mathscr{D}^{1}_{i}$ & $\mathscr{D}^{2}_{i}$ & $\mathscr{D}^{3}_{i}$ & $\mathscr{D}^{4}_{i}$ & $\mathscr{D}^{5}_{i}$ & $\mathscr{D}^{6}_{i}$ & $\mathscr{D}^{7}_{i}$ & $\mathscr{D}^{8}_{i}$ & $\mathscr{D}^{9}_{i}$ & $\mathscr{D}^{10}_{i}$  \\ \hline
\multirow{4}*{$(U_{1},\mathscr{D}^{j}_{1})$} & NIS & 0.294 & 0.291 & 0.308 & 0.305 & 0.303 & 0.326 & 0.302 & 0.315 & 0.333 & 0.333 \\
& NIX & 0.433 & 0.452 & 0.470 & 0.487 & 0.505 & 0.515 & 0.538 & 0.555 & 0.563 & 0.581 \\
& IS & 0.019 & 0.019 & 0.019 & 0.019 & 0.019 & 0.019 & 0.019 & 0.019 & 0.019 & 0.019 \\
& IX & 0.044 & 0.044 & 0.044 & 0.045 & 0.045 & 0.046 & 0.046 & 0.046 & 0.045 & 0.044 \\
\hline
\multirow{4}*{$(U_{2},\mathscr{D}^{j}_{2})$} & NIS & 1.392 & 1.239 & 1.392 & 1.488 & 1.414 & 1.318 & 1.261 & 1.508 & 1.446 & 1.290 \\
& NIX & 1.789 & 1.844 & 1.871 & 1.896 & 1.957 & 2.018 & 2.051 & 2.067 & 2.114 & 2.153 \\
& IS & 0.116 & 0.116 & 0.116 & 0.115 & 0.115 & 0.114 & 0.114 & 0.115 & 0.113 & 0.114 \\
& IX & 0.276 & 0.276 & 0.279 & 0.279 & 0.277 & 0.277 & 0.278 & 0.278 & 0.277 & 0.278 \\
\hline
\multirow{4}*{$(U_{3},\mathscr{D}^{j}_{3})$} & NIS & 2.815 & 2.743 & 2.950 & 3.004 & 2.978 & 3.276 & 2.860 & 3.058 & 3.427 & 3.091 \\
& NIX & 4.123 & 4.185 & 4.255 & 4.270 & 4.391 & 4.421 & 4.553 & 4.592 & 4.592 & 4.685 \\
& IS & 0.281 & 0.280 & 0.282 & 0.282 & 0.280 & 0.281 & 0.280 & 0.281 & 0.280 & 0.280 \\
& IX & 0.693 & 0.692 & 0.689 & 0.691 & 0.691 & 0.693 & 0.692 & 0.688 & 0.686 & 0.687 \\
\hline
\multirow{4}*{$(U_{4},\mathscr{D}^{j}_{4})$} & NIS & 4.812 & 4.911 & 5.758 & 5.260 & 5.441 & 5.773 & 5.546 & 5.216 & 5.085 & 5.683 \\
& NIX & 7.356 & 7.437 & 7.474 & 7.583 & 7.771 & 7.855 & 7.985 & 8.060 & 8.128 & 8.171 \\
& IS & 0.536 & 0.534 & 0.534 & 0.535 & 0.535 & 0.535 & 0.534 & 0.533 & 0.534 & 0.534 \\
& IX & 1.312 & 1.306 & 1.307 & 1.307 & 1.308 & 1.310 & 1.308 & 1.307 & 1.308 & 1.306 \\
\hline
\multirow{4}*{$(U_{5},\mathscr{D}^{j}_{5})$} & NIS & 7.816 & 8.327 & 8.747 & 8.482 & 9.026 & 9.564 & 9.596 & 9.171 & 9.096 & 8.673 \\
& NIX & 11.695 & 11.841 & 11.794 & 12.020 & 12.065 & 12.161 & 12.315 & 12.551 & 12.595 & 12.828 \\
& IS & 0.899 & 0.900 & 0.901 & 0.901 & 0.901 & 0.900 & 0.899 & 0.898 & 0.901 & 0.901 \\
& IX & 2.220 & 2.217 & 2.217 & 2.217 & 2.217 & 2.215 & 2.217 & 2.215 & 2.215 & 2.219 \\
\hline
\multirow{4}*{$(U_{6},\mathscr{D}^{j}_{6})$} & NIS & 11.815 & 11.974 & 11.925 & 13.100 & 11.836 & 11.959 & 12.498 & 13.341 & 13.524 & 13.960 \\
& NIX & 16.855 & 17.095 & 17.211 & 17.254 & 17.646 & 17.956 & 17.972 & 18.145 & 18.271 & 18.487 \\
& IS & 1.311 & 1.310 & 1.310 & 1.311 & 1.312 & 1.314 & 1.314 & 1.312 & 1.313 & 1.313 \\
& IX & 3.232 & 3.232 & 3.234 & 3.233 & 3.237 & 3.235 & 3.234 & 3.233 & 3.235 & 3.235 \\
\hline
\multirow{4}*{$(U_{7},\mathscr{D}^{j}_{7})$} & NIS & 15.390 & 17.084 & 18.474 & 16.966 & 17.148 & 17.170 & 15.884 & 19.168 & 16.926 & 17.414 \\
& NIX & 23.261 & 23.363 & 23.450 & 23.800 & 24.187 & 24.420 & 25.003 & 24.783 & 25.168 & 25.532 \\
& IS & 1.816 & 1.812 & 1.816 & 1.815 & 1.816 & 1.814 & 1.814 & 1.816 & 1.813 & 1.813 \\
& IX & 4.489 & 4.493 & 4.487 & 4.487 & 4.482 & 4.491 & 4.484 & 4.486 & 4.493 & 4.488 \\
\hline
\multirow{4}*{$(U_{8},\mathscr{D}^{j}_{8})$} & NIS & 19.268 & 20.399 & 20.427 & 25.125 & 24.195 & 21.648 & 22.758 & 25.579 & 23.050 & 24.523 \\
& NIX & 30.730 & 31.684 & 31.922 & 31.874 & 32.383 & 32.862 & 33.181 & 33.333 & 34.418 & 34.292 \\
& IS & 2.329 & 2.348 & 2.334 & 2.339 & 2.335 & 2.333 & 2.331 & 2.335 & 2.336 & 2.329 \\
& IX & 5.739 & 5.730 & 5.730 & 5.736 & 5.726 & 5.733 & 5.738 & 5.735 & 5.722 & 5.727 \\
\hline
\multirow{4}*{$(U_{9},\mathscr{D}^{j}_{9})$} & NIS & 27.727 & 24.680 & 27.103 & 27.168 & 28.197 & 29.591 & 27.240 & 28.775 & 30.566 & 33.012 \\
& NIX & 38.314 & 40.551 & 40.245 & 40.752 & 41.291 & 41.870 & 42.196 & 42.385 & 42.256 & 42.956 \\
& IS & 3.081 & 3.079 & 3.077 & 3.084 & 3.080 & 3.077 & 3.082 & 3.082 & 3.077 & 3.076 \\
& IX & 7.632 & 7.634 & 7.626 & 7.639 & 7.634 & 7.633 & 7.634 & 7.632 & 7.632 & 7.645 \\
\hline
\multirow{4}*{$(U_{10},\mathscr{D}^{j}_{10})$} & NIS & 38.748 & 31.240 & 35.121 & 35.774 & 37.235 & 36.723 & 38.385 & 37.174 & 40.780 & 36.097 \\
& NIX & 47.874 & 49.295 & 50.355 & 52.001 & 50.068 & 53.765 & 52.696 & 54.178 & 53.650 & 55.917 \\
& IS & 3.728 & 3.730 & 3.725 & 3.727 & 3.719 & 3.723 & 3.724 & 3.721 & 3.722 & 3.721 \\
& IX & 9.219 & 9.223 & 9.230 & 9.217 & 9.215 & 9.218 & 9.219 & 9.225 & 9.217 & 9.220 \\
\hline
\end{tabular}
\end{center}
\end{table}

(3) We compute the variance of ten experimental
results for computing the approximations of sets in each dynamic
covering information system and show all variance values in Table 4.
According to the experimental results, we see that Algorithms 3.4, 3.5, 3.11, and 3.12 are stable to compute
the second and sixth lower and upper approximations of sets in
dynamic covering information systems with the immigration of
coverings. Especially, Algorithms 3.5 and 3.12 are more stable to compute the second and sixth lower and upper approximations of sets than Algorithms 3.4 and 3.11, respectively, in dynamic covering information systems.

\begin{table}[H]\renewcommand{\arraystretch}{1.09}
\caption{Variance values of computational times using NIS, NIX, IS, and IX in $(U_{i},\mathscr{D}^{j}_{i})$} \tabcolsep0.08in
\label{bigtable}
\begin{center}
\begin{tabular}{cccccccccccc}
\hline
$(U,\mathscr{D})$ & Algo. & $\mathscr{D}^{1}_{i}$ & $\mathscr{D}^{2}_{i}$ & $\mathscr{D}^{3}_{i}$ & $\mathscr{D}^{4}_{i}$ & $\mathscr{D}^{5}_{i}$ & $\mathscr{D}^{6}_{i}$ & $\mathscr{D}^{7}_{i}$ & $\mathscr{D}^{8}_{i}$ & $\mathscr{D}^{9}_{i}$ & $\mathscr{D}^{10}_{i}$  \\ \hline
\multirow{4}*{$(U_{1},\mathscr{D}^{j}_{1})$} & NIS & 0.005 & 0.004 & 0.002 & 0.001 & 0.001 & 0.002 & 0.001 & 0.001 & 0.001 & 0.001 \\
& NIX & 0.004 & 0.001 & 0.001 & 0.001 & 0.001 & 0.001 & 0.002 & 0.001 & 0.001 & 0.001 \\
& IS & 0.001 & 0.000 & 0.000 & 0.000 & 0.000 & 0.000 & 0.000 & 0.000 & 0.000 & 0.000 \\
& IX & 0.000 & 0.001 & 0.001 & 0.000 & 0.001 & 0.001 & 0.000 & 0.000 & 0.000 & 0.001 \\
\hline
\multirow{4}*{$(U_{2},\mathscr{D}^{j}_{2})$} & NIS & 0.001 & 0.002 & 0.002 & 0.002 & 0.002 & 0.002 & 0.001 & 0.003 & 0.002 & 0.002 \\
& NIX & 0.003 & 0.002 & 0.002 & 0.003 & 0.003 & 0.002 & 0.004 & 0.003 & 0.002 & 0.003 \\
& IS & 0.001 & 0.000 & 0.001 & 0.000 & 0.000 & 0.000 & 0.001 & 0.000 & 0.000 & 0.001 \\
& IX & 0.001 & 0.001 & 0.001 & 0.001 & 0.002 & 0.001 & 0.001 & 0.001 & 0.001 & 0.002 \\
\hline
\multirow{4}*{$(U_{3},\mathscr{D}^{j}_{3})$} & NIS & 0.003 & 0.003 & 0.004 & 0.004 & 0.003 & 0.004 & 0.002 & 0.004 & 0.004 & 0.002 \\
& NIX & 0.005 & 0.004 & 0.008 & 0.004 & 0.004 & 0.013 & 0.005 & 0.017 & 0.005 & 0.003 \\
& IS & 0.002 & 0.001 & 0.001 & 0.001 & 0.001 & 0.001 & 0.001 & 0.001 & 0.001 & 0.002 \\
& IX & 0.006 & 0.003 & 0.002 & 0.001 & 0.003 & 0.003 & 0.002 & 0.004 & 0.003 & 0.002 \\
\hline
\multirow{4}*{$(U_{4},\mathscr{D}^{j}_{4})$} & NIS & 0.003 & 0.013 & 0.009 & 0.003 & 0.006 & 0.004 & 0.004 & 0.006 & 0.004 & 0.006 \\
& NIX & 0.019 & 0.004 & 0.005 & 0.007 & 0.004 & 0.006 & 0.007 & 0.004 & 0.004 & 0.014 \\
& IS & 0.003 & 0.002 & 0.002 & 0.002 & 0.003 & 0.003 & 0.002 & 0.002 & 0.002 & 0.003 \\
& IX & 0.004 & 0.004 & 0.002 & 0.004 & 0.003 & 0.007 & 0.004 & 0.003 & 0.004 & 0.002 \\
\hline
\multirow{4}*{$(U_{5},\mathscr{D}^{j}_{5})$} & NIS & 0.005 & 0.019 & 0.014 & 0.003 & 0.009 & 0.006 & 0.008 & 0.008 & 0.006 & 0.006 \\
& NIX & 0.020 & 0.022 & 0.009 & 0.010 & 0.020 & 0.014 & 0.023 & 0.009 & 0.021 & 0.013 \\
& IS & 0.003 & 0.003 & 0.004 & 0.004 & 0.002 & 0.003 & 0.003 & 0.003 & 0.004 & 0.001 \\
& IX & 0.008 & 0.007 & 0.007 & 0.007 & 0.005 & 0.005 & 0.005 & 0.003 & 0.004 & 0.005 \\
\hline
\multirow{4}*{$(U_{6},\mathscr{D}^{j}_{6})$} & NIS & 0.008 & 0.007 & 0.007 & 0.032 & 0.012 & 0.008 & 0.007 & 0.007 & 0.025 & 0.008 \\
& NIX & 0.024 & 0.015 & 0.019 & 0.034 & 0.021 & 0.045 & 0.028 & 0.033 & 0.042 & 0.031 \\
& IS & 0.005 & 0.004 & 0.005 & 0.003 & 0.005 & 0.007 & 0.005 & 0.003 & 0.003 & 0.005 \\
& IX & 0.004 & 0.004 & 0.004 & 0.007 & 0.006 & 0.007 & 0.005 & 0.008 & 0.005 & 0.005 \\
\hline
\multirow{4}*{$(U_{7},\mathscr{D}^{j}_{7})$} & NIS & 0.023 & 0.008 & 0.057 & 0.012 & 0.009 & 0.021 & 0.022 & 0.032 & 0.025 & 0.030 \\
& NIX & 0.110 & 0.036 & 0.027 & 0.037 & 0.031 & 0.096 & 0.071 & 0.060 & 0.074 & 0.124 \\
& IS & 0.007 & 0.003 & 0.003 & 0.005 & 0.003 & 0.006 & 0.006 & 0.004 & 0.004 & 0.004 \\
& IX & 0.008 & 0.011 & 0.007 & 0.009 & 0.003 & 0.006 & 0.005 & 0.005 & 0.014 & 0.005 \\
\hline
\multirow{4}*{$(U_{8},\mathscr{D}^{j}_{8})$} & NIS & 0.043 & 0.039 & 0.034 & 0.075 & 0.041 & 0.126 & 0.053 & 0.035 & 0.042 & 0.087 \\
& NIX & 0.211 & 0.402 & 0.411 & 0.510 & 0.278 & 0.313 & 0.433 & 0.437 & 0.808 & 0.492 \\
& IS & 0.005 & 0.013 & 0.006 & 0.007 & 0.004 & 0.007 & 0.006 & 0.003 & 0.003 & 0.012 \\
& IX & 0.011 & 0.011 & 0.010 & 0.007 & 0.010 & 0.011 & 0.010 & 0.012 & 0.006 & 0.007 \\
\hline
\multirow{4}*{$(U_{9},\mathscr{D}^{j}_{9})$} & NIS & 0.050 & 0.059 & 0.070 & 0.105 & 0.072 & 0.247 & 0.092 & 0.102 & 0.116 & 0.191 \\
& NIX & 0.234 & 0.830 & 0.794 & 0.729 & 0.880 & 0.828 & 0.426 & 0.176 & 0.722 & 0.761 \\
& IS & 0.010 & 0.005 & 0.005 & 0.008 & 0.008 & 0.005 & 0.010 & 0.010 & 0.010 & 0.010 \\
& IX & 0.006 & 0.009 & 0.008 & 0.013 & 0.006 & 0.015 & 0.011 & 0.007 & 0.007 & 0.019 \\
\hline
\multirow{4}*{$(U_{10},\mathscr{D}^{j}_{10})$} & NIS & 0.495 & 0.088 & 0.167 & 0.167 & 0.143 & 0.592 & 0.210 & 0.124 & 0.308 & 0.459 \\
& NIX & 0.608 & 1.353 & 1.380 & 1.841 & 0.396 & 1.083 & 0.747 & 1.435 & 1.754 & 1.877 \\
& IS & 0.011 & 0.011 & 0.009 & 0.009 & 0.007 & 0.012 & 0.005 & 0.009 & 0.007 & 0.005 \\
& IX & 0.008 & 0.008 & 0.017 & 0.013 & 0.011 & 0.011 & 0.015 & 0.016 & 0.016 & 0.018 \\
\hline
\end{tabular}
\end{center}
\end{table}

\subsection{The influence of the cardinality of object set}

In this section, we analyze the influence of the cardinality of
object set on time of computing the second and sixth lower and upper
approximations of sets using Algorithms 3.4, 3.5, 3.11, and 3.12 in
dynamic covering information systems with the covering immigration.

There are ten sub-covering information systems with the same
cardinality of covering sets. First, we compare the times of
computing the second lower and upper approximations of sets using
Algorithm 3.4 with those using Algorithm 3.5 in dynamic covering
information systems with the same cardinality of covering sets. From
the results in Table 3, we see that the computing times are
increasing with the increasing cardinality of object sets using
Algorithms 3.4 and 3.5. We also find that Algorithm 3.5 executes
faster than Algorithm 3.5 in dynamic covering information systems.
Second, we also compare the times of computing the sixth lower and
upper approximations of sets using Algorithm 3.11 with those using
Algorithm 3.12 in dynamic covering information systems with the same
cardinality of covering sets. From the results in Table 3, we see
that the computing times are increasing with the increasing
cardinality of object sets using Algorithms 3.11 and 3.12. We also
find that Algorithm 3.12 executes faster than Algorithm 3.11 in
dynamic covering information systems. Third, to illustrate the
effectiveness of Algorithms 3.5 and 3.12, we show these results in
Figures 1-10. In each figure, $NIS,
IS, NIX,$ and $ IX$ mean Algorithms 3.4, 3.5, 3.11, and 3.12,
respectively; $i$ stands for the cardinality of object set in $X$
Axis, while the y-coordinate stands for the time to construct the
approximations of concepts. Therefore, Algorithms 3.5 and 3.12 are
more effective to compute the second and sixth lower and upper
approximations of sets, respectively, in dynamic covering
information systems.

\begin{figure}[H]
\begin{center}
\includegraphics[width=8cm]{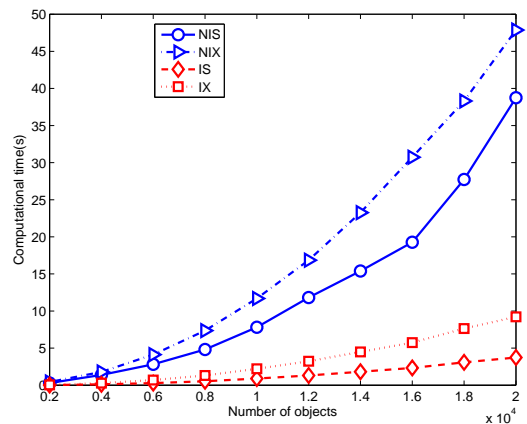}\\
\caption{Computational times using Algorithms 3.4, 3.5, 3.11, and 3.12 in
$(U_{i},\mathscr{D}^{1}_{i})$.}
\end{center}
\end{figure}

\begin{figure}[H]
\begin{center}
\includegraphics[width=8cm]{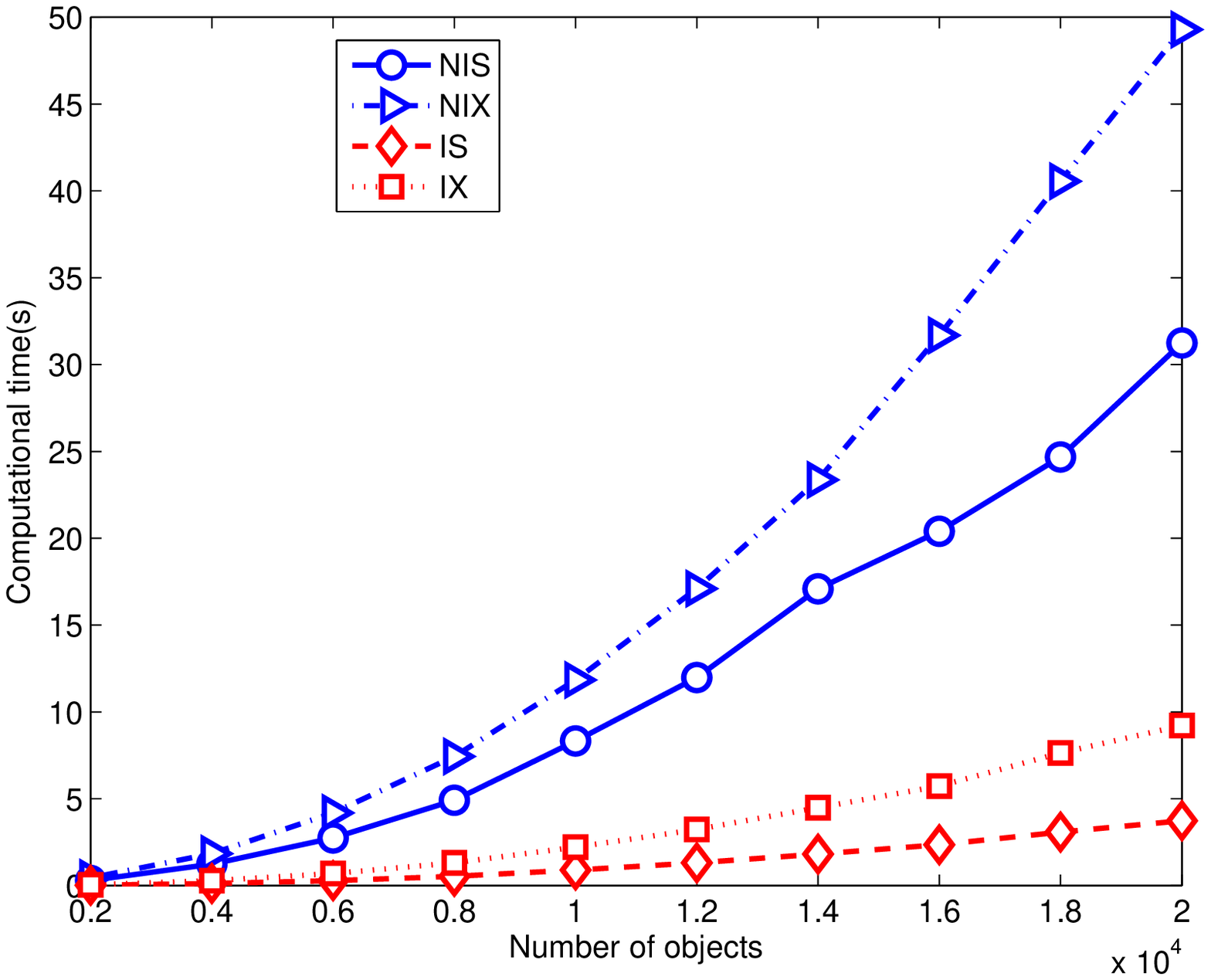}\\
\caption{Computational times using Algorithms 3.4, 3.5, 3.11, and 3.12 in
$(U_{i},\mathscr{D}^{2}_{i})$.}
\end{center}
\end{figure}

\begin{figure}[H]
\begin{center}
\includegraphics[width=8cm]{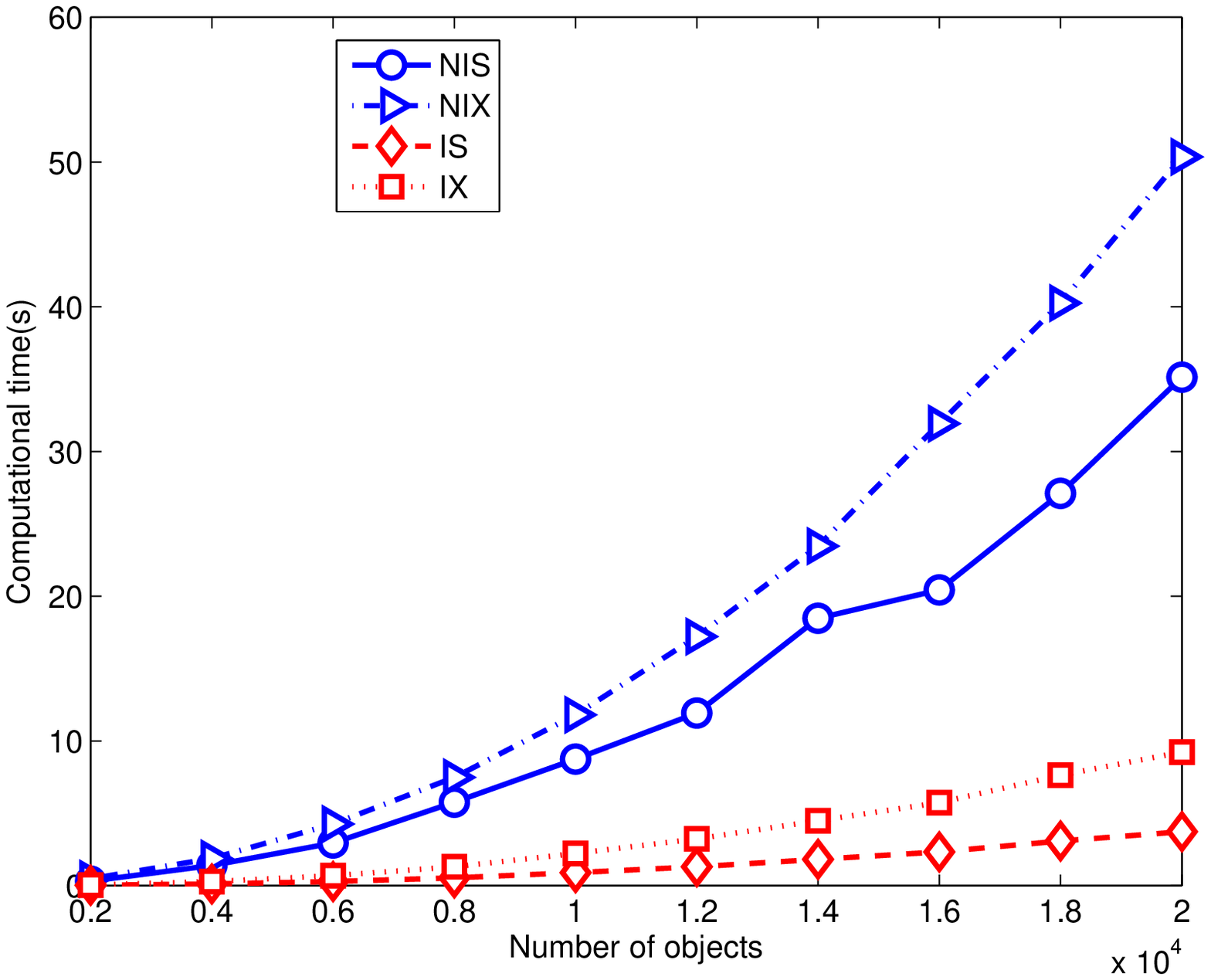}\\
\caption{Computational times using Algorithms 3.4, 3.5, 3.11, and 3.12 in
$(U_{i},\mathscr{D}^{3}_{i})$.}
\end{center}
\end{figure}

\begin{figure}[H]
\begin{center}
\includegraphics[width=8cm]{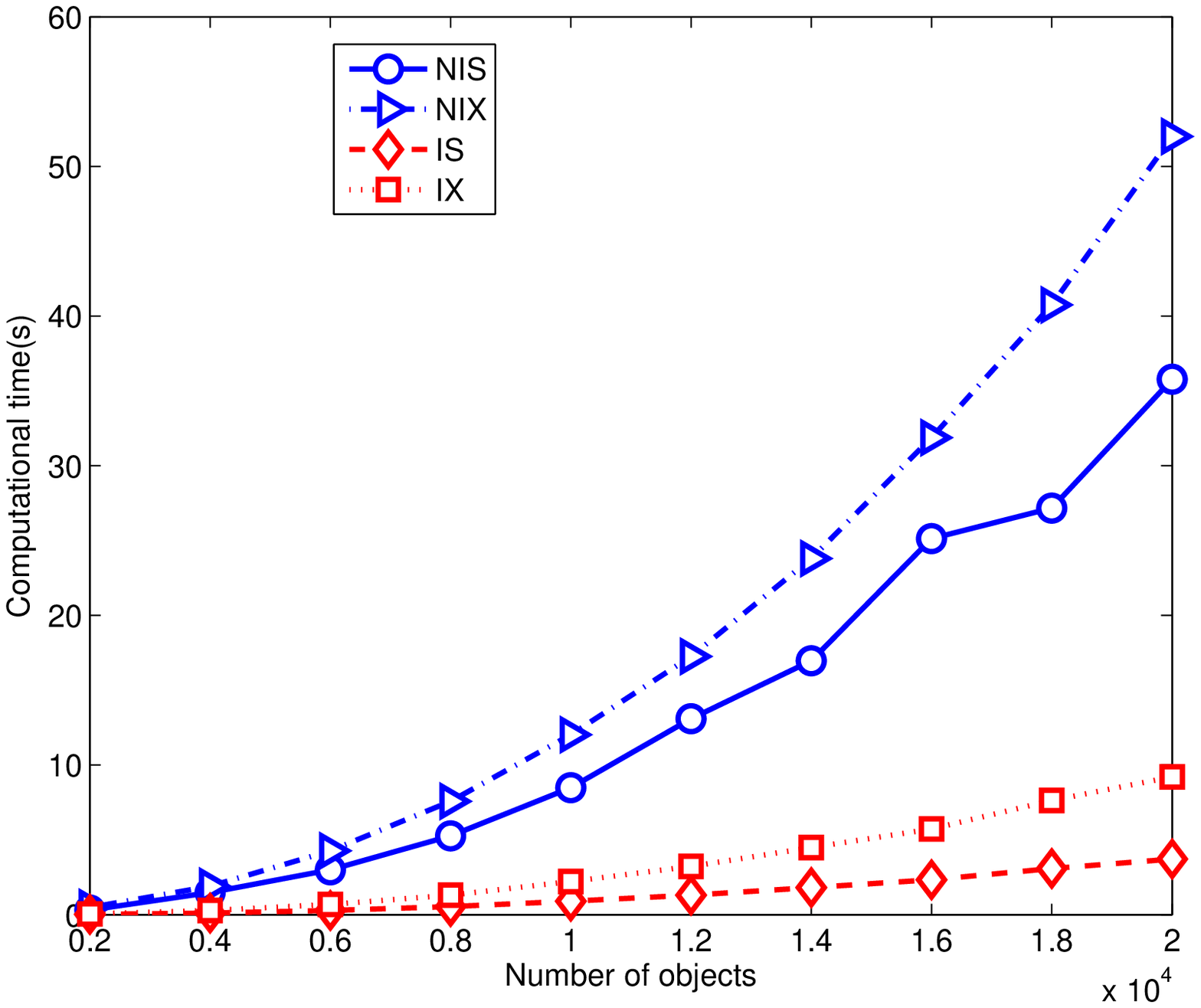}\\
\caption{Computational times using Algorithms 3.4, 3.5, 3.11, and 3.12 in
$(U_{i},\mathscr{D}^{4}_{i})$.}
\end{center}
\end{figure}

\begin{figure}[H]
\begin{center}
\includegraphics[width=8cm]{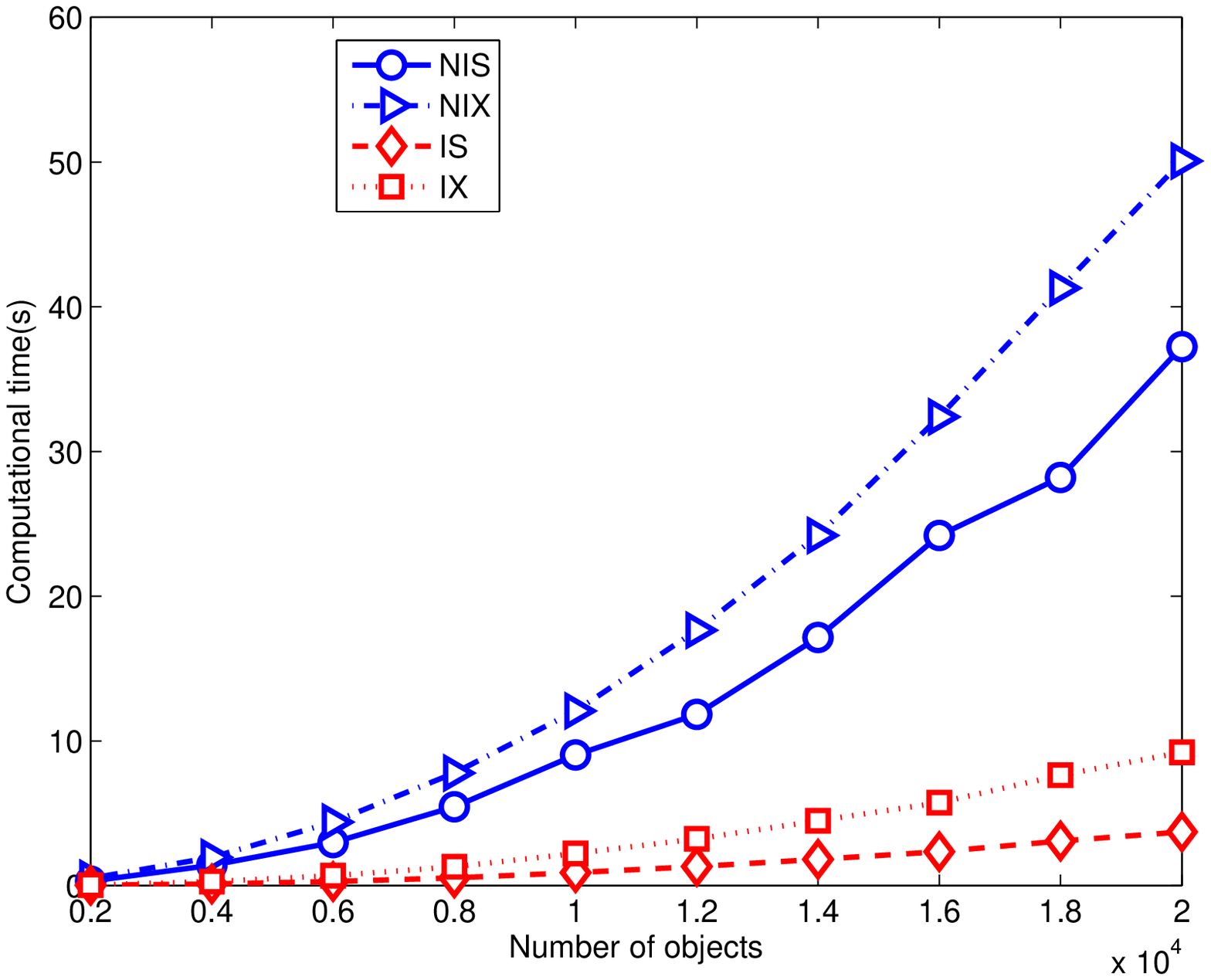}\\
\caption{Computational times using Algorithms 3.4, 3.5, 3.11, and 3.12 in
$(U_{i},\mathscr{D}^{5}_{i})$.}
\end{center}
\end{figure}

\begin{figure}[H]
\begin{center}
\includegraphics[width=8cm]{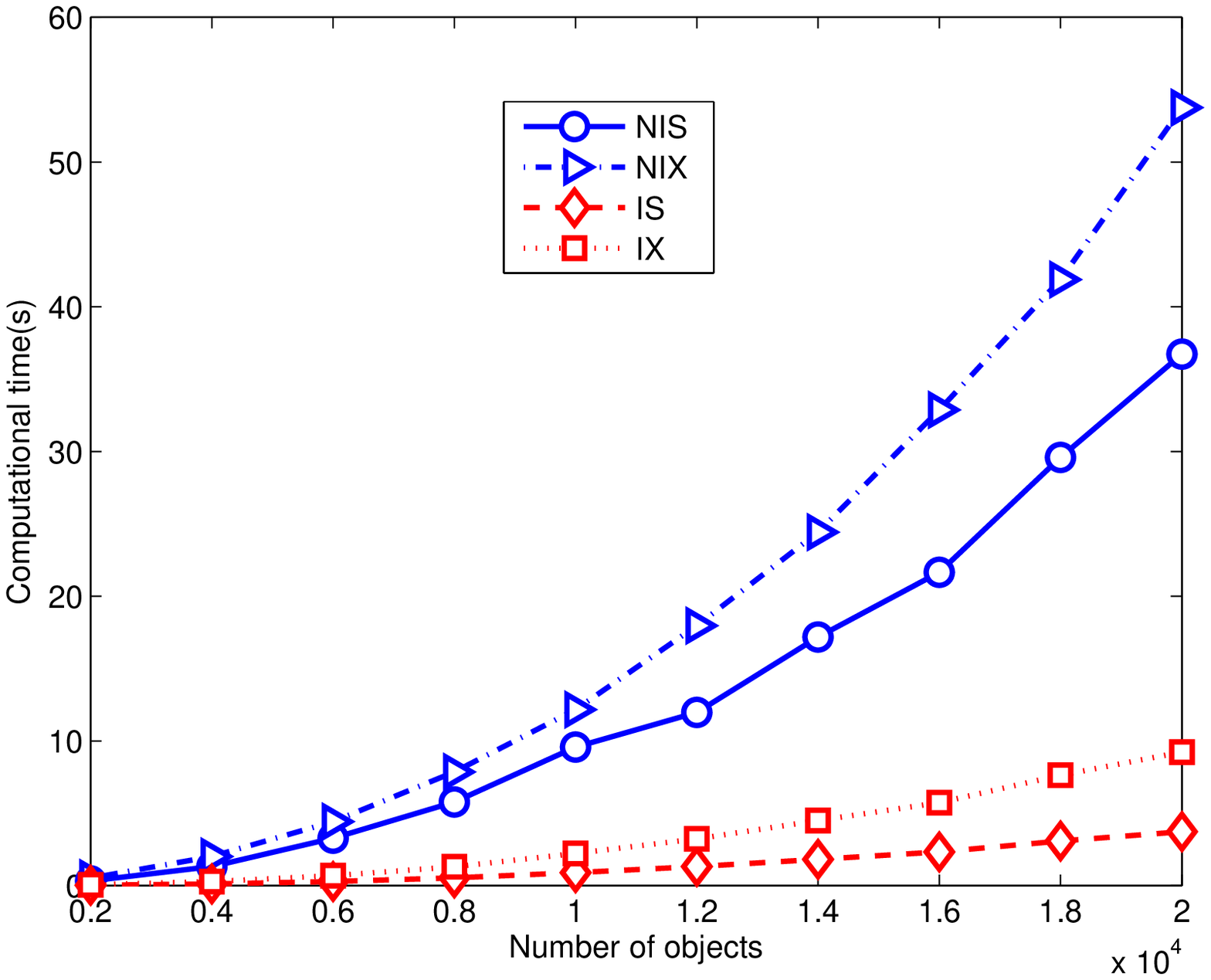}\\
\caption{Computational times using Algorithms 3.4, 3.5, 3.11, and 3.12 in
$(U_{i},\mathscr{D}^{6}_{i})$.}
\end{center}
\end{figure}

\begin{figure}[H]
\begin{center}
\includegraphics[width=8cm]{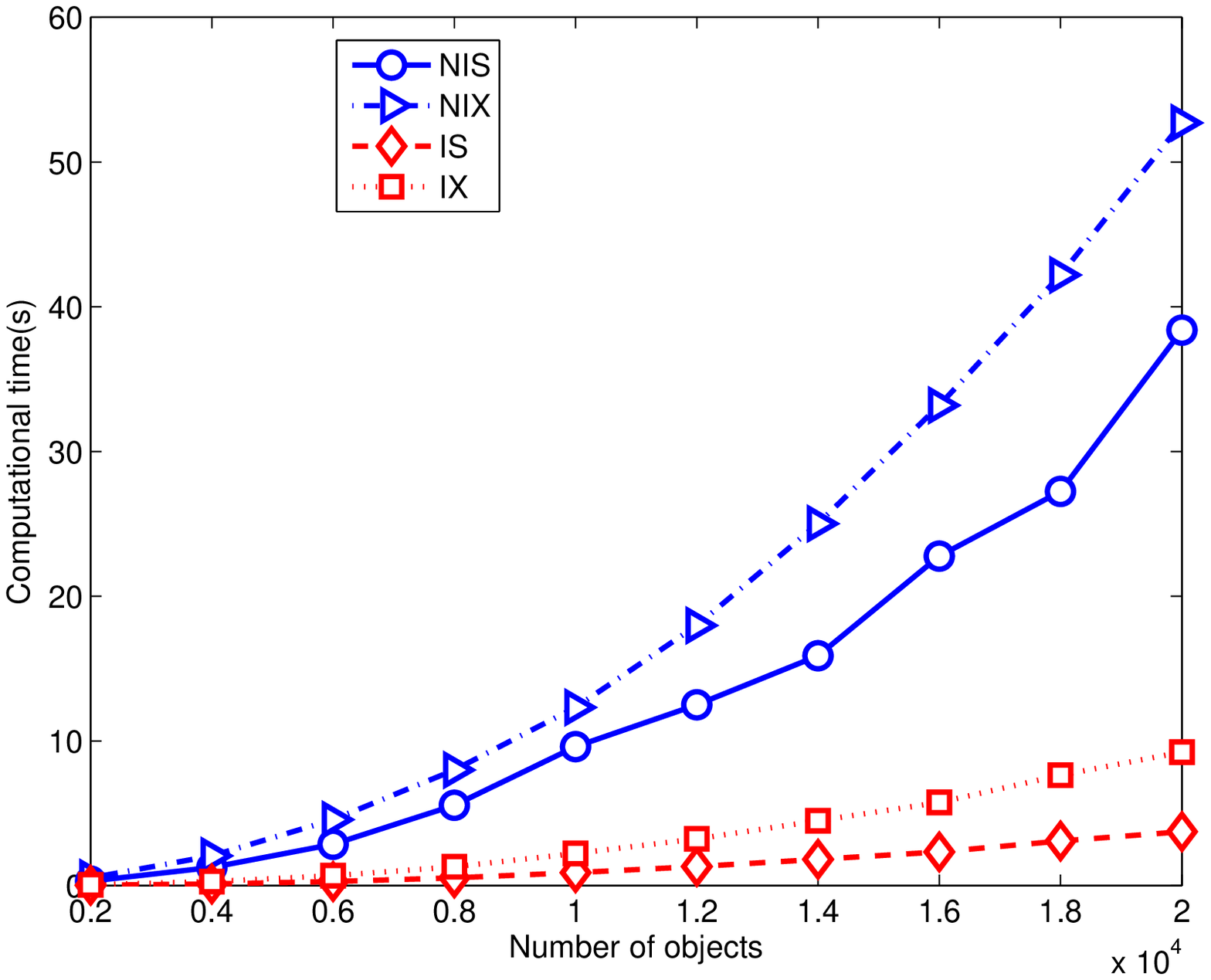}\\
\caption{Computational times using Algorithms 3.4, 3.5, 3.11, and 3.12 in
$(U_{i},\mathscr{D}^{7}_{i})$.}
\end{center}
\end{figure}

\begin{figure}[H]
\begin{center}
\includegraphics[width=8cm]{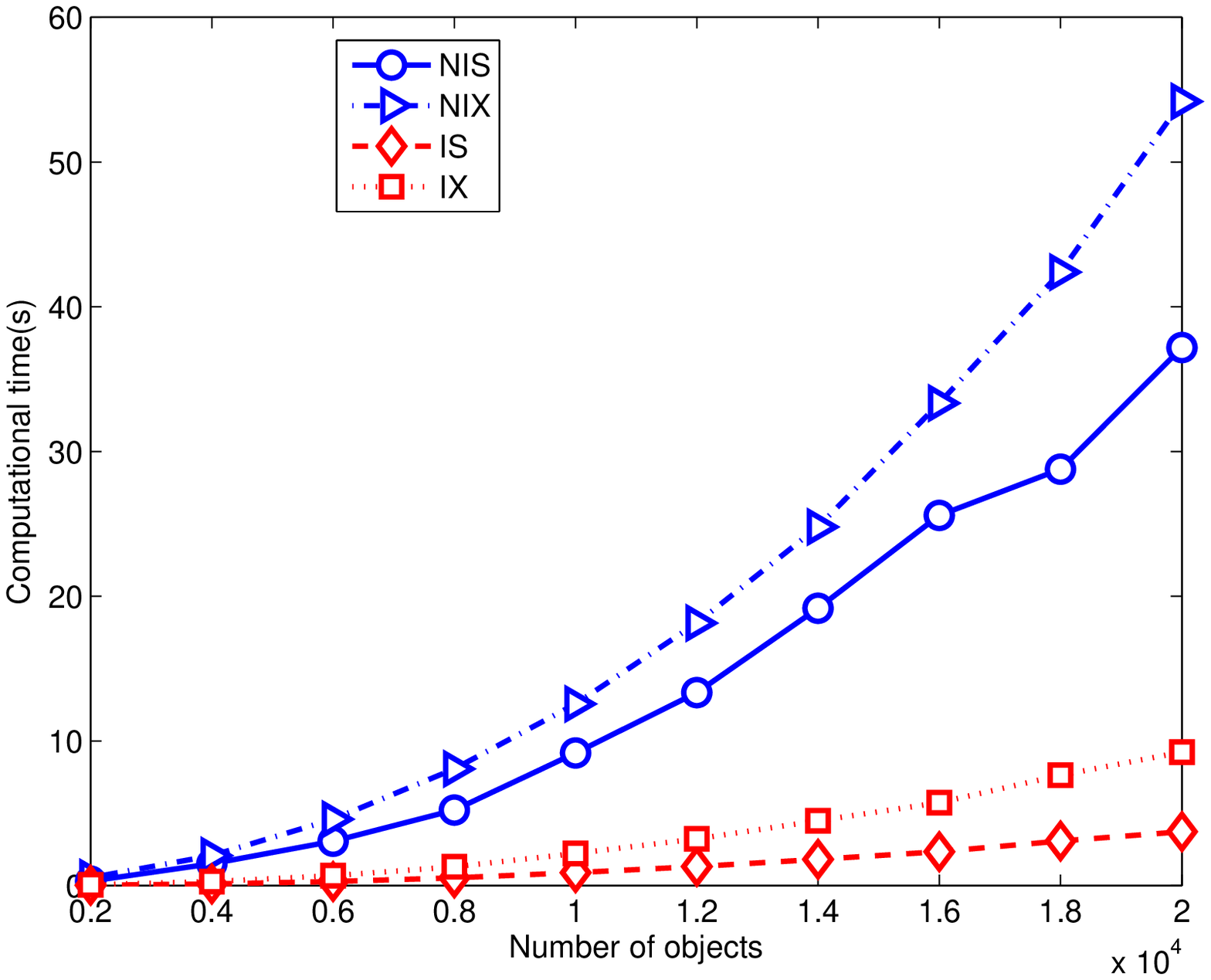}\\
\caption{Computational times  using Algorithms 3.4, 3.5, 3.11, and 3.12 in
$(U_{i},\mathscr{D}^{8}_{i})$.}
\end{center}
\end{figure}

\begin{figure}[H]
\begin{center}
\includegraphics[width=8cm]{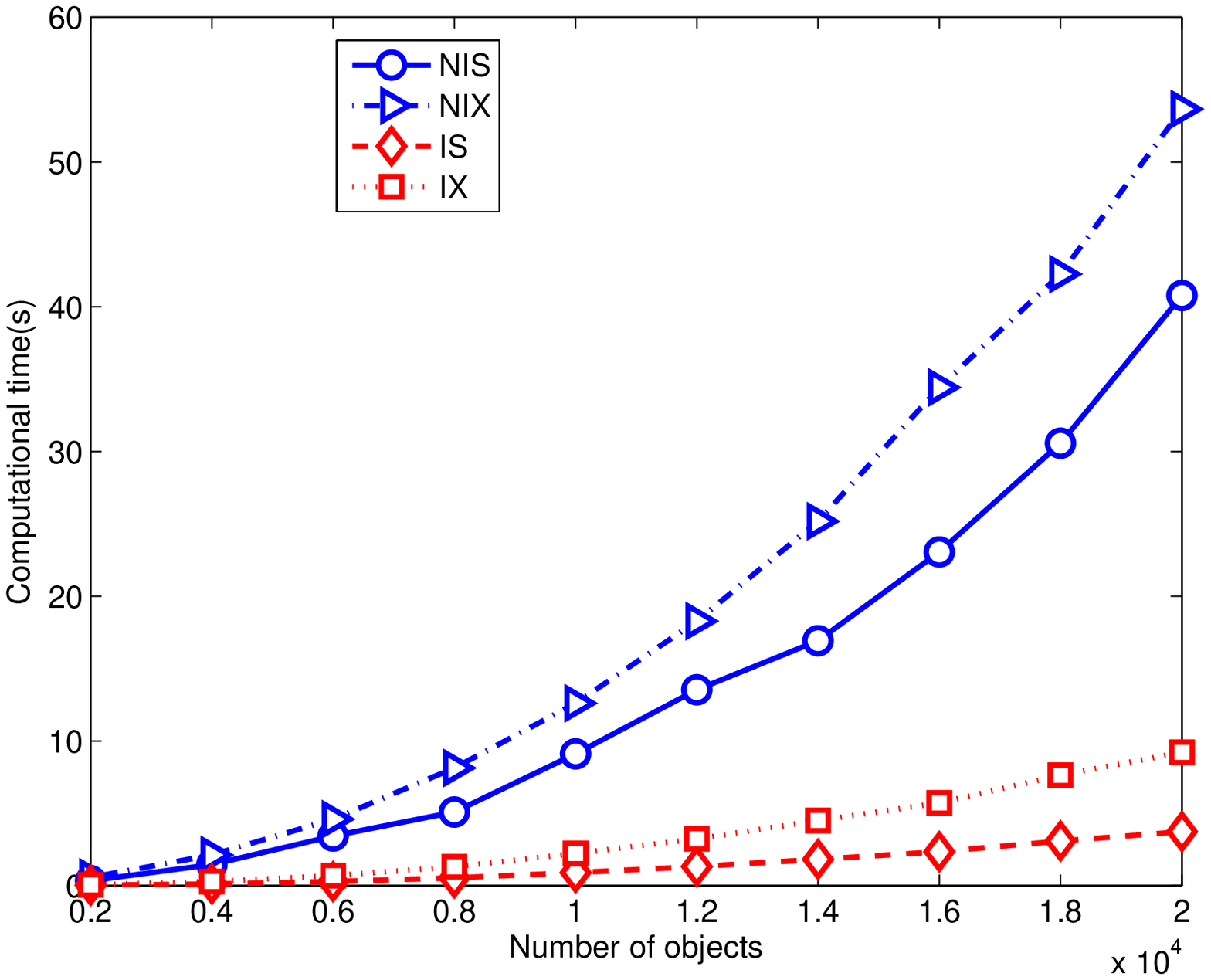}\\
\caption{Computational times using Algorithms 3.4, 3.5, 3.11, and 3.12 in
$(U_{i},\mathscr{D}^{9}_{i})$.}
\end{center}
\end{figure}

\begin{figure}[H]
\begin{center}
\includegraphics[width=8cm]{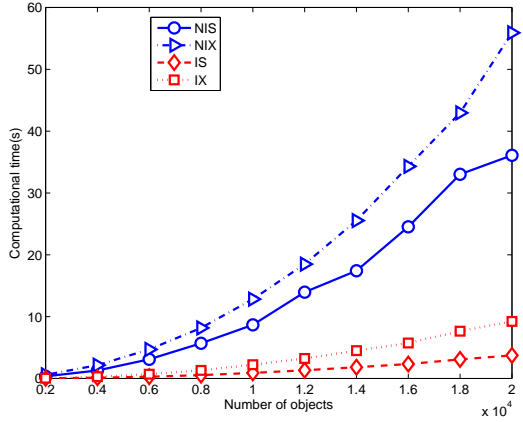}\\
\caption{Computational times using Algorithms 3.4, 3.5, 3.11, and 3.12 in
$(U_{i},\mathscr{D}^{10}_{i})$.}
\end{center}
\end{figure}

\subsection{The influence of the cardinality of covering set}

In this section, we analyze the influence of the cardinality of
covering set on time of computing the second and sixth lower and
upper approximations of sets using Algorithms 3.4, 3.5, 3.11, and 3.12
in dynamic covering information systems with the covering immigration.

In Table 2, there also exist ten sub-covering information systems
with the same cardinality of object sets. First, we compare the
times of computing the second lower and upper approximations of sets
using Algorithm 3.4 with those using Algorithm 3.5 in dynamic
covering information systems with the same cardinality of object
sets. According to the experimental results in Table 3, we see that the computing times
are almost not increasing with the increasing cardinality of covering sets
using Algorithms 3.4 and 3.5. We also find that Algorithm 3.5
executes faster than Algorithm 3.4 in dynamic covering information
systems. Second, we compare the times of computing the sixth lower
and upper approximations of sets using Algorithm 3.11 with those
using Algorithm 3.12 in dynamic covering information systems with
the same cardinality of object sets. From the results in Table 3, we
see that the computing times are increasing with the increasing
cardinality of covering sets using Algorithms 3.11. But the computing times are almost not increasing with the increasing
cardinality of covering sets using Algorithms 3.12. We also
find that Algorithms 3.12 executes faster than Algorithm 3.11 in
dynamic covering information systems. Third, to illustrate the
effectiveness of Algorithms 3.5 and 3.12, we show these results in
Figures 11-20. In each figure, $NIS,
IS, NIX,$ and $ IX$ mean Algorithms 3.4, 3.5, 3.11 and 3.12,
respectively; $i$ stands for the cardinality of covering set in $X$
Axis, while the y-coordinate stands for the time to construct the
approximations of concepts. Therefore, Algorithms 3.5 and 3.12 are
more effective to compute the second and sixth lower and upper
approximations of sets, respectively, in dynamic covering
information systems with the immigration of coverings.

According to the experimental results, we see that Algorithms 3.5 and 3.12 are more effective to
compute the second and sixth lower and upper approximations of sets than Algorithms 3.4 and 3.11, respectively, in dynamic covering information systems with the immigrations of objects and coverings.

\begin{figure}[H]
\begin{center}
\includegraphics[width=8cm]{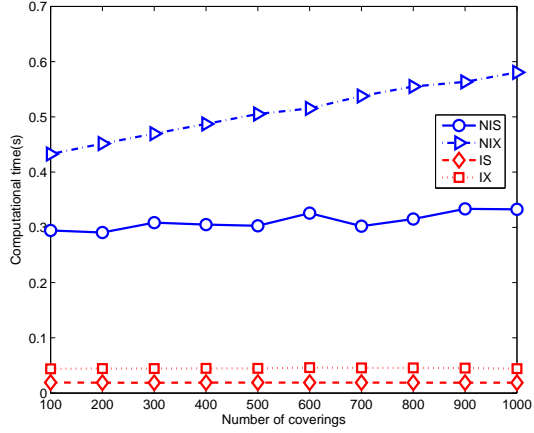}\\
\caption{Computational times using Algorithms 3.4, 3.5, 3.11, and 3.12 in
$(U_{1},\mathscr{D}^{i}_{1})$.}
\end{center}
\end{figure}

\begin{figure}[H]
\begin{center}
\includegraphics[width=8cm]{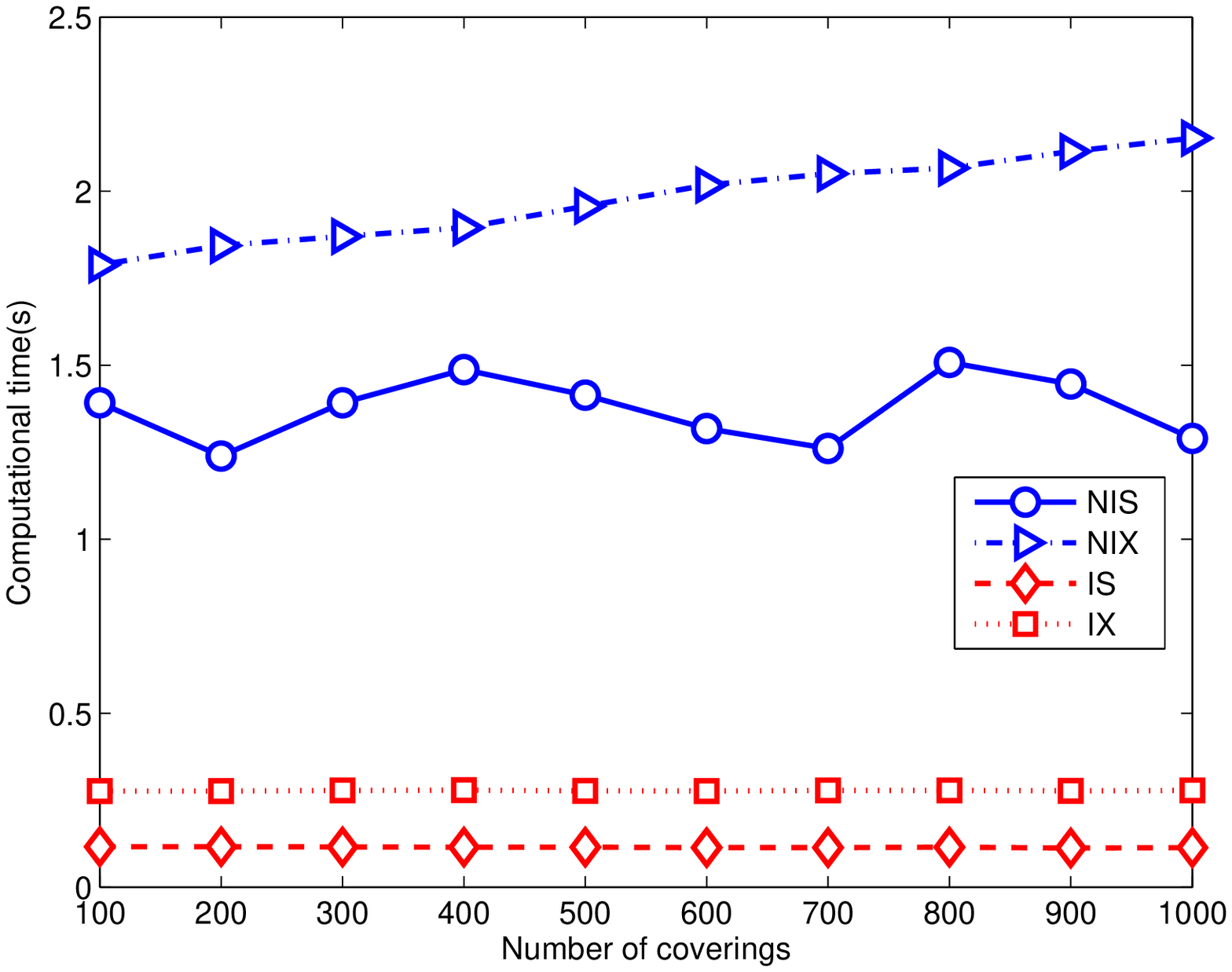}\\
\caption{Computational times using Algorithms 3.4, 3.5, 3.11, and 3.12 in
$(U_{2},\mathscr{D}^{i}_{2})$.}
\end{center}
\end{figure}

\begin{figure}[H]
\begin{center}
\includegraphics[width=8cm]{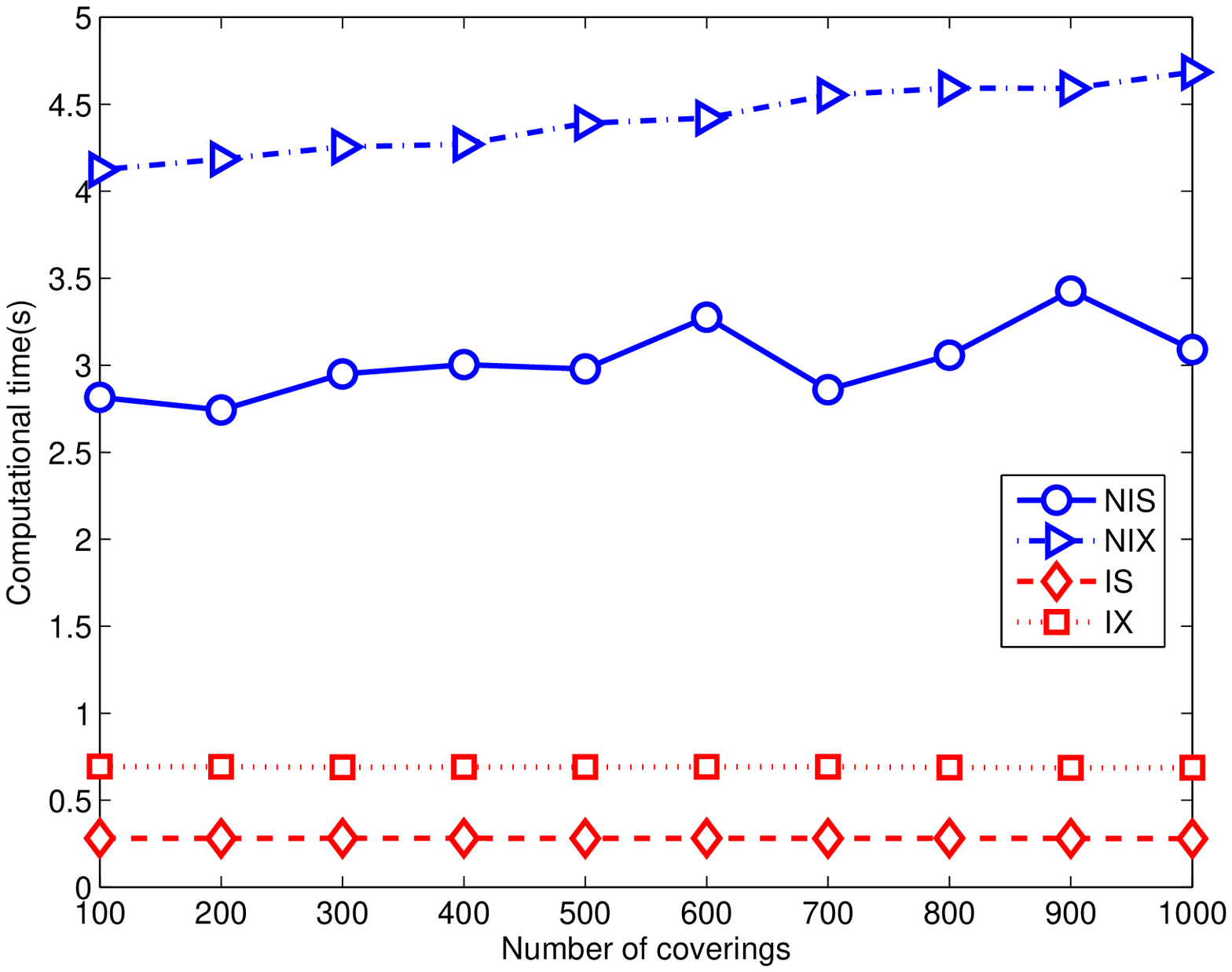}\\
\caption{Computational times using Algorithms 3.4, 3.5, 3.11, and 3.12 in
$(U_{3},\mathscr{D}^{i}_{3})$.}
\end{center}
\end{figure}

\begin{figure}[H]
\begin{center}
\includegraphics[width=8cm]{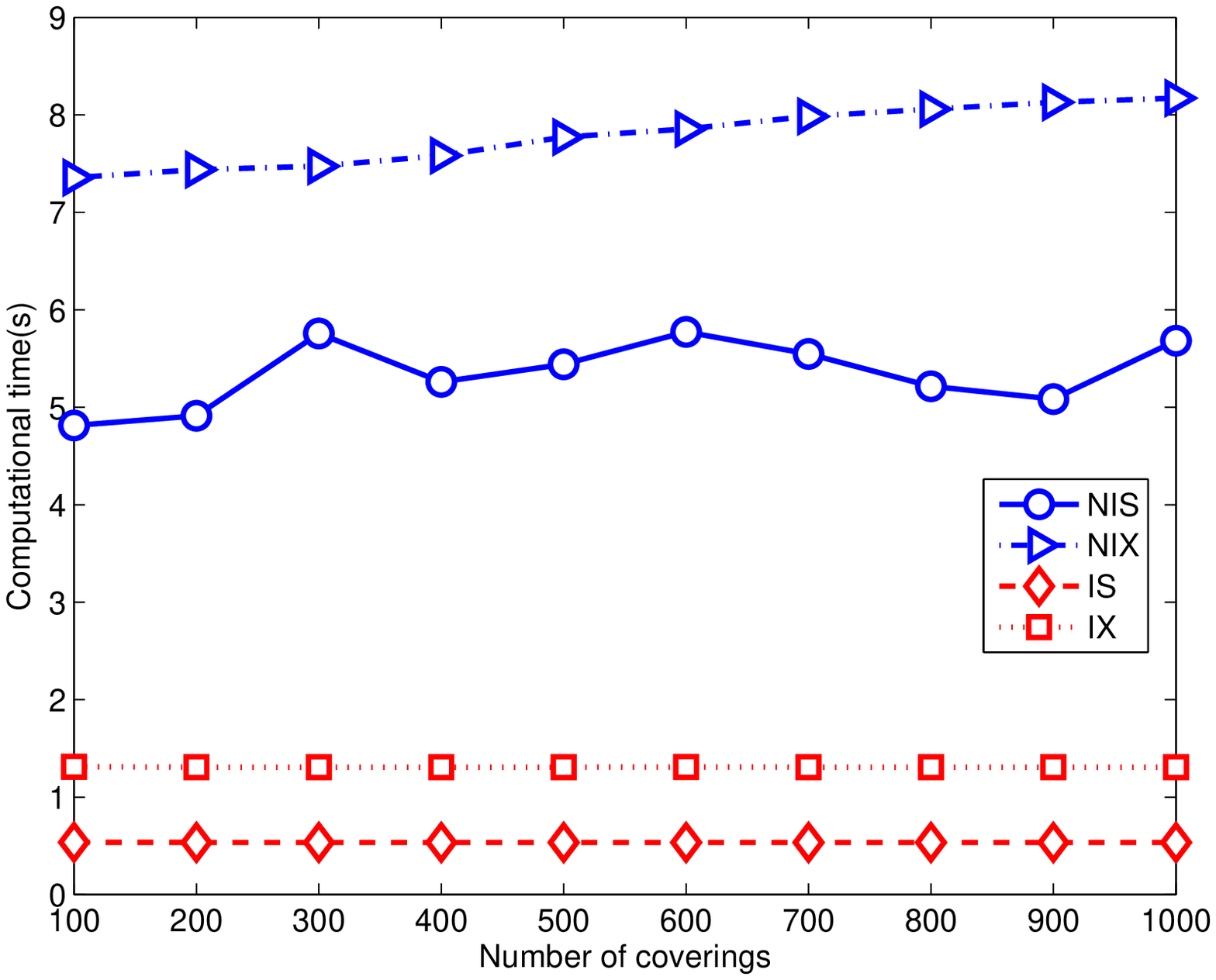}\\
\caption{Computational times using Algorithms 3.4, 3.5, 3.11, and 3.12 in
$(U_{4},\mathscr{D}^{i}_{4})$.}
\end{center}
\end{figure}

\begin{figure}[H]
\begin{center}
\includegraphics[width=8cm]{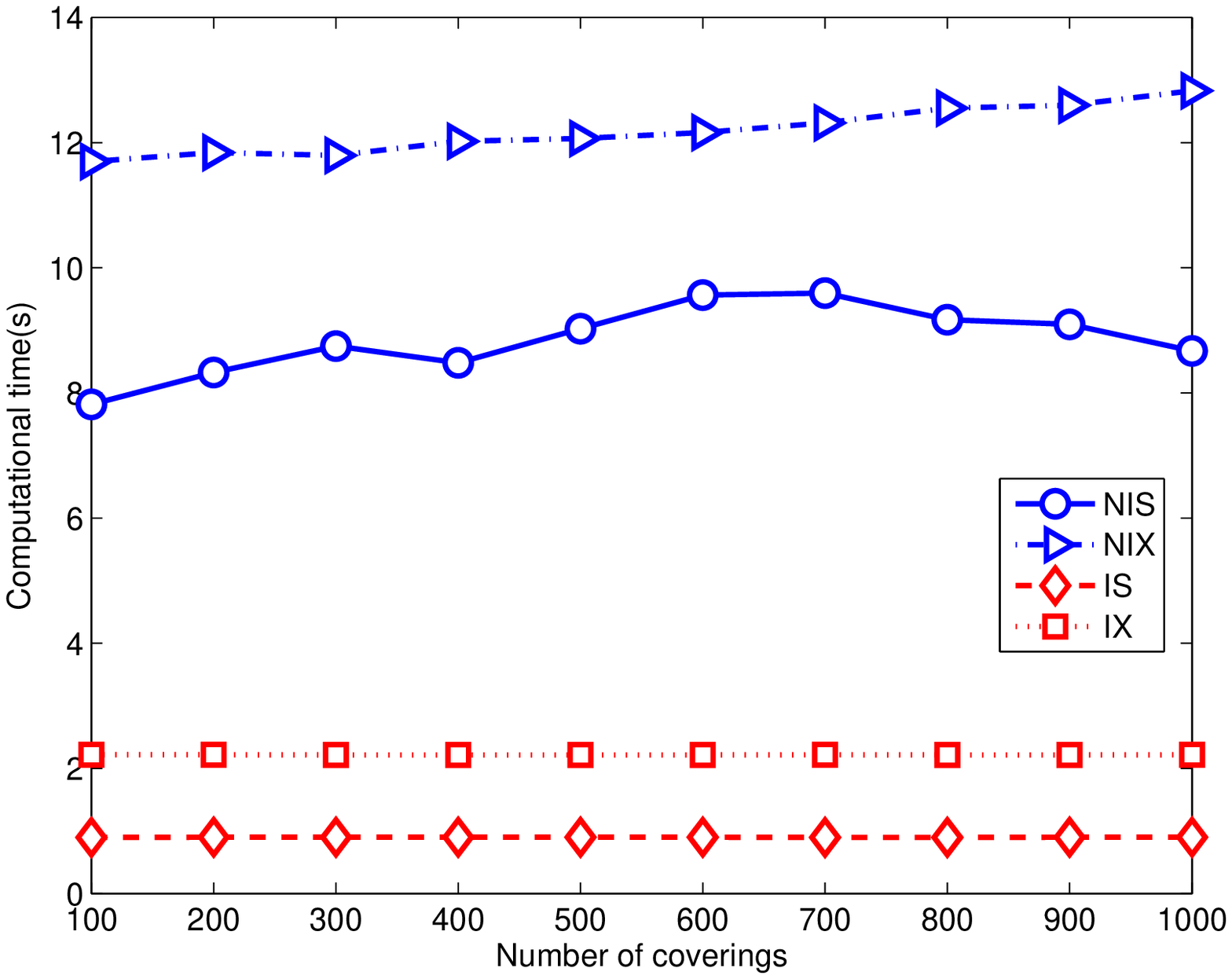}\\
\caption{Computational times using Algorithms 3.4, 3.5, 3.11, and 3.12 in
$(U_{5},\mathscr{D}^{i}_{5})$.}
\end{center}
\end{figure}

\begin{figure}[H]
\begin{center}
\includegraphics[width=8cm]{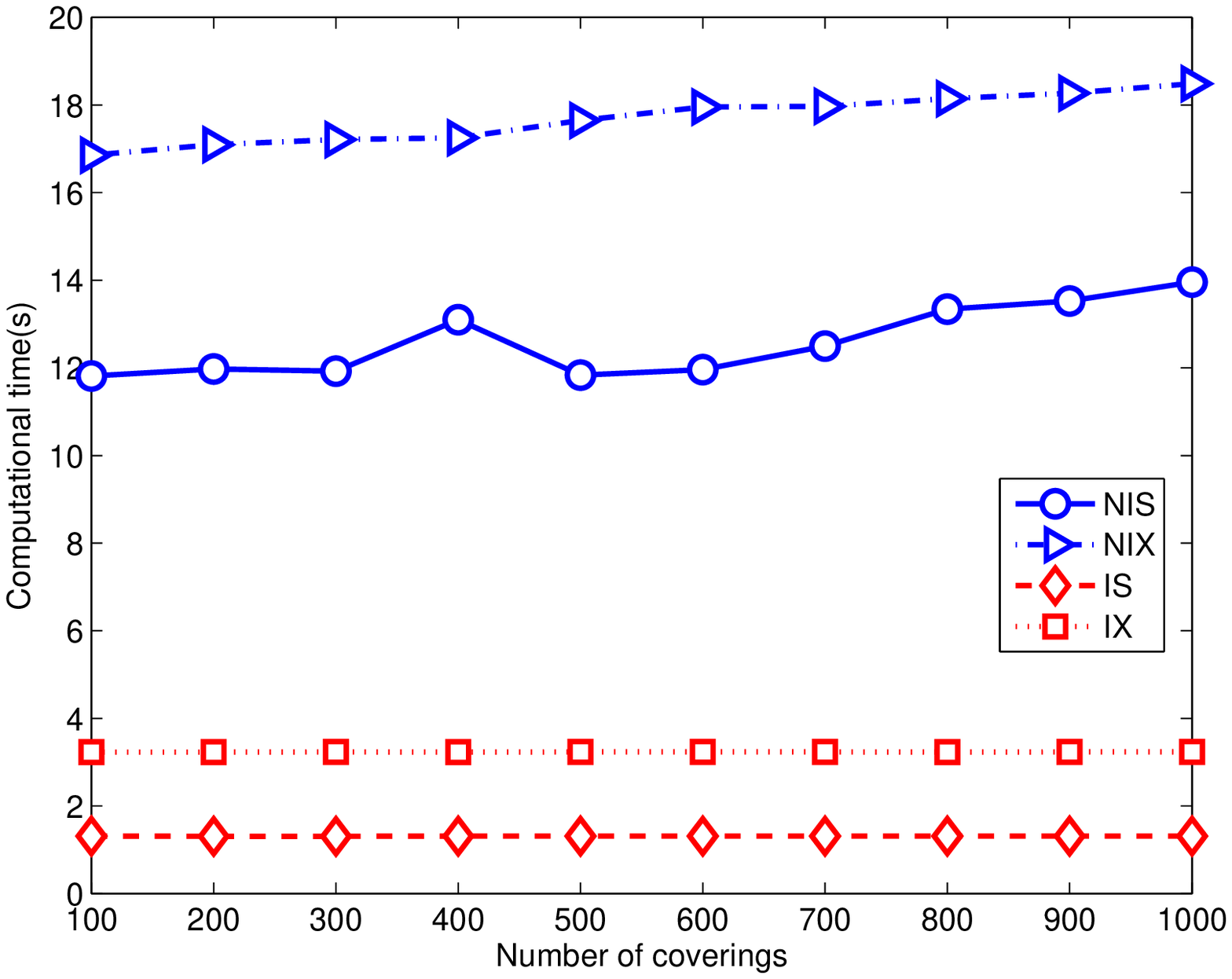}\\
\caption{Computational times using Algorithms 3.4, 3.5, 3.11, and 3.12 in
$(U_{6},\mathscr{D}^{i}_{6})$.}
\end{center}
\end{figure}

\begin{figure}[H]
\begin{center}
\includegraphics[width=8cm]{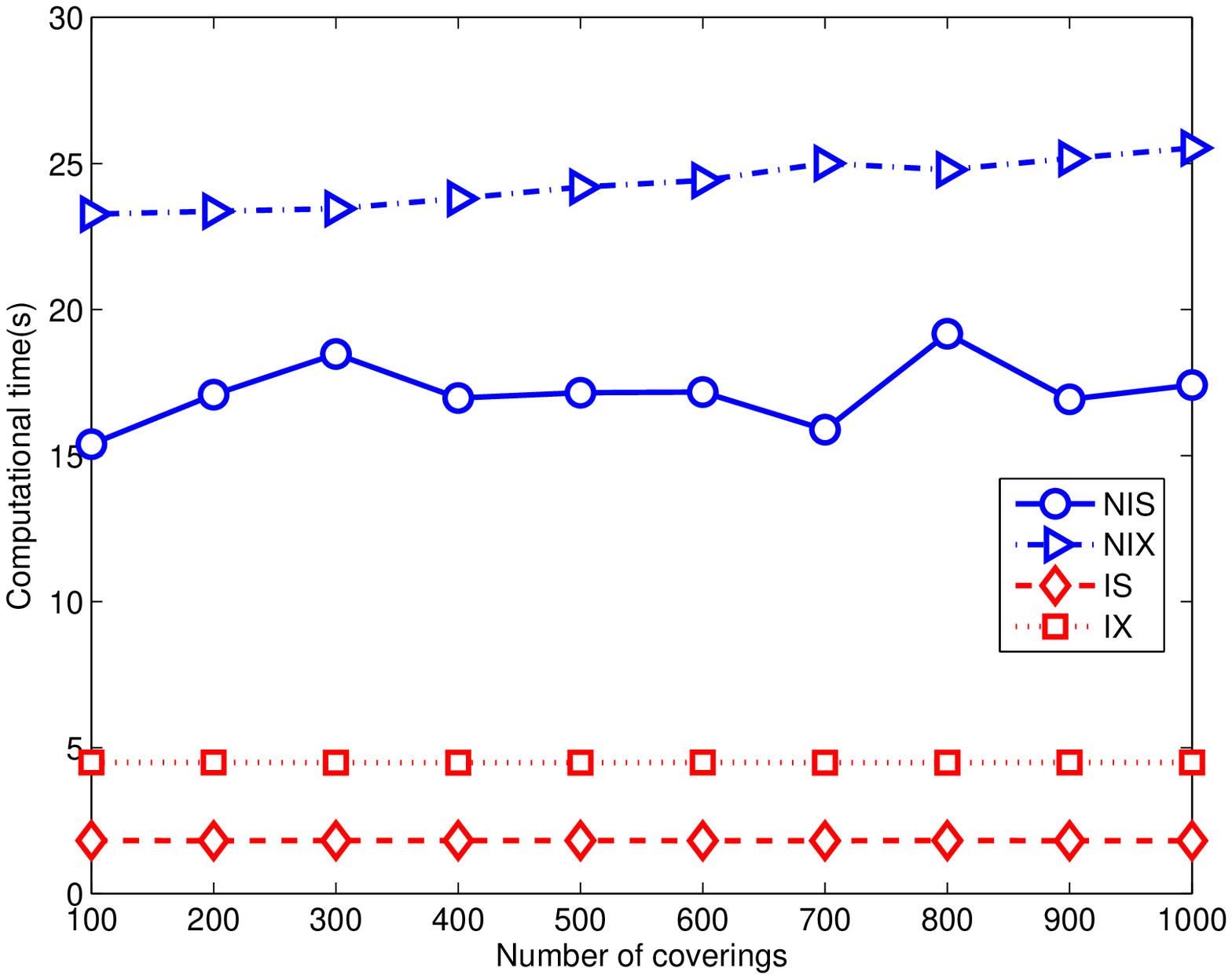}\\
\caption{Computational times using Algorithms 3.4, 3.5, 3.11, and 3.12 in
$(U_{7},\mathscr{D}^{i}_{7})$.}
\end{center}
\end{figure}

\begin{figure}[H]
\begin{center}
\includegraphics[width=8cm]{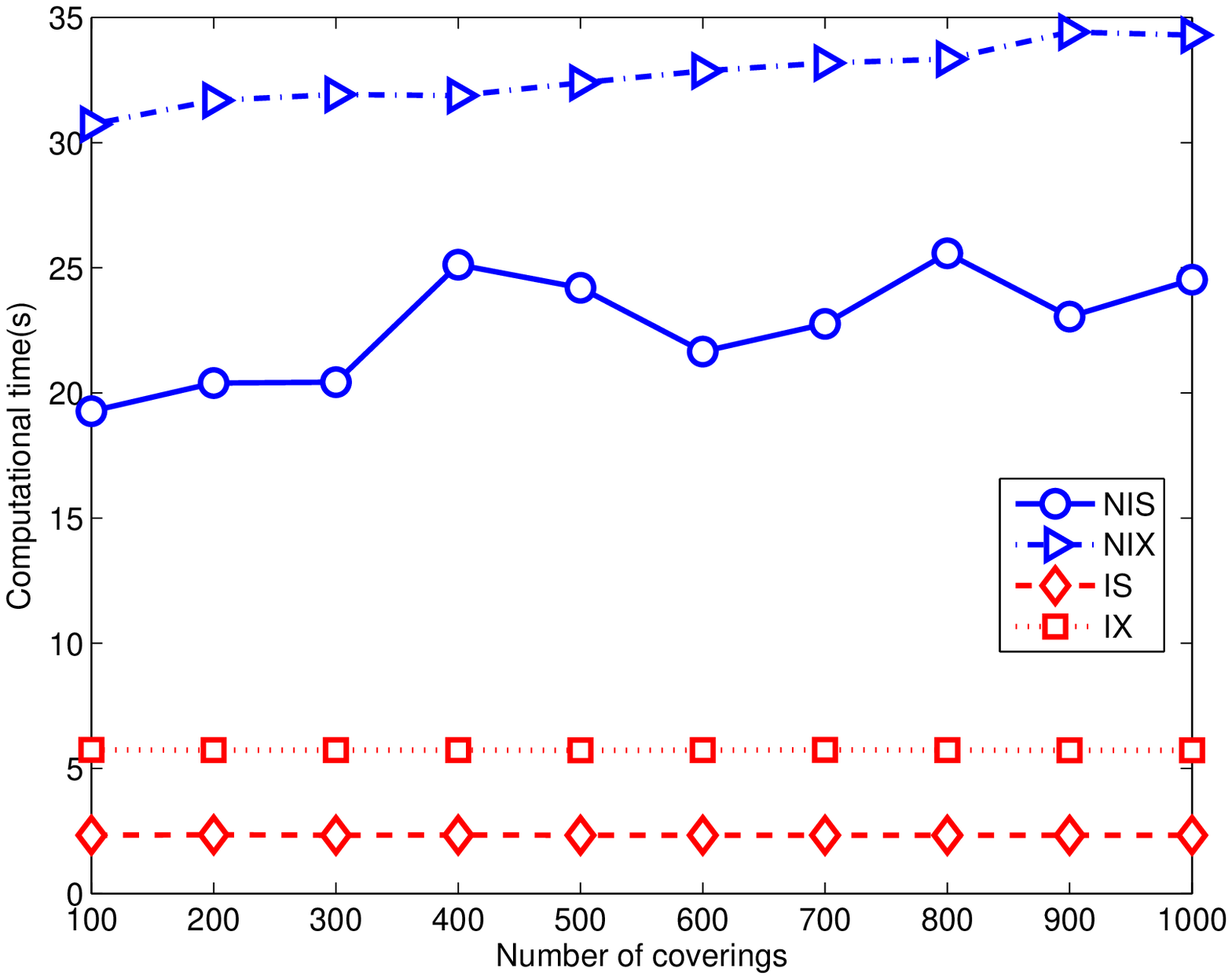}\\
\caption{Computational times using Algorithms 3.4, 3.5, 3.11, and 3.12 in
$(U_{8},\mathscr{D}^{i}_{8})$.}
\end{center}
\end{figure}

\begin{figure}[H]
\begin{center}
\includegraphics[width=8cm]{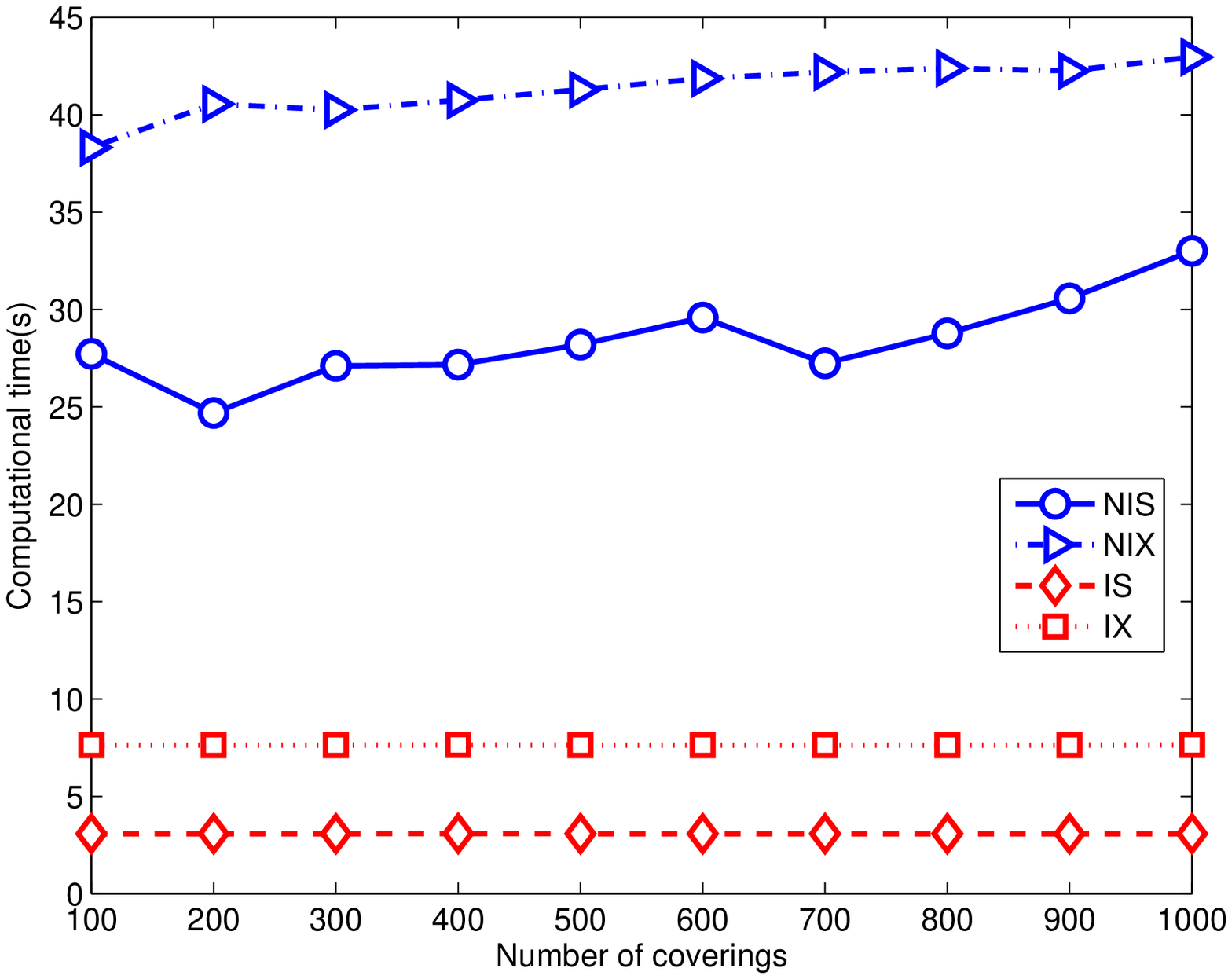}\\
\caption{Computational times using Algorithms 3.4, 3.5, 3.11, and 3.12 in
$(U_{9},\mathscr{D}^{i}_{9})$.}
\end{center}
\end{figure}

\begin{figure}[H]
\begin{center}
\includegraphics[width=8cm]{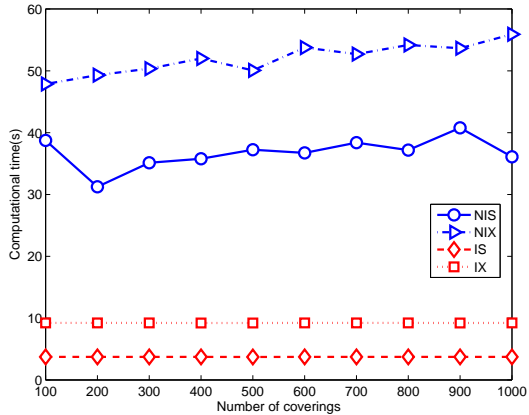}\\
\caption{Computational times using Algorithms 3.4, 3.5, 3.11, and 3.12 in
$(U_{10},\mathscr{D}^{i}_{10})$.}
\end{center}
\end{figure}

%
%To illustrate the efficiency of Algorithms 3.5 and 3.12, we also
%show the experimental results in Table 3 by Figure 3. We see that
%the time of computing the second and sixth lower and upper
%approximations of sets increase with the increasing of the
%cardinalities of object set and covering set. Especially, the time
%increases more quickly with the increasing of the cardinality of
%object set than that with the increasing of the cardinality of
%covering sets. Therefore, we see that Algorithms 3.5 and 3.12
%perform faster than Algorithms 3.4 and 3.11 when computing the
%second and sixth lower and upper approximations of sets in dynamic
%covering information systems with the immigration of coverings,
%respectively.

\textbf{Remark.} In the experiment, we can transform data sets
downloaded from the University of California at Irvine(UCI)'s
repository of machine learning databases into covering information
systems. For example, we can transform the Balance Scale Weight $\&$
Distance Database with four conditional attributes into the covering
information system $(U,\mathscr{D})$, where $|U|=625$ and
$|\mathscr{D}|=4$. Concretely, since there are five attribute values
for each conditional attribute, we can obtain a covering with five
elements for each conditional attribute. Subsequently, based on
Left-Weight, Left-Distance, Right-Weight, and Right-Distance, we have
the covering information system $(U,\mathscr{D})$, where $|U|=625$
and $|\mathscr{D}|=4$. Furthermore, we can obtain a decision
covering information system $(U,\mathscr{D}^{\ast})$ by constructing
a covering based on the decision attribute, where $|U|=625$ and
$|\mathscr{D}^{\ast}|=5$. Therefore, we can obtain covering
information systems and decision covering information systems by
transforming Irvine(UCI)'s repository of machine learning databases.
Since the purpose of the experiment is to test the effectiveness of
Algorithms 3.5 and 3.12 and the transformation process costs more
time, we generated randomly ten artificial covering information systems
$(U_{i},\mathscr{D}_{i})$ to test the designed algorithms in the
experiments.

\section{Knowledge reduction of covering decision information systems with the covering immigration}

In this section, we employ examples to illustrate how to conduct knowledge reduction of covering decision information systems with the covering immigration.

\begin{example}
Let $(U,\mathscr{D}_{C}\cup \mathscr{D}_{D} )$ be a covering decision information
system, where
$\mathscr{D}_{C}=\{\mathscr{C}_{1},\mathscr{C}_{2},\mathscr{C}_{3}\}$,
$\mathscr{C}_{1}=\{\{x_{1},x_{2},x_{3},x_{4}\},\{x_{5}\}\}$,
$\mathscr{C}_{2}=\{\{x_{1},x_{2}\},\{x_{3},x_{4},x_{5}\}\}$,
$\mathscr{C}_{3}=\{\{x_{1},x_{2},x_{5}\},\{x_{3},x_{4}\}\}$,
$\mathscr{D}_{D}=\{D_{1},D_{2}\}$, $D_{1}=\{x_{1},x_{2}\},$ and $D_{2}=\{x_{3},x_{4},x_{5}\}$. First, according to Definition 2.3, we obtain
\begin{eqnarray*}
\Gamma(\mathscr{D}_{C})=M_{\mathscr{D}_{C}}\bullet
M_{\mathscr{D}_{C}}^{T}=\left[
\begin{array}{cccccc}
1 & 1 & 1 & 1 &1 \\
1 & 1 & 1 & 1 &1 \\
1 & 1 & 1 & 1 &1 \\
1 & 1 & 1 & 1 &1 \\
1 & 1 & 1 & 1 &1 \\
\end{array}
\right].
\end{eqnarray*}

Second, by Definition 2.4, we have
\begin{eqnarray*}
\Gamma(\mathscr{D}_{C})\bullet M_{\mathscr{D}_{D}} &=&\left[
\begin{array}{cccccc}
1 & 1 & 1 & 1 &1 \\
1 & 1 & 1 & 1 &1 \\
1 & 1 & 1 & 1 &1 \\
1 & 1 & 1 & 1 &1 \\
1 & 1 & 1 & 1 &1 \\
\end{array}
\right]\bullet \left[
\begin{array}{cc}
1 & 0  \\
1 & 0  \\
0 & 1  \\
0 & 1  \\
0 & 1  \\
\end{array}
\right]= \left[
\begin{array}{cc}
1 & 1  \\
1 & 1  \\
1 & 1  \\
1 & 1  \\
1 & 1  \\
\end{array}
\right],\\
\Gamma(\mathscr{D}_{C})\odot M_{\mathscr{D}_{D}} &=&\left[
\begin{array}{cccccc}
1 & 1 & 1 & 1 &1 \\
1 & 1 & 1 & 1 &1 \\
1 & 1 & 1 & 1 &1 \\
1 & 1 & 1 & 1 &1 \\
1 & 1 & 1 & 1 &1 \\
\end{array}
\right]\odot \left[
\begin{array}{cc}
1 & 0  \\
1 & 0  \\
0 & 1  \\
0 & 1  \\
0 & 1  \\
\end{array}
\right]= \left[
\begin{array}{cc}
0 & 0  \\
0 & 0  \\
0 & 0  \\
0 & 0  \\
0 & 0  \\
\end{array}
\right].
\end{eqnarray*}

Third, according to Definition 2.3, we get
\begin{eqnarray*}
\Gamma(\{\mathscr{C}_{1},\mathscr{C}_{3}\})=\left[
\begin{array}{cccccc}
1 & 1 & 1 & 1 &1 \\
1 & 1 & 1 & 1 &1 \\
1 & 1 & 1 & 1 &0 \\
1 & 1 & 1 & 1 &0 \\
1 & 1 & 0 & 0 &1 \\
\end{array}
\right],
\Gamma(\mathscr{C}_{1})=\left[
\begin{array}{cccccc}
1 & 1 & 1 & 1 &0 \\
1 & 1 & 1 & 1 &0 \\
1 & 1 & 1 & 1 &0 \\
1 & 1 & 1 & 1 &0 \\
0 & 0 & 0 & 0 &1 \\
\end{array}
\right],
\Gamma(\mathscr{C}_{3})=\left[
\begin{array}{cccccc}
1 & 1 & 0 & 0 &1 \\
1 & 1 & 0 & 0 &1 \\
0 & 0 & 1 & 1 &0 \\
0 & 0 & 1 & 1 &0 \\
1 & 1 & 0 & 0 &1 \\
\end{array}
\right].
\end{eqnarray*}

Fourth, by Definition 2.4, we derive
\begin{eqnarray*}
\Gamma(\{\mathscr{C}_{1},\mathscr{C}_{3}\})\bullet M_{\mathscr{D}_{D}} &=&\left[
\begin{array}{cccccc}
1 & 1 & 1 & 1 &1 \\
1 & 1 & 1 & 1 &1 \\
1 & 1 & 1 & 1 &0 \\
1 & 1 & 1 & 1 &0 \\
1 & 1 & 0 & 0 &1 \\
\end{array}
\right]\bullet \left[
\begin{array}{cc}
1 & 0  \\
1 & 0  \\
0 & 1  \\
0 & 1  \\
0 & 1  \\
\end{array}
\right]= \left[
\begin{array}{cc}
1 & 1  \\
1 & 1  \\
1 & 1  \\
1 & 1  \\
1 & 1  \\
\end{array}
\right],\\
\Gamma(\{\mathscr{C}_{1},\mathscr{C}_{3}\})\odot M_{\mathscr{D}_{D}} &=&\left[
\begin{array}{cccccc}
1 & 1 & 1 & 1 &1 \\
1 & 1 & 1 & 1 &1 \\
1 & 1 & 1 & 1 &0 \\
1 & 1 & 1 & 1 &0 \\
1 & 1 & 0 & 0 &1 \\
\end{array}
\right]\odot \left[
\begin{array}{cc}
1 & 0  \\
1 & 0  \\
0 & 1  \\
0 & 1  \\
0 & 1  \\
\end{array}
\right]= \left[
\begin{array}{cc}
0 & 0  \\
0 & 0  \\
0 & 0  \\
0 & 0  \\
0 & 0  \\
\end{array}
\right],\\
\Gamma(\mathscr{C}_{1})\bullet M_{\mathscr{D}_{D}} &=&\left[
\begin{array}{cccccc}
1 & 1 & 1 & 1 &0 \\
1 & 1 & 1 & 1 &0 \\
1 & 1 & 1 & 1 &0 \\
1 & 1 & 1 & 1 &0 \\
0 & 0 & 0 & 0 &1 \\
\end{array}
\right]\bullet \left[
\begin{array}{cc}
1 & 0  \\
1 & 0  \\
0 & 1  \\
0 & 1  \\
0 & 1  \\
\end{array}
\right]= \left[
\begin{array}{cc}
1 & 1  \\
1 & 1  \\
1 & 1  \\
1 & 1  \\
0 & 1  \\
\end{array}
\right],\\
\Gamma(\mathscr{C}_{1})\odot M_{\mathscr{D}_{D}} &=&\left[
\begin{array}{cccccc}
1 & 1 & 1 & 1 &0 \\
1 & 1 & 1 & 1 &0 \\
1 & 1 & 1 & 1 &0 \\
1 & 1 & 1 & 1 &0 \\
0 & 0 & 0 & 0 &1 \\
\end{array}
\right]\odot \left[
\begin{array}{cc}
1 & 0  \\
1 & 0  \\
0 & 1  \\
0 & 1  \\
0 & 1  \\
\end{array}
\right]= \left[
\begin{array}{cc}
0 & 0  \\
0 & 0  \\
0 & 0  \\
0 & 0  \\
0 & 1  \\
\end{array}
\right],\\
\Gamma(\mathscr{C}_{3})\bullet M_{\mathscr{D}_{D}} &=&\left[
\begin{array}{cccccc}
1 & 1 & 0 & 0 &1 \\
1 & 1 & 0 & 0 &1 \\
0 & 0 & 1 & 1 &0 \\
0 & 0 & 1 & 1 &0 \\
1 & 1 & 0 & 0 &1 \\
\end{array}
\right]\bullet \left[
\begin{array}{cc}
1 & 0  \\
1 & 0  \\
0 & 1  \\
0 & 1  \\
0 & 1  \\
\end{array}
\right]= \left[
\begin{array}{cc}
1 & 1  \\
1 & 1  \\
0 & 1  \\
0 & 1  \\
1 & 1  \\
\end{array}
\right],\\
\Gamma(\mathscr{C}_{3})\odot M_{\mathscr{D}_{D}} &=&\left[
\begin{array}{cccccc}
1 & 1 & 0 & 0 &1 \\
1 & 1 & 0 & 0 &1 \\
0 & 0 & 1 & 1 &0 \\
0 & 0 & 1 & 1 &0 \\
1 & 1 & 0 & 0 &1 \\
\end{array}
\right]\odot \left[
\begin{array}{cc}
1 & 0  \\
1 & 0  \\
0 & 1  \\
0 & 1  \\
0 & 1  \\
\end{array}
\right]= \left[
\begin{array}{cc}
0 & 0  \\
0 & 0  \\
0 & 1  \\
0 & 1  \\
0 & 0  \\
\end{array}
\right].
\end{eqnarray*}

Therefore, according to Definition 2.5, $\{\mathscr{C}_{1},\mathscr{C}_{3}\}$ is a type-1 reduct of $(U,\mathscr{D}_{C}\cup \mathscr{D}_{D} )$.
\end{example}

In Example 5.1, we must compute $
\Gamma(\{\mathscr{C}_{1},\mathscr{C}_{3}\})\bullet M_{\mathscr{D}_{D}},
\Gamma(\{\mathscr{C}_{1},\mathscr{C}_{3}\})\odot M_{\mathscr{D}_{D}},
\Gamma(\mathscr{C}_{1})\bullet M_{\mathscr{D}_{D}},
\Gamma(\mathscr{C}_{1})\odot M_{\mathscr{D}_{D}},
\Gamma(\mathscr{C}_{3})\bullet M_{\mathscr{D}_{D}}$, and $
\Gamma(\mathscr{C}_{3})\odot M_{\mathscr{D}_{D}}
$ for constructing the type-1 reducts of covering decision information system $(U,\mathscr{D}_{C}\cup \mathscr{D}_{D} )$.

In what follows, we employ an example to illustrate how to compute the type-1 reducts of dynamic covering decision information systems with the immigration of coverings.

\begin{example} (Continued from Example 5.1)
Let $(U,\mathscr{D}^{+}_{C}\cup \mathscr{D}_{D} )$ be a dynamic covering decision information
system of  $(U,\mathscr{D}_{C}\cup \mathscr{D}_{D} )$, where
$\mathscr{D}^{+}_{C}=\{\mathscr{C}_{1},\mathscr{C}_{2},\mathscr{C}_{3},\mathscr{C}_{4}\}$,
$\mathscr{C}_{1}=\{\{x_{1},x_{2},x_{3},x_{4}\},\{x_{5}\}\}$,
$\mathscr{C}_{2}=\{\{x_{1},x_{2}\},\{x_{3},x_{4},\\x_{5}\}\}$,
$\mathscr{C}_{3}=\{\{x_{1},x_{2},x_{5}\},\{x_{3},x_{4}\}\}$,
$\mathscr{C}_{4}=\{\{x_{1},x_{2}\},\{x_{3},x_{4}\},\{x_{5}\}\}$,
$\mathscr{D}_{D}=\{D_{1},D_{2}\}$, $D_{1}=\{x_{1},x_{2}\},$ and $D_{2}=\{x_{3},x_{4},x_{5}\}$. First, by Theorem 3.3 and Example 5.1, we obtain
\begin{eqnarray*}
\Gamma(\mathscr{D}^{+}_{C})=\Gamma(\mathscr{D}_{C})\bigvee\Gamma(\mathscr{C}_{4})=\left[
\begin{array}{cccccc}
1 & 1 & 1 & 1 &1 \\
1 & 1 & 1 & 1 &1 \\
1 & 1 & 1 & 1 &1 \\
1 & 1 & 1 & 1 &1 \\
1 & 1 & 1 & 1 &1 \\
\end{array}
\right].
\end{eqnarray*}

Second, by Definition 2.4, we have
\begin{eqnarray*}
\Gamma(\mathscr{D}^{+}_{C})\bullet M_{\mathscr{D}_{D}} &=&\left[
\begin{array}{cccccc}
1 & 1 & 1 & 1 &1 \\
1 & 1 & 1 & 1 &1 \\
1 & 1 & 1 & 1 &1 \\
1 & 1 & 1 & 1 &1 \\
1 & 1 & 1 & 1 &1 \\
\end{array}
\right]\bullet \left[
\begin{array}{cc}
1 & 0  \\
1 & 0  \\
0 & 1  \\
0 & 1  \\
0 & 1  \\
\end{array}
\right]= \left[
\begin{array}{cc}
1 & 1  \\
1 & 1  \\
1 & 1  \\
1 & 1  \\
1 & 1  \\
\end{array}
\right],\\
\Gamma(\mathscr{D}^{+}_{C})\odot M_{\mathscr{D}_{D}} &=&\left[
\begin{array}{cccccc}
1 & 1 & 1 & 1 &1 \\
1 & 1 & 1 & 1 &1 \\
1 & 1 & 1 & 1 &1 \\
1 & 1 & 1 & 1 &1 \\
1 & 1 & 1 & 1 &1 \\
\end{array}
\right]\odot \left[
\begin{array}{cc}
1 & 0  \\
1 & 0  \\
0 & 1  \\
0 & 1  \\
0 & 1  \\
\end{array}
\right]= \left[
\begin{array}{cc}
0 & 0  \\
0 & 0  \\
0 & 0  \\
0 & 0  \\
0 & 0  \\
\end{array}
\right].
\end{eqnarray*}

Third, by Example 5.1, we get
\begin{eqnarray*}
\Gamma(\mathscr{D}_{C})\bullet M_{\mathscr{D}_{D}} &=&\left[
\begin{array}{cc}
1 & 1  \\
1 & 1  \\
1 & 1  \\
1 & 1  \\
1 & 1  \\
\end{array}
\right],
\Gamma(\mathscr{D}_{C})\odot M_{\mathscr{D}_{D}} =\left[
\begin{array}{cc}
0 & 0  \\
0 & 0  \\
0 & 0  \\
0 & 0  \\
0 & 0  \\
\end{array}
\right],
\Gamma(\{\mathscr{C}_{1},\mathscr{C}_{3}\})\bullet M_{\mathscr{D}_{D}} =\left[
\begin{array}{cc}
1 & 1  \\
1 & 1  \\
1 & 1  \\
1 & 1  \\
1 & 1  \\
\end{array}
\right],\\
\Gamma(\{\mathscr{C}_{1},\mathscr{C}_{3}\})\odot M_{\mathscr{D}_{D}} &=&\left[
\begin{array}{cc}
0 & 0  \\
0 & 0  \\
0 & 0  \\
0 & 0  \\
0 & 0  \\
\end{array}
\right],
\Gamma(\mathscr{C}_{1})\bullet M_{\mathscr{D}_{D}} =\left[
\begin{array}{cc}
1 & 1  \\
1 & 1  \\
1 & 1  \\
1 & 1  \\
0 & 1  \\
\end{array}
\right],
\Gamma(\mathscr{C}_{1})\odot M_{\mathscr{D}_{D}} =\left[
\begin{array}{cc}
0 & 0  \\
0 & 0  \\
0 & 0  \\
0 & 0  \\
0 & 1  \\
\end{array}
\right],
\\
\Gamma(\mathscr{C}_{3})\bullet M_{\mathscr{D}_{D}} &=&\left[
\begin{array}{cc}
1 & 1  \\
1 & 1  \\
0 & 1  \\
0 & 1  \\
1 & 1  \\
\end{array}
\right],
\Gamma(\mathscr{C}_{3})\odot M_{\mathscr{D}_{D}} =\left[
\begin{array}{cc}
0 & 0  \\
0 & 0  \\
0 & 1  \\
0 & 1  \\
0 & 0  \\
\end{array}
\right].
\end{eqnarray*}

Therefore, according to Definition 2.5, $\{\mathscr{C}_{1},\mathscr{C}_{3}\}$ is a type-1 reduct of $(U,\mathscr{D}^{+}_{C}\cup \mathscr{D}_{D} )$.
\end{example}

In Example 5.2, we must compute $ \Gamma(\mathscr{D}^{+}_{C})\bullet M_{\mathscr{D}_{D}},
\Gamma(\mathscr{D}^{+}_{C})\odot M_{\mathscr{D}_{D}},
\Gamma(\mathscr{D}_{C})\bullet M_{\mathscr{D}_{D}},
\Gamma(\mathscr{D}_{C})\odot M_{\mathscr{D}_{D}},
\Gamma(\{\mathscr{C}_{1},\mathscr{C}_{3}\})\bullet M_{\mathscr{D}_{D}},
\Gamma(\{\mathscr{C}_{1},\mathscr{C}_{3}\})\odot M_{\mathscr{D}_{D}},
\Gamma(\mathscr{C}_{1})\bullet M_{\mathscr{D}_{D}},
\Gamma(\mathscr{C}_{1})\odot M_{\mathscr{D}_{D}},
\Gamma(\mathscr{C}_{3})\bullet M_{\mathscr{D}_{D}}$, and $
\Gamma(\mathscr{C}_{3})\odot M_{\mathscr{D}_{D}}
$  if we construct the type-1 reducts of dynamic covering decision information system $(U,\mathscr{D}^{+}_{C}\cup \mathscr{D}_{D} )$ with the non-incremental approach. But we only need to compute $ \Gamma(\mathscr{D}^{+}_{C})\bullet M_{\mathscr{D}_{D}}$ and $
\Gamma(\mathscr{D}^{+}_{C})\odot M_{\mathscr{D}_{D}}$ for constructing the type-1 reducts of $(U,\mathscr{D}^{+}_{C}\cup \mathscr{D}_{D} )$ with the incremental approach. Therefore, the designed algorithm is effective to conduct knowledge reduction of dynamic covering decision information systems with the immigration of coverings.

\section{Conclusions}

In this paper, we have updated the type-1 and type-2 characteristic
matrices and designed effective algorithms for computing the second
and sixth lower and upper approximations of sets in dynamic covering
information systems with variations of coverings. We have employed
examples to illustrate how to calculate the second and sixth lower
and upper approximations of sets. We have employed experimental
results to illustrate the designed algorithms are effective to
calculate the second and sixth lower and upper approximations of
sets in dynamic covering information systems with the immigration of
coverings. We have explored two examples to demonstrate how to
conduct knowledge reduction of dynamic covering decision information
systems with the immigration of coverings.

In the future, we will investigate the calculation of
approximations of sets in other dynamic covering information
systems and propose effective algorithms for knowledge reduction
of dynamic covering decision information systems. Furthermore, we
will provide parallel algorithms for knowledge reduction of dynamic
covering decision information systems using the type-1 and type-2
characteristic matrices.

\section*{ Acknowledgments}

We would like to thank the anonymous reviewers very much for their
professional comments and valuable suggestions. This work is
supported by the National Natural Science Foundation of China (NO.
61273304, 61573255,11201490,11371130,11401052,11401195), the Postdoctoral Science Foundation of China (NO.2013M542558,2015M580353), the Scientific
Research Fund of Hunan Provincial Education Department(No.14C0049,15B004).


\begin{thebibliography}{00}

\bibitem{Chen1}
H.M. Chen, T.R. Li, S.J. Qiao, D. Ruan, A rough set based dynamic
maintenance approach for approximations in coarsening and refining
attribute values, International Journal of Intelligent Systems
25(10) (2010) 1005-1026.\vskip.10in

\bibitem{Chen2}
H.M. Chen, T.R. Li, D. Ruan, Maintenance of approximations in
incomplete ordered decision systems while attribute values
coarsening or refining, Knowledge-Based Systems 31 (2012)
140-161.\vskip.10in

\bibitem{Chen3}
H.M. Chen, T.R. Li, D. Ruan, J.H. Lin, C.X. Hu,  A rough-set based
incremental approach for updating approximations under dynamic
maintenance environments, IEEE Transactions on Knowledge and Data
Engineering 25(2) (2013) 174-184.\vskip.10in

\bibitem{Chen4}
D.G. Chen, X.X. Zhang, W.L. Li, On measurements of covering rough sets based on granules and evidence theory,
Information Sciences (2015), doi: http://dx.doi.org/10.1016/j.ins.2015.04.051.\vskip.10in

\bibitem{Lang1}
G.M. Lang, Q.G. Li, M.J. Cai, T. Yang, Characteristic matrices-based knowledge reduction in dynamic covering decision information systems, Knowledge-Based Systems (2015),
doi: http://dx.doi.org/10.1016/j.knosys.2015.03.021.\vskip.10in

\bibitem{Lang2}
G.M. Lang, Q.G. Li, M.J. Cai, T. Yang, Q.M. Xiao, Incremental approaches to constructing approximations of sets based on characteristic matrices, International Journal of Machine Learning and Cybernetics (2014), doi: http://dx. doi:10. 1007/s 13042-014-0315-4.\vskip.10in

\bibitem{Lang3}
G.M. Lang, Q.G. Li, L.K. Guo, Homomorphisms between covering approximation spaces, Fundamenta Informaticae 137 (2015) 351-371.\vskip.10in

\bibitem{Lang4}
G.M. Lang, Q.G. Li, L.K. Guo, Homomorphisms-based attribute reduction of dynamic fuzzy covering information systems, International Journal of General Systems 44(7-8) (2015) 791-811.\vskip.10in

\bibitem{Lang5}
G.M. Lang, Q.G. Li, T. Yang, An incremental approach to attribute reduction of dynamic set-valued information systems, International Journal of Machine Learning and Cybernetics 5 (2014) 775-788.\vskip.10in

\bibitem{Li1}
S.Y. Li, T.R. Li, D. Liu, Incremental updating approximations in
dominance-based rough sets approach under the variation of the
attribute set, Knowledge-Based Systems 40 (2013) 17-26.\vskip.10in

\bibitem{Li2}
S.Y. Li, T.R. Li, D. Liu, Dynamic maintenance of approximations in
dominance-based rough set approach under the variation of the object
set, International Journal of Intelligent Systems 28(8) (2013)
729-751.\vskip.10in

\bibitem{Li3}
T.R. Li, D. Ruan, W. Geert, J. Song, Y. Xu, A rough sets based
characteristic relation approach for dynamic attribute
generalization in data mining, Knowledge-Based Systems 20(5) (2007)
485-494.\vskip.10in

\bibitem{Li4}
T.R. Li, D. Ruan, J. Song, Dynamic maintenance of decision rules
with rough set under characteristic relation, Wireless
Communications, Networking and Mobile Computing (2007)
3713-3716.\vskip.10in

\bibitem{Liang1}
J.Y. Liang, F. Wang, C.Y. Dang, Y.H. Qian, A group incremental
approach to feature selection applying rough set technique, IEEE
Transactions on Knowledge and Data Engineering 26(2) (2014)
294-308.\vskip.10in

\bibitem{Liu1}
G.L. Liu, The axiomatization of the rough set upper approximation operations,
Fundamenta Informaticae 69(3) (2006) 331-342.\vskip.10in

\bibitem{Liu2}
D. Liu, T.R. Li, D. Ruan, J.B. Zhang, Incremental learning
optimization on knowledge discovery in dynamic business intelligent
systems, Journal of Global Optimization 51(2) (2011)
325-344.\vskip.10in

\bibitem{Liu3}
D. Liu, T.R. Li, D. Ruan, W.L. Zou, An incremental approach for
inducing knowledge from dynamic information systems, Fundamenta
Informaticae 94(2) (2009) 245-260.\vskip.10in

\bibitem{Liu4}
D. Liu, T.R. Li, J.B.Zhang, A rough set-based incremental approach
for learning knowledge in dynamic incomplete information systems,
International Journal of Approximate Reasoning 55(8) (2014)
1764-1786. \vskip.10in

\bibitem{Liu5}
D. Liu, T.R. Li, J.B. Zhang, Incremental updating approximations in probabilistic rough sets under the variation of attributes, Knowledge-based Systems 73 (2015) 81-96.\vskip.10in

\bibitem{Liu6}
D. Liu, D.C. Liang, C.C. Wang, A novel three-way decision model based on incomplete information
system, Knowledge-Based Systems (2015),
http://dx.doi.org/10.1016/j.knosys.2015.07.036. \vskip.10in

\bibitem{Liu7}
C.H. Liu, D.Q. Miao, J. Qian, On multi-granulation covering rough sets,
International Journal of Approximate Reasoning  55(6) (2014) 1404-1418.\vskip.10in

\bibitem{Luo1}
C. Luo, T.R. Li, H.M. Chen, Dynamic maintenance of approximations in
set-valued ordered decision systems under the attribute
generalization, Information Sciences 257 (2014) 210-228.\vskip.10in

\bibitem{Luo2}
C. Luo, T.R. Li, H.M. Chen, D. Liu, Incremental approaches for
updating approximations in set-valued ordered information systems,
Knowledge-Based Systems 50 (2013) 218-233.\vskip.10in

\bibitem{Luo3}
C. Luo, T.R. Li, H.M. Chen, L.X. Lu, Fast algorithms for computing rough approximations in set-valued decision systems while updating criteria values,
Information Sciences 299 (2015) 221-242.\vskip.10in

\bibitem{Miao}
D.Q. Miao, C. Gao, N. Zhang, Z.F. Zhang, Diverse reduct subspaces based co-training for partially labeled data,
International Journal of Approximate Reasoning 52(8) (2011) 1103-1117.\vskip.10in

\bibitem{Qian}
Y.H. Qian, J.Y. Liang, D.Y. Li, F. Wang, N.N. Ma, Approximation
reduction in inconsistent incomplete decision tables,
Knowledge-Based Systems 23(5) (2010) 427-433.\vskip.10in

\bibitem{Sang}
Y.L. Sang, J.Y. Liang, Y.H. Qian, Decision-theoretic rough sets under dynamic granulation,
Knowledge-Based Systems (2015), doi: http://dx.doi.org/10.1016/j.knosys.2015.08.001.\vskip.10in

\bibitem{Shu1}
W.H. Shu, H. Shen, Updating attribute reduction in incomplete
decision systems with the variation of attribute set, International
Journal of Approximate Reasoning 55(3) (2013) 867-884.\vskip.10in

\bibitem{Shu2}
W.H. Shu, H. Shen, Incremental feature selection based on rough set
in dynamic incomplete data, Pattern Recognition 47(12) (2014)
3890-3906.\vskip.10in

\bibitem{Shu3}
W.H. Shu, W.B. Qian, An incremental approach to attribute reduction from dynamic incomplete decision
systems in rough set theory,  Data and Knowledge Engineering (2015), doi:  http://dx.doi.org/10.1016/j.datak.2015.06.009.\vskip.10in

\bibitem{Tan1}
A.H. Tan, J.J. Li, Y.J. Lin, G.P. Lin,
Matrix-based set approximations and reductions in covering decision information systems,
International Journal of Approximate Reasoning 59 (2015) 68-80.\vskip.10in

\bibitem{Tan2}
A.H. Tan, J.J. Li, G.P. Lin, Y.J. Lin,
Fast approach to knowledge acquisition in covering information systems using matrix operations,
Knowledge-Based Systems 79(2015) 90-98.\vskip.10in

\bibitem{Wang1}
F. Wang, J.Y. Liang, C.Y. Dang, Attribute reduction for dynamic data
sets, Applied Soft Computing 13 (2013) 676-689. \vskip.10in

\bibitem{Wang2}
F. Wang, J.Y. Liang, Y.H. Qian, Attribute reduction: A dimension
incremental strategy, Knowledge-Based Systems 39 (2013) 95-108.
\vskip.10in

\bibitem{Wang3}
S.P. Wang, W. Zhu, Q.H. Zhu, F. Min, Characteristic matrix of
covering and its application to boolean matrix decomposition and
axiomatization, Information Sciences 263(1) (2014)
186-197.\vskip.10in

\bibitem{Yang1}
T. Yang, Q.G. Li, Reduction about approximation spaces of covering
generalized rough sets, International Journal of Approximate
Reasoning 51(3) (2010) 335-345.\vskip.10in

\bibitem{Yang2}
X.B. Yang, M. Zhang, H.L. Dou, J. Y. Yang, Neighborhood systems-based rough sets
in incomplete information system, Knowledge-Based Systems 24(6)
(2011) 858-867.\vskip.10in

\bibitem{Yang3}
X.B. Yang, Y. Qi, H.L. Yu, X.N. Song, J.Y. Yang, Updating
multigranulation rough approximations with increasing of granular
structures, Knowledge-Based Systems 64 (2014) 59-69.\vskip.10in

\bibitem{Yao1}
Y.Y. Yao, Relational interpretations of neighborhood operators and
rough set approximation operators, Information Sciences 111(1)
(1998) 239-259.\vskip.10in

\bibitem{Yao2}
Y.Y. Yao, B.X. Yao, Covering based rough set approximations, information Sciences 200 (2012) 91-107.\vskip.10in

\bibitem{Zakowski}
W. Zakowski, Approximations in the space $(u, \pi)$, Demonstratio
Mathematics 16 (1983) 761-769.\vskip.10in

\bibitem{Zhang1}
J.B. Zhang, T.R. Li, H.M. Chen, Composite rough sets for dynamic
data mining, Information Sciences 257 (2014) 81-100.\vskip.10in

\bibitem{Zhang2}
J.B. Zhang, T.R. Li, D. Ruan, D. Liu, Rough sets based matrix
approaches with dynamic attribute variation in set-valued
information systems, International Journal of Approximate Reasoning
53(4) (2012) 620-635.\vskip.10in

\bibitem{Zhang3}
J.B. Zhang, T.R. Li, D. Ruan, D. Liu, Neighborhood rough sets for
dynamic data mining, International Journal of Intelligent Systems 27(4) (2012) 317-342.\vskip.10in

\bibitem{Zhang4}
J.B. Zhang, T.R. Li, D. Ruan, Z.Z. Gao, C.B. Zhao, A parallel method for computing rough set approximations, Information Sciences 194(0) (2012) 209-223.\vskip.10in

\bibitem{Zhang5}
J.B. Zhang, J.S. Wong, T.R. Li, Y. Pan, A comparison of parallel large-scale knowledge acquisition using rough set theory on different mapreduce runtime
systems, International Journal of Approximate Reasoning 55(3) (2014) 896-907.\vskip.10in

\bibitem{Zhang6}
J.B. Zhang, J.S. Wong, Y. Pan, T.R. Li, A parallel matrix-based method for computing approximations in incomplete information systems, IEEE Transactions on Knowledge and Data Engineering 27(2) (2015) 326-339.\vskip.10in

\bibitem{Zhu1}
P. Zhu, Covering rough sets based on neighborhoods: an approach
without using neighborhoods, International Journal of Approximate
Reasoning 52(3) (2011) 461-472.\vskip.10in

\bibitem{Zhu2}
W. Zhu, Relationship among basic concepts in covering-based
rough sets, Information Sciences 179(14) (2009) 2478-2486.\vskip.10in

\bibitem{Zhu3}
W. Zhu, Relationship between generalized rough sets based on binary
relation and coverings, Information Sciences 179(3) (2009) 210-225.
\vskip.10in

\end{thebibliography}
\end{document}